\documentclass[twocolumn,prd,floats,preprintnumbers,showpacs,nofootinbib]{revtex4}  

\usepackage[english]{babel}
\usepackage{graphicx}
\usepackage{amsmath,amssymb}
\usepackage{color}

\begin{document}

\newcommand{\etal}{et al.\ }

 \newcommand{\nc}{\newcommand}
 \nc{\8}{\"{a}}
 \nc{\bea}{\begin{eqnarray}} \nc{\eea}{\end{eqnarray}}
 \nc{\beq}{\begin{equation}} \nc{\eeq}{\end{equation}}
 \nc{\bi}{\begin{itemize}}\nc{\ei}{\end{itemize}}
 \nc{\ben}{\begin{enumerate}}\nc{\een}{\end{enumerate}}
 \nc{\nn}{\nonumber}
 \nc{\exercise}{{\bf (exercise)}}
 \nc{\const}{\mbox{const.}}
 
 \nc{\G}[3]{\Gamma^#1_{#2#3}}
 \nc{\bG}[3]{\bar{\Gamma}^#1_{#2#3}}
 \nc{\dG}[3]{\delta\Gamma^#1_{#2#3}}
 \nc{\Gd}[4]{\Gamma^#1_{#2#3,#4}}
 \nc{\bGd}[4]{\bar{\Gamma}^#1_{#2#3,#4}}
 \nc{\dGd}[4]{\delta\Gamma^#1_{#2#3,#4}}
 \nc{\piG}{\pi G}

 \nc{\zth}{$0^\mathrm{th}$\ }
 \nc{\fst}{$1^\mathrm{st}$\ }
 \nc{\snd}{$2^\mathrm{nd}$\ }

 \nc{\half}{{\textstyle \frac12}}
 \nc{\third}{{\textstyle \frac13}}
 \nc{\quarter}{{\textstyle \frac14}}
 \nc{\fifth}{{\textstyle \frac15}}
 \nc{\sixth}{{\textstyle \frac16}}
 \nc{\seventh}{{\textstyle \frac17}}
 \nc{\eighth}{{\textstyle \frac18}}
 \nc{\nineth}{{\textstyle \frac19}}
 \nc{\tenth}{{\textstyle \frac{1}{10}}}
 \nc{\threehalves}{{\textstyle \frac32}} 
 \nc{\ninehalves}{{\textstyle \frac92}}
 \nc{\twothirds}{{\textstyle \frac23}}
 \nc{\fourthirds}{{\textstyle \frac43}}
 \nc{\eightthirds}{{\textstyle \frac83}}
 \nc{\tenthirds}{{\textstyle \frac{10}{3}}}
 \nc{\threequarters}{{\textstyle \frac34}}
 \nc{\ninequarters}{{\textstyle \frac94}}
 \nc{\twofifths}{{\textstyle \frac25}}
 \nc{\threefifths}{{\textstyle \frac35}}
 \nc{\sixfifths}{{\textstyle \frac65}}
 
 \nc{\pardx}[3]{\biggl(\frac{\partial#1}{\partial#2}\biggr)_#3}
 \nc{\pard}[2]{\frac{\partial#1}{\partial#2}}
 \nc{\divg}{\nabla\cdot}
 \nc{\rot}{\nabla\times}
 \nc{\curl}{\nabla\times}

 \nc{\br}{{\bm r}}
 \nc{\bx}{{\bm x}}
 \nc{\bu}{{\bm u}}
 \nc{\bv}{{\bm v}}
 \nc{\bk}{{\bm k}}
 \nc{\ki}{{{\bm k} i}}

 \nc{\vk}{{\vec{k}}}
 \nc{\vx}{{\vec{x}}}
 \nc{\vv}{{\vec{v}}}

 \nc{\bgg}{\bar{g}_{\mu\nu}}
 \nc{\bgrho}{\bar{\rho}}
 \nc{\bgp}{\bar{p}}
 \nc{\bgphi}{\bar{\varphi}}
 \nc{\delphi}{{\delta\varphi}}
 \nc{\ddelphi}{\delta\dot{\varphi}}
 \nc{\dddelphi}{\delta\ddot{\varphi}}

 \nc{\keq}{k_\mathrm{eq}}
 \nc{\kdec}{k_\mathrm{dec}}
 \nc{\aeq}{a_\mathrm{eq}}
 \nc{\adec}{a_\mathrm{dec}}
 \nc{\aend}{a_\mathrm{end}} 
 \nc{\Heq}{H_\mathrm{eq}}
 \nc{\Hdec}{H_\mathrm{dec}}
 \nc{\teq}{t_\mathrm{eq}}
 \nc{\tdec}{t_\mathrm{dec}}
 \nc{\tend}{t_\mathrm{end}} 
 \nc{\etaeq}{\eta_\mathrm{eq}}
 \nc{\Hubeq}{{\cal H}_\mathrm{eq}}
 
 \nc{\Hub}{{\cal H}}
 \nc{\R}{{\cal R}}
 \nc{\Ord}{{\cal O}}
 \nc{\Pow}{{\cal P}}
 \nc{\Cov}{{\cal C}}
 \nc{\Se}{{\cal S}}
 \nc{\La}{{\cal L}}
 \nc{\Vol}{{\cal V}}
\newcommand{\h}{\mathcal{H}} 

 \nc{\koHub}{\left(\frac{k}{\Hub}\right)}
 \nc{\koHubeq}{\left(\frac{k}{\Hubeq}\right)}

 \nc{\M}{M_\mathrm{Pl}}
 \nc{\V}{V(\varphi)}
 \nc{\GeV}{\mbox{ GeV}}
 
 \nc{\ad}{{\hat{a}^\dagger}}
 \nc{\an}{{\hat{a}}}
 \nc{\vac}{{|0\rangle}}
 \nc{\vacd}{{\langle 0|}}
 \nc{\ophi}{{\hat{\varphi}}}
 \nc{\odelphi}{{\delta\hat{\varphi}}}
 
 \nc{\qand}{\quad\mbox{and}\quad}
 \nc{\qqand}{\qquad\mbox{and}\qquad}
 \nc{\qor}{\quad\mbox{or}\quad}
 \nc{\qqor}{\qquad\mbox{or}\qquad}
 \nc{\So}{\quad\Rightarrow\quad}
 
 \nc{\vfi}{\varphi}
 \nc{\bfi}{\bar{\varphi}}
 \nc{\delfi}{{\delta\varphi}}
 \nc{\veps}{\varepsilon}
 \nc{\dfi}{\dot{\varphi}}
 \nc{\ddfi}{\ddot{\varphi}}
 \nc{\dsigma}{\dot{\sigma}}
 \nc{\ddsigma}{\ddot{\sigma}} 
 \nc{\dtheta}{\dot{\theta}}
 \nc{\ddtheta}{\ddot{\theta}}
 \nc{\dchi}{\dot{\chi}}
 \nc{\ddchi}{\ddot{\chi}}

 \nc{\Vz}{V_{\sigma}}
 \nc{\Vzz}{V_{\sigma\sigma}}
 \nc{\Vzs}{V_{\sigma s}}
 \nc{\Vss}{V_{ss}} 
 \nc{\etzz}{\eta_{\sigma\sigma}}
 \nc{\etzs}{\eta_{\sigma s}}
 \nc{\etss}{\eta_{ss}}
 \nc{\xizzs}{\xi_{\sigma\sigma s}}
 \nc{\delsigma}{{\delta\sigma}}
 \nc{\dels}{\delta s}

 \nc{\snt}{\sin\theta}
 \nc{\cst}{\cos\theta}
 \nc{\tnt}{\tan\theta}
 
 \nc{\sgn}{\mbox{sign}}
 
\newcommand{\mr}[1]{\mathrm{#1}}
\newcommand{\mtc}[1]{\mathcal{#1}}
\newcommand{\mtr}[1]{\mathrm{#1}}
\newcommand{\mtb}[1]{\mathbf{#1}}
\newcommand{\nad}{n_{\mr{ad}}}
\newcommand{\nadI}{n_{\mr{ar}}}
\newcommand{\nadII}{n_{\mr{as}}}
\newcommand{\niso}{n_{\mr{iso}}}
\newcommand{\ncor}{n_{\mr{cor}}}
\newcommand{\acor}{\alpha_{\mr{cor}}}
\newcommand{\abs}[1]{\vert #1 \vert}
\newcommand{\comment}[1]{}
\nc{\Mpci}{{\mbox{Mpc}^{-1}}}

\definecolor{darkbrown}{rgb}{0.5,0.1,0.8}
\newcommand{\jvc}[1]{{{\textcolor{blue}{#1}}}}
\newcommand{\msc}[1]{{\bf{\textcolor{red}{#1}}}}
\newcommand{\jvco}[1]{{\bf{\textcolor{darkbrown}{JV: #1}}}}
\newcommand{\hxc}[1]{{\bf{\textcolor{blue}{HX: #1}}}}

\newcommand{\new}[1]{{\textcolor{red}{#1}}}

\title{Constraints on scalar and tensor perturbations in phenomenological and two-field inflation models: Bayesian evidences for primordial isocurvature and tensor modes}

\author{Jussi V\"{a}liviita}
\email{jussi.valiviita@astro.uio.no}
\affiliation{Institute of Theoretical Astrophysics, University of Oslo, P.O.
Box 1029 Blindern, N-0315 Oslo, Norway}

\author{Matti Savelainen}

\author{Marianne Talvitie}

\author{Hannu Kurki-Suonio}
\email{hannu.kurki-suonio@helsinki.fi}

\author{Stanislav Rusak}

\affiliation{Department of Physics and Helsinki Institute of Physics, University of Helsinki,
P.O. Box 64, FIN-00014 University of Helsinki, Finland}

\date{13th February 2012}

\begin{abstract}
We constrain cosmological models where the primordial perturbations have both an adiabatic and a (possibly correlated) cold dark matter (CDM) or baryon isocurvature component. We use both a phenomenological approach, where the power spectra of primordial perturbations are parametrized with amplitudes and spectral indices, and a slow-roll two-field inflation approach where slow-roll parameters are used as primary parameters determining the spectral indices  and the tensor-to-scalar ratio. In the phenomenological case, with cosmic microwave background (CMB) data, the upper limit to the CDM isocurvature fraction $\alpha$ is 6.4\% at $k = 0.002$\ Mpc$^{-1}$ and 15.4\% at $k = 0.01$\ Mpc$^{-1}$. At smaller scales (larger $k$) larger isocurvature fractions are allowed, and therefore large values of the isocurvature spectral index, $\niso\approx 2$, are formally favored. The median 95\% range for the non-adiabatic contribution to the CMB temperature variance is $-0.030 < \alpha_T < 0.049$. Including the supernova (or large-scale structure, LSS) data, these limits become: $\alpha <$ 7.0\%, 13.7\%, and $-0.048 < \alpha_T < 0.042$ (or $\alpha<$ 10.2\%, 16.0\%, and  $-0.071 < \alpha_T < 0.024$). The CMB constraint on the tensor-to-scalar ratio, $r\lesssim 0.26$ at $k = 0.01$\ Mpc$^{-1}$, is not affected by the nonadiabatic modes. In the slow-roll two-field inflation approach, the spectral indices are constrained close to $1$. This leads to tighter limits on the isocurvature fraction, with the CMB data $\alpha < 2.6\%$ at $k = 0.01$\ Mpc$^{-1}$, but since the non-adiabatic contribution to the CMB temperature variance comes mostly from larger scales its median 95\% range is not much affected, $-0.058 < \alpha_T < 0.045$. Including supernova (or LSS) data, these limits become: $\alpha<$ 3.2\% and $-0.056 < \alpha_T < 0.030$ (or $\alpha<$ 3.4\% and $-0.063 < \alpha_T < -0.008$). When all spectral indices are close to each other the isocurvature fraction is somewhat degenerate with the tensor-to-scalar ratio. In addition to the generally correlated models, we study also special cases where the adiabatic and isocurvature modes are uncorrelated or fully (anti)correlated. We calculate Bayesian evidences (model probabilities) in 21 different cases for our nonadiabatic models and for the corresponding adiabatic models, and find that in all cases the current data support the pure adiabatic model. 
\end{abstract}

\preprint{HIP-2012-03/TH}

\maketitle

%
%

\section{Introduction}
\label{sec:intro}

The nature of primordial perturbations is the key observation window to the physics of the very early Universe and the 
very-high-energy physics that is beyond the reach of particle accelerators. 
The simplest model for the generation of primordial perturbations, 
single-field slow-roll inflation, predicts adiabatic Gaussian nearly scale-invariant primordial scalar perturbations,
and tensor perturbations at a level that may or may not be observable, depending on the detailed model.
Current cosmological data are consistent with this kind of perturbations, with
so far no evidence for primordial tensor perturbations.
There is some evidence of a small deviation from scale invariance.    More complicated models, like
multi-field inflation, may give rise to deviations from adiabaticity or Gaussianity that improved observations may be able to detect.

In this paper we study what information currently available data on the cosmic microwave background \cite{Larson:2010gs, Reichardt:2008ay, Brown:2009uy}
give on possible deviations from adiabaticity, and check how the results change if we add supernova \cite{Amanullah:2010vv} or large-scale structure \cite{Reid:2009xm} (that includes baryon acoustic oscillations, BAO \cite{Percival:2009xn}) data into the analysis.
While adiabatic primordial perturbations are fully determined by one physical perturbation quantity, which we may take to be the curvature perturbation in the comoving gauge, $\R$, deviations from adiabaticity represent additional, isocurvature, degrees of freedom that can be given in terms of entropy perturbations, which represent 
how perturbations of different particle species differ from each other. For example, the primordial cold dark matter (CDM) entropy perturbation can be written as
\beq
S_{c\gamma} = \frac{\delta(n_c/n_\gamma)}{n_c/n_\gamma} = \frac{\delta n_c}{n_c} - \frac{\delta n_\gamma}{n_\gamma}\,,
\label{eq:Seaydef}
\eeq
where $n_c$ and $n_\gamma$ are the number densities of CDM particles and photons at early time, deep in the radiation dominated era. 
This kind of perturbations do not (initially) perturb the spatial curvature of comoving slices ($\R = 0$), hence the name ``isocurvature perturbations''. 
There is so far no evidence for an isocurvature perturbation component, and observations require it to be subdominant compared to the adiabatic component 
\cite{Beltran:2005xd,Beltran:2005gr,Keskitalo:2006qv,Trotta:2006ww,Seljak:2006bg,Lewis:2006ma,Bean:2006qz,Kawasaki:2007mb,Beltran:2008ei,Sollom:2009vd,Valiviita:2009bp,Castro:2009ej,Li:2010yb}. Even if no isocurvature component is found, improving upper limits to it will help to constrain the particle theory related to the origin of perturbations \cite{LiddleAndLythBook}. 

Out of four possible isocurvature modes \cite{Bucher:2000kb,Bucher:1999re}, we focus here on cold dark matter (CDM) and baryon isocurvature perturbations, which are observationally indistinguishable from each other (see however  \cite{Kawasaki:2011ze} for the prospects of using the Hydrogen 21 cm lines), and do not consider the less well motivated  neutrino isocurvature perturbations \cite{Kasanda:2011np}.
In our earlier work \cite{Valiviita:2009bp,Keskitalo:2006qv,KurkiSuonio:2004mn} we assumed that tensor perturbations were negligible.  Now we allow for the presence of tensor perturbations. 
We assume a power-law power spectrum for the primordial curvature and entropy perturbations and allow for correlations between them. We also study special cases with no correlation or with $\pm$100\% correlation.

Assuming spatially flat geometry of the Universe, we perform full parameter scans of this \emph{mixed} (adiabatic and isocurvature) model, as well as the standard pure adiabatic $\Lambda$CDM model for comparison.
We present posterior probability densities of the standard cosmological parameters and the extra isocurvature parameters, and report the \emph{Bayesian evidences} for these models calculated with the  MultiNest nested sampling package \cite{Feroz:2008xx}, using the CMB data alone or CMB and supernova data or CMB and large-scale structure data.

We use two different approaches: 1) A phenomenological approach, where we make no reference to the origin of the primordial perturbations, and just try to determine or constrain their amplitudes and spectral indices from the data. 
2) A slow-roll two-field inflation approach, where we assume the perturbations were generated by quantum fluctuations during two-field inflation, and the spectral indices are determined by the slow-roll parameters at the time the cosmological scales exited horizon during inflation.

\section{Curvature and entropy (isocurvature) perturbations}

In this section we define the curvature and isocurvature perturbations, as well as our notation and sign convention for them.

In general, the metric perturbed by scalar perturbations is given by the line element
\bea
\label{eq:gengauge} ds^2 & = & a^2\Big\{ -(1+2\phi)
d\tau^2+2\partial_iB\,d\tau dx^i \nn\\ 
& & + \Big[(1-2\psi)\delta_{ij}+
2\partial_i\partial_j E\Big]dx^idx^j\Big\}\,.
\eea 
From the above metric perturbations one can construct gauge-invariant Bardeen
potentials \cite{Bardeen:1980kt}:
\bea
\label{eq.PhiandPsi}
\Phi & \equiv & \phi + \h \left( B-E' \right) + \left( B-E'\right)'\,,\\
\Psi & \equiv & \psi - \h(B-E')\,,
\eea
where $\h = a'/a$ is the conformal Hubble parameter, $a$ is the
scale factor of the Universe, and prime denotes a derivative with respect
to conformal time $\tau$.
In the longitudinal (Newtonian) gauge, where $B = E = 0$, the Bardeen potentials match the Newtonian metric perturbations.

The comoving gauge is defined by requiring both comoving slicing, $B=0$, and comoving threading, $v = 0$, where $v$ is the fluid velocity potential. We denote the comoving gauge by superscript $C$, and define the comoving curvature perturbation $\mtc{R}$ by
\beq
\mtc{R} \equiv -\psi^C = -\psi^G - \h(v^G - B^G).
\label{eq:Rdef}
\eeq
Here we used the gauge transformation properties of $\psi$ in order to relate $\psi^C$ to the perturbation quantities in any other gauge $G$. 

Our sign convention for $\mtc{R}$ is such that a positive $\mtc{R}$ corresponds to positive curvature of the 3-dimensional $\tau$ = constant slice.
If we go to a spatially flat gauge (superscript $F$) where $\psi^F = E^F = 0$, we find
\beq
\mtc{R} = \frac{1}{3(1+w)} \left[ \frac{\delta\rho^F}{\rho} + \frac{2}{3} \left(\frac{k}{\h} \right)^2 \Psi \right]\,,
\eeq
where $\rho$ is the background (average, unperturbed) energy density, and $w=p/\rho$ with $p$ the background pressure. A positive $\mtc{R}$ corresponds to an overdensity, $\delta\rho^F>0$, on large scales $k/\h \ll 1$.

Finally, the definition (\ref{eq:Rdef}) can be expressed in various forms in terms of the gauge invariant Bardeen potentials:
\bea
\mtc{R} & = & -\Psi - \frac{2}{3(1+w)}\left( \h^{-1}\Psi' + \Phi \right)\\
& = & -\Psi + \frac{\h}{\h'}\left( \Psi' + \h\Phi \right)\,.
\eea
It is important to note that our sign convention is opposite to, e.g., \cite{Gordon:2000hv,Gordon:2001ph,Byrnes:2006fr}, and whether $\psi$/$\phi$ appear in the time or space part of the metric and what sign they have in Eq. (\ref{eq:gengauge}) varies a lot in the literature. 
Furthermore, often the definitions of $\Psi$ and $\Phi$ are swapped and also their sign convention varies. Our sign convention matches with our previous works \cite{KurkiSuonio:2004mn,Keskitalo:2006qv,Valiviita:2009bp} and, e.g., with \cite{Li:2010yb}. (Ref.~\cite{LiddleAndLythBook} uses our sign convention for $\R$, but has $\Psi$ and $\Phi$ swapped.)

We define the total entropy perturbation by
\beq
\mtc{S} \equiv \h\left( \frac{\delta p}{p'} - \frac{\delta\rho}{\rho'}\right)\,,
\eeq
where $p'$ and $\rho'$ are the conformal time derivatives of the background pressure and energy density, respectively. Further, we define an entropy perturbation between particle species $x$ and $y$ by
\beq
S_{xy} \equiv -3\h\left( \frac{\delta\rho_x}{\rho'_x} - \frac{\delta\rho_y}{\rho'_y}\right)\,.
\label{eq:Sxy}
\eeq
Using the continuity equation, this can be written as $S_{xy} = \delta_x / (1+w_x) - \delta_y / (1+w_y)$, where $\delta = \delta\rho / \rho$. This leads easily to Eq.~(\ref{eq:Seaydef}) for CDM and photons. Both $\mtc{S}$ and $S_{xy}$ are gauge invariants, and hence we did not need to specify the gauge above.

It can be shown that on super-Hubble scales ($k/\h \ll 1$) $\h^{-1}\mtc{R}' \propto \mtc{S}$ \cite{GarciaBellido:1995qq,Wands:2000dp,Gordon:2000hv,Gordon:2001ph,Amendola:2001ni}. 
Therefore, in the absence of entropy perturbation, the comoving curvature perturbation remains constant on super-Hubble scales. 
In the contrary, $\h^{-1} \mtc{S}' \propto \mtc{S}$, and hence the total entropy perturbation evolves with time even on super-Hubble scales. 
Moreover, these differential equations imply that (part of) $\mtc{S}$ can be ``converted'' into $\mtc{R}$, but $\mtc{R}$ cannot be converted into $\mtc{S}$, and even if $\mtc{R}$ and $\mtc{S}$ were initially (e.g. at horizon exit during inflation) uncorrelated, 
the later evolution can lead to a correlation between them: there will be a part of $\mtc{R}$ that has been ``created'' from the initial $\mtc{S}$, and hence these are fully correlated or anti-correlated \cite{Amendola:2001ni}.

If we assume that the Universe consists of photons ($\gamma$), neutrinos ($\nu$), baryons ($b$), and cold dark matter ($c$), and that the entropy perturbation between neutrinos and photons vanishes, then the total entropy perturbation during radiation domination can be written as
\beq
\mtc{S} = \frac{a}{4a_{eq} + 3a} \left( f_c S_{c\gamma} + f_b S_{b\gamma} \right)\,,
\label{eq:totalS}
\eeq
where $a_{eq}$ is the scale factor at radiation-matter equality, $f_c = \rho_c/(\rho_c + \rho_b)$, and $f_b = \rho_b/(\rho_c + \rho_b)$.

In this paper, we perform all the isocurvature analysis by specifying a non-zero primordial $S_{c\gamma}$, i.e., a primordial cold dark matter isocurvature mode, while keeping all the other relative entropy perturbations $S_{xy}$ zero.
From now on we shorten our notation, and denote
\beq
S\equiv S_{c\gamma}
\eeq
at primordial time $t_{\rm rad}$, deep in the radiation dominated era, on super-Hubble scales.

Finally, 
we comment once more on our sign convention. The ordinary Sachs-Wolfe (SW) effect in the CMB temperature anisotropy, in the direction $\mathbf{n}$ at the sky, is
\beq
\frac{\delta T}{T}(\mathbf{n}) = 
-\frac{1}{5} \left[ \mtc{R}(t_{\rm rad}, \mathbf{n}) + 2f_c S(t_{\rm rad}, \mathbf{n})\right]\,.
\eeq
(Note that we have $-1/5\mtc{R}$ for the adiabatic part whereas equation (3) in \cite{Amendola:2001ni} shows $+1/5\mtc{R}$ due to their different sign convention for $\mtc{R}$.)
Note that $\mtc{R}$ and $S$ appear with the same sign in this expression. Hence, a positive primordial correlation, $\mtc{C}_{\mtc{R}S} > 0$, 
leads to extra power at low multipoles in the CMB temperature angular power spectrum, $C_\ell \propto \langle |\delta T/T|^2 \rangle$, compared to the uncorrelated case. 
Vice versa, a negative primordial correlation leads to a negative contribution to the SW effect, and hence reduces the angular power at low multipoles.

\section{Mechanisms that produce isocurvature}
\label{sec:mechanisms}

\vspace{-5mm}
There are various mechanisms that can produce (correlated) isocurvature and adiabatic perturbations. Compared to the single-field slow-roll inflation which can stimulate only the adiabatic perturbation mode, at least one extra degree of freedom is needed. Inflationary models with multiple (scalar) fields are a natural extension. 
Production of isocurvature perturbations
in this type of models has been studied, e.g., in  Refs.
\cite{Choi:2008et,Hamazaki:2007eq,DiMarco:2007pb,Lalak:2007vi,Choi:2007su,Byrnes:2006fr,Rigopoulos:2005us,Bassett:2005xm,Hattori:2005ac,DiMarco:2005nq,Parkinson:2004yx,Gruzinov:2004ad,Bartolo:2003ad,Vernizzi:2003vs,Lee:2003ed,vanTent:2003mn,Mazumdar:2003iy,Malik:2002jb,DiMarco:2002eb,Ashcroft:2002vj,Bernardeau:2002jy,Wands:2002bn,Tsujikawa:2002nf,Starobinsky:2001xq,Bartolo:2001rt,GrootNibbelink:2001qt,Bartolo:2001vw,Hwang:2001fb,Hwang:2000jh,Gordon:2000hv,Taylor:2000ze,Finelli:2000ya,Liddle:1999pr,Bassett:1999ta,Pierpaoli:1999zj,Langlois:1999dw,Felder:1999pv,Perrotta:1998vf,Chiba:1997ij,Nakamura:1996da},
some references going to a non-linear level \cite{Langlois:2006vv,Rigopoulos:2002mc,Tsujikawa:2002qx}.
Many of these references derive so called consistency relations (there exists actually an infinite hierarchy of them \cite{Lidsey:1995np,Cortes:2006ap}), or analytical expressions for the spectral tilts and amplitudes of adiabatic and isocurvature components in specific inflationary set-ups. We will use the notation and the leading order results of \cite{Byrnes:2006fr} for two-field slow-roll inflation.

Various types of axions carry isocurvature perturbations  \cite{Hertzberg:2008wr,Kawasaki:2008jc,Sikivie:2006ni,Beltran:2006sq,Lazarides:2006jw,Dine:2004cq,Bozza:2002fp,Vernizzi:2000vc,Kanazawa:1998nk,Kanazawa:1998pa,Kasuya:1997td,Kawasaki:1997ct,Kasuya:1996ns,Kawasaki:1995ta}, and it is possible to construct axion models which lead to a large isocurvature spectral index, $\niso \sim 2$---$4$, \cite{Kasuya:2009up}. (These values are formally preferred in our phenomenological approach.) Most axion-dilaton-moduli models give rise to isocurvature perturbations \cite{Copeland:1997nw}. Moduli fields and isocurvature have been studied in \cite{Lemoine:2009is,Lemoine:2009yu,Yang:2008ns,Lalak:2007aj}. One solid candidate for producing correlated mixtures of adiabatic and isocurvature perturbations is the curvaton model \cite{Lyth:2001nq,Enqvist:2001zp}, or curvaton-type models, such as late-decaying scalar condensations \cite{Moroi:2001ct}. Isocurvature perturbations in these models have been discussed in Refs. \cite{Lemoine:2008qj,Matsuda:2007ax,Choi:2007fya,Lemoine:2006sc,Lazarides:2005ek,Lazarides:2004we,Moroi:2004rs,Ferrer:2004nv,Chun:2004gx,Gupta:2003jc,BasteroGil:2003tj,Hamaguchi:2003dc,Lyth:2003ip,Dimopoulos:2003ii,Moroi:2002vx,BasteroGil:2002xr,Lyth:2002my,Bartolo:2002vf,Moroi:2002rd}. A generic feature of (minimally) supersymmetric (standard) model is the existence of extra fields that can carry isocurvature perturbations \cite{McDonald:2008rv,McDonald:2006if,Feldstein:2004xi,Enqvist:2003gh}. Refs. \cite{Kasuya:2008xp,Kamada:2008sv,Charng:2008ke,Kawasaki:2001in,Enqvist:1999hv,Enqvist:1998pf,Koyama:1998hk} focus on the Affleck-Dine mechanism which typically leads to isocurvature perturbations. In brane models isocurvature also arises naturally \cite{Contaldi:2008hr,Copeland:2006hv,Burgess:2004kv,Koyama:2003be,Koyama:2001ct} as well as in the ekpyrotic scenario \cite{Koyama:2007mg,Notari:2002yc}.

Dark energy isocurvature perturbations have been studied in \cite{Gordon:2004ez}. In some models with interaction between CDM and dark energy, an isocurvature mode is rapidly growing in some regions of the parameter space overtaking the usual adiabatic mode, and hence these regions can be ruled out \cite{Valiviita:2008iv,Majerotto:2009np,Valiviita:2009nu,Clemson:2011an}. Isocurvature in quintessence cosmologies has been studied in \cite{Moroi:2003pq,Li:2001st,Hwang:2001uaa,Kawasaki:2001nx,Kawasaki:2001bq,Abramo:2001mv}. Large-scale magnetic fields can also induce a magnetic isocurvature component \cite{Giovannini:2006gz,Tsagas:1999ft,Giovannini:1997gp}. Finally, cosmic strings and other topological defects generically create isocurvature perturbations, but in this case they may not be ``primordial'' but instead continuously created during the evolution of Universe \cite{Takahashi:2006yc,Battye:1998xe,Deruelle:1997py,Durrer:1997rh,Hu:1996yt}. In order to match cosmic string models with observations one needs another mechanism to produce the predominant adiabatic component: in \cite{Bevis:2007gh} a model with a mixture of adiabatic perturbations from inflaton, and isocurvature perturbations from strings was observationally constrained.



After introducing our model and its parametrization in the next section, we give a few specific examples in Sec.~\ref{sec:TheoryExamples}.

\section{The model}

The general perturbation, here presented in Fourier space, is a superposition of adiabatic and isocurvature perturbations, $\mtc{R}(\mathbf{k})$ and $S(\mathbf{k})$. Its power spectrum
$\Pow =  \Pow_{\mtc{R}}  +  \mtc{C}_{\mtc{R}S}  +   \mtc{C}_{S\mtc{R}} + \Pow_{S}$, where $\mtc{C}$ represents the correlation, is defined by the expectation value
\begin{multline}
\Big\langle [\mtc{R}(\mathbf{k}) + S(\mathbf{k})]^\ast [\mtc{R}(\mathbf{\tilde k}) + S(\mathbf{\tilde k})] \Big\rangle \equiv (2\pi)^3 \delta^{(3)}(\mathbf{k}-\mathbf{\tilde{k}}) \times \\
\frac{2\pi^2}{k^3}\left[ \Pow_{\mtc{R}}(k) + \mtc{C}_{\mtc{R}S}(k) +  \mtc{C}_{S\mtc{R}}(k) + \Pow_{S}(k)  \right]\,.
\end{multline}
Note that we used a calligraphic letter ${\cal P}$, while $P$ is reserved for $P(k) \equiv \frac{2\pi^2}{k^3} {\cal P}(k)$.

\begin{table*}
\begin{tabular}{|l|l|r|}
\hline
Parameter & Explanation & range (min, max)\\
\hline
 & \multicolumn{1}{c|}{Primary background parameters (common for both parametrizations)} & \\
\hline
$\omega_b$ & physical baryon density; $\omega_b = h^2\Omega_b$ & (0.010, 0.050)\\
$\omega_c$ & physical cold dark matter density; $\omega_c = h^2\Omega_c$ & (0.02, 0.30)\\
$100\theta$ & $\theta$ is the sound horizon angle; $\theta = r_s(z_\ast)/D_A(z_\ast)$ & (0.5, 2.2)\\
 $\tau$ & optical depth to reionization & (0.02, 0.30)\\
\hline
\hline
 & \multicolumn{1}{c|}{Primary perturbation parameters in {\bf amplitude parametrization}} & \\
\hline
$\ln (10^{10} A_1^2)$ & $A_1^2$ is the overall primordial perturbation power at $k = k_1 = 0.002\,$Mpc$^{-1}$ &  (1.0, 7.0) \\
$\ln (10^{10} A_2^2)$ & $A_2^2$ is the overall primordial perturbation power at $k = k_2 = 0.05\,$Mpc$^{-1}$ & (1.0, 7.0)\\
$\gamma_1$ & primordial ratio of correlated adiabatic component to total adiabatic power at $k = k_1$ & (-1.0, 1.0) \\
$|\gamma_2|$ & primordial ratio of correlated adiabatic component to total adiabatic power at $k = k_2$ &  (0, 1.0) \\
$\alpha_1$ & primordial isocurvature fraction at $k = k_1$; $A_1^2\alpha_1$ = primord. isocurvature power at $k_1$ &  (0, 1.0) \\
$\alpha_2$ & primordial isocurvature fraction at $k = k_2$; $A_2^2\alpha_2$ = primord. isocurvature power at $k_2$ &  (0, 1.0)  \\
($\ \ r_0\ \ $)   & primordial ratio of tensor and scalar power spectra at  $k = k_0 = 0.01\,$Mpc$^{-1}$ & (0, 0.75)\\
$A_{SZ}$ & amplitude of the Sunyaev-Zel'dovich (SZ) template & (0, 2) \\
\hline
 & \multicolumn{1}{c|}{Primary perturbation parameters in {\bf slow-roll parametrization}} & \\
\hline
$\etzz$ & slow-roll parameter (when scale $k = k_0 = 0.01\,$Mpc$^{-1}$ exited the horizon); $\etzz=\textstyle \frac{1}{8\pi G}\frac{\partial_\sigma\partial_\sigma V}{V}$ & (-0.075, 0.075)\\
$\etzs$ & slow-roll parameter (when scale $k = k_0$ exited the horizon); $\etzs=\textstyle \frac{1}{8\pi G}\frac{\partial_\sigma\partial_s V}{V}$, see Ref.~\cite{Byrnes:2006fr}  & (-0.075, 0.075)\\
$\etss$ & slow-roll parameter (when scale $k = k_0$ exited the horizon); $\etss=\textstyle \frac{1}{8\pi G}\frac{\partial_s\partial_s V}{V}$, see Ref.~\cite{Byrnes:2006fr} & (-0.075, 0.075)\\
($\ \ \veps\ \ $) & slow-roll parameter (when scale $k = k_0$ exited the horizon); $\veps=\textstyle \frac{1}{16\pi G}(\frac{\partial_\sigma V}{V})^2$, see Ref.~\cite{Byrnes:2006fr} & (0, 0.075)\\
$\ln (10^{10} A_0^2)$ & $A_0^2$ is the overall primordial perturbation power at $k = k_0$ &  (1.0, 7.0) \\
$\gamma_0$ &  primordial ratio of correlated adiabatic component to total adiabatic power at $k = k_0$ & (-1.0, 1.0) \\
& [$\gamma_0 \equiv \sgn(\cos\Delta_0)\cos^2\Delta_0$, where $\Delta_0$ is the ``correlation angle'' at horizon exit of scale $k_0$] & \\
$\alpha_0$ &  primordial isocurvature fraction at $k = k_0$; $A_0^2\alpha_0$ = primord. isocurvature power at $k_0$ & (0, 1.0) \\
$A_{SZ}$ & amplitude of the SZ template & (0, 2) \\
\hline
\hline
 & \multicolumn{1}{c|}{Derived parameters} & \\
\hline
$H_0$ & Hubble parameter [km$\,$s$^{-1}$Mpc$^{-1}$]; calculated from $\omega_b$, $\omega_c$, and $\theta$ & tophat (40, 100) \\
$h$ & $h=H_0/(100\,\mbox{km$\,$s$^{-1}$Mpc$^{-1}$})$ & (0.40, 1.00) \\
$\Omega_m$ & matter density parameter; $\Omega_m = (\omega_b + \omega_c)/h^2$ & \\
$\Omega_\Lambda$ & vacuum energy density parameter; $\Omega_\Lambda = 1 - \Omega_m$ & \\ 
$\nadI$ & spectral index of primordial uncorrelated adiabatic part; $\nadI - 1 = d\ln({\cal P}_{\mr{ar}})/d\ln k|_{k=k_0}$ & \\
$\nadII$ & spectral index of primordial correlated adiabatic part; $\nadII - 1 = d\ln({\cal P}_{\mr{as}})/d\ln k|_{k=k_0}$ & \\
$\niso$ & spectral index of primordial isocurvature part; $\niso - 1 = d\ln({\cal P}_{S})/d\ln k|_{k=k_0}$ & \\
$n_{\mathrm{ad}}^{\mathrm{eff}}$ & effective single adiabatic spectral index at  $k = k_0 = 0.01\,$Mpc$^{-1}$, Eq.~(\ref{eqn:nadeffk}) & \\
$\alpha_{\mr{cor0}}$ & $\alpha_{\mr{cor0}} = \mathrm{sign} (\gamma_0) \sqrt {\alpha_0 (1- \alpha_0) |\gamma_0|}$; $A_0^2 \alpha_{\mr{cor0}} = {\mtc C}_{\mtc{RS}}(k_0) = {\mtc C}_{\mtc{SR}}(k_0)$ = primord.\ correlation ampl. & \\ 
$\alpha_T $ & total non-adiabatic contribution to the CMB temperature variance, Eq.~(\ref{eq:alphaT})
&\\
\hline
\end{tabular}
\caption{{\bf Parameters and parametrizations.} 
Our 11 (12 with primordial tensor perturbations) primary nested sampling parameters in two different parametrizations (amplitude and slow-roll) and a selection of derived parameters. 
In addition, in the amplitude parametrization, parameters $A_0$, $\gamma_0$, and $\alpha_0$ can be derived from $A_i$, $\gamma_i$, and $\alpha_i$ ($i=1,2$). 
In the slow-roll parametrization, $r$ is a derived parameter: $r_0= 16\veps(1-|\gamma_0|)$.
\label{tab:parameters}}
\end{table*}

Following \cite{Valiviita:2003ty,KurkiSuonio:2004mn,Keskitalo:2006qv,Kawasaki:2007mb,Valiviita:2009bp} we divide the comoving curvature perturbations into
an uncorrelated part (``$\mr{ar}$''), and a part fully correlated with the entropy perturbation (``$\mr{as}$''), and assume power-law forms for the spectra:
 \beq
 	\Pow_\R(k) = \Pow_{\mr{ar}}(k) + \Pow_{\mr{as}}(k) \,,
 \eeq
where
 \bea
 	\Pow_{\mr{ar}}(k) & = & A_{r0}^2\left( k/k_0 \right)^{\nadI-1} \,,\nn\\
	\Pow_{\mr{as}}(k) & = & A_{s0}^2\left( k/k_0 \right)^{\nadII-1} \,,
 \eea
and
 \beq
 	\Pow_S(k) = B_0^2\left( k/k_0 \right)^{\niso-1} \,.
 \eeq
The covariance is given by
 \beq
 	\Cov_{\R S}(k) = \Cov_{S \R}(k) = A_{s0}B_0\left( k/k_0 \right)^{\ncor-1} \,,
 \eeq
where
 \beq
 \ncor = (\nadII+\niso) / 2\,.
 \eeq
In general, we denote $A_{ri}^2 \equiv \Pow_{\mr{ar}}(k_i)$, $A_{si}^2 \equiv \Pow_{\mr{as}}(k_i)$,
and $B_i^2 \equiv \Pow_S(k_i)$.

For fitting the different models to the data, we choose our reference scales as
 \bea
 	k_1 & = & 0.002\, \mbox{Mpc}^{-1} \nn\\ 
	k_0 & = & 0.010\, \mbox{Mpc}^{-1} \nn\\ 
	k_2 & = & 0.050\, \mbox{Mpc}^{-1} \,.
 \eea
 
We further define
 \beq
 	A_i^2 \equiv A_{ri}^2+A_{si}^2+B_i^2 \,,\qquad 
	\alpha_i \equiv B_i^2 / A_i^2 
 \eeq
 and
 \beq
 	\gamma_i \equiv \sgn(A_{si}B_i)\frac{A_{si}^2}{A_{ri}^2+A_{si}^2} \,,
 \label{eq:a2frac}
 \eeq
so that
 \bea
	A_{ri}^2 & = & (1-|\gamma_i|)(A_{ri}^2+A_{si}^2) \ = \ (1-|\gamma_i|)(1-\alpha_i)A_i^2 \nn\\
 	A_{si}^2 & = & |\gamma_i|(A_{ri}^2+A_{si}^2) \ = \ |\gamma_i|(1-\alpha_i)A_i^2 \nn\\
	B_i^2 & = & \alpha_i A_i^2 \nn\\
        A_{si}B_i & = &  \alpha_{\mr{cor}i} A_i^2 =  {\mtc C}_{\mtc{R}S}(k_i) = {\mtc C}_{S\mtc{R}}(k_i)\,,
 \label{eq:frac2a}
 \eea
where we defined the relative amplitude of the primordial correlation between the adiabatic and CDM isocurvature perturbations by
\beq
\alpha_{\mr{cor}i} \equiv  \mathrm{sign} (\gamma_i) \sqrt {\alpha_i (1- \alpha_i) |\gamma_i|}\,.
\eeq
The total CMB temperature angular power spectrum can be written as
\begin{align}
  \label{eq:totCl}
  C_{\ell} &= A_0^{2} \bigl[ (1-\alpha_0)(1-\abs{\gamma_0})\hat{C}^{\mr{ar}}_{\ell} +
  (1-\alpha_0)\abs{\gamma_0}\hat{C}^{\mr{as}}_{\ell} \nonumber \\
  &\quad + \alpha_0 \hat{C}^{\mr{iso}}_{\ell} +
\alpha_{\mr{cor0}} \hat{C}^{\mr{cor}}_{\ell} 
+ (1-\alpha_0)r_0\hat{C}^{T}_{\ell} \bigr] \nonumber \\
  & \equiv
  C^{\mr{ar}}_{\ell} + C^{\mr{as}}_{\ell} + C^{\mr{iso}}_{\ell} + C^{\mr{cor}}_{\ell} + C^{T}_{\ell}
  \,,
\end{align}
where the $\hat{C}_{\ell}$ represent the different contributions to the 
angular power spectrum that would result from a corresponding primordial spectrum with a unit amplitude at the pivot scale $k=k_0$ (see \cite{KurkiSuonio:2004mn}). The last term, $C^{T}_{\ell}$, is the possible contribution from primordial tensor perturbations, see the next subsection.

From Eq.~(\ref{eq:totalS}) we notice that in order to lead to the same observational effects we would need a primordial baryon isocurvature perturbation that is $S_{b\gamma} = (f_c/f_b)S_{c\gamma} = (\omega_c/\omega_b) S$.
So, if we find a constraint $B_c$ for $B_i$ (in case of cold dark matter), this can be converted to the constraint $B_b = (\omega_c/\omega_b) B_c$ for the baryon isocurvature. Therefore a constraint $\alpha_c$ for $\alpha_i$ can be converted to a constraint
\beq
\alpha_b = \frac{(\omega_c/\omega_b)^2 \alpha_c}{1-\alpha_c+(\omega_c/\omega_b)^2\alpha_c}\,.
\eeq
For example, if  $\omega_b = 0.02$ and $\omega_c=0.10$, then $\alpha_c = 0.10$ corresponds to $\alpha_b \approx 0.74$. Unless otherwise stated, $\alpha$s in the following refer to the CDM isocurvature fraction. The constraints on the primordial baryon isocurvature fraction reported in this paper have been obtained from our runs with CDM isocurvature by the above mapping and by weighting the posterior by a Jacobian determinant of the mapping (in this case simply $|\partial\alpha_b / \partial\alpha_c|$), in order to simulate a situation where $\alpha_b$ would have had a flat prior \cite{KurkiSuonio:2004mn,Keskitalo:2006qv,Valiviita:2009bp}.

Additional derived parameters can be defined. 
For example, a parameter $\nad^{\mr{eff}}$
represents the spectral index for adiabatic modes obtained expressing
the adiabatic contribution as a single power law:
\begin{multline}
  n_{\mr{ad}}^{\mr{eff}}(k) - 1 \equiv
  \frac{d\ln{\cal P}_{\cal R}(k)}{d\ln k}\\
  = \frac{(\nadI \! - \! 1) (1 \! - \! |\gamma_0|) \bar k^{\nadI-1} +
    (\nadII \! - \! 1) |\gamma| \bar k^{\nadII-1}}{(1 \! - \! |\gamma_0|)\bar
    k^{\nadI-1} + |\gamma_0| \bar k^{\nadII-1}}\,,
\label{eqn:nadeffk}
\end{multline}
where $\bar k \equiv k / k_0$.

Our \emph{pivot-scale free measure of the non-adiabaticity} will be
the total non-adiabatic contribution to the CMB temperature variance
\bea
\alpha_T &\equiv& \frac {\langle ( \delta T^{\mathrm{non-ad}})^2 \rangle} {\langle ( \delta T^{\mathrm{total\ from\ scalar\ perturbations}})^2 \rangle} \nonumber \\
&=& \frac {\sum_{\ell=2}^{2100} (2\ell + 1) (C_\ell^{\mathrm{iso}} + C_\ell^{\mathrm{cor}})     } {\sum_{\ell=2}^{2100} (2\ell + 1) ( C^{\mr{ar}}_{\ell} + C^{\mr{as}}_{\ell} + C^{\mr{iso}}_{\ell} + C^{\mr{cor}}_{\ell})}.
\label{eq:alphaT}
\eea

\subsection{Phenomenological parametrizations}

The above-described model has six independent (scalar) perturbation parameters.  We can choose the \emph{spectral index parametrization}, where they are
 \beq
 	\ln A_0^2,\ \alpha_0,\ \gamma_0,\ \nadI,\ \nadII,\ \niso \,,
 \eeq
or the \emph{amplitude parametrization}, where they are
 \beq
 	\ln A_1^2,\ \ln A_2^2,\ \alpha_1,\ \alpha_2,\ \gamma_1,\ |\gamma_2| \,.
 \eeq
The model does not allow correlation to change sign as a function of scale, and therefore the sign of $\gamma_2$ is not an independent quantity. 

In addition, we can have primordial tensor perturbations whose power spectrum we write as
\beq
 	\Pow_T(k) \ = \ \Pow_T(k_0)\left( k/k_0 \right)^{n_T} \,.
\eeq
We parametrize the primordial tensor power in the traditional way by the tensor-to-scalar ratio
\beq
 	r(k) \equiv \ \frac{\Pow_T(k)}{\Pow_\R(k)} \,,
\eeq
which may depend on the scale, as we see by substituting the power spectra into the above definition
\bea
r(k) & = & \frac{\Pow_T(k_0)\left(\frac{k}{k_0}\right)^{n_T}}
	{A_{r0}^2\left(\frac{k}{k_0}\right)^{\nadI-1} + A_{s0}^2\left(\frac{k}{k_0}\right)^{\nadII-1}} \nn \\
	& = &
	r_0\frac{\left(\frac{k}{k_0}\right)^{n_T}}
	{(1-|\gamma_0|)\left(\frac{k}{k_0}\right)^{\nadI-1}\!\!\!+|\gamma_0|\left(\frac{k}{k_0}\right)^{\nadII-1}} \,,
\eea
where $r_0 \equiv r(k_0)$.
We assume the first consistency relation (see, e.g., \cite{Lidsey:1995np,Bartolo:2001rt,Wands:2002bn, Cortes:2006ap,Kawasaki:2007mb})
\beq
 	n_{T} \ = \ -\frac{r}{8(1-|\gamma|)}\,, 
\label{eq:firstconsistency}
\eeq
to be approximately valid at all scales of interest. This requires (moderate) ``running'' of tensor spectral index $n_T$, since $r$ and $\gamma$ depend on $k$. Hence we define $n_T$ up to first order as
\beq
n_T(k) = n_T(k_0) + \half q_{T0} \ln(k/k_0)\,,
\eeq
where the running is
\beq
q_{T0} = \frac{d^2 \ln\Pow_T(k)}{d(\ln k)^2}\Big|_{k=k_0} = \frac{dn_T}{d\ln k}\Big|_{k=k_0}.
\eeq
Demanding Eq.~(\ref{eq:firstconsistency}) to be valid at leading order in $\ln(k/k_0)$ leads to
\beq
q_{T0} = n_{T0}\left[n_{T0} - (\nadI - 1)\right]\,.
\label{eq:running}
\eeq
This is the second consistency relation \cite{Cortes:2007ak,Cortes:2006ap}, and we impose this together with the first consistency relation as in \cite{Brown:2009uy}  --- see discussion after Eq.~(\ref{eq:Deltarelations}) on how Eqs.~(\ref{eq:firstconsistency}) and (\ref{eq:running}) follow from slow-roll inflation. 
Therefore, the possible primordial tensor perturbations add only one extra parameter to the model, and we choose it to be tensor-to-scalar ratio $r_0$ at $k_0= 0.010\, \mbox{Mpc}^{-1}$. 
This then gives the spectral index as a derived parameter,
\beq
n_{T0}  = -\frac{r_0}{8(1-|\gamma_0|)},
\eeq
and the running via Eq.~(\ref{eq:running}). 
(In the phenomenological approach it would have been logical to keep $r$ and $n_T$ as independent free parameters,
and thus use the data to test the consistency relation.  However, as neither tensor perturbations, nor deviations from adiabaticity have so far been detected, the data currently have not much power for this, and we choose to focus 
on the question of adiabaticity in this paper.)

In phenomenological studies we choose the amplitude parametrization rather than the spectral index parametrization, since the amplitude parametrization leads to faster convergence of the parameter scan, 
leads to tighter and more reliable constraints on the parameters, and does not have the ambiguity of arbitrarily chosen prior ranges \cite{Trotta:2006ww} in Bayesian model selection that the spectral indices suffer from \cite{Sollom:2009vd,Valiviita:2009bp}. 
We list the primary (sampling) parameters of the amplitude parametrization in the first two blocks of Table~\ref{tab:parameters}. 
In the last block we list some interesting derived parameters, e.g., the spectral indices and amplitudes at $k=k_0$ that are easy to calculate from the amplitudes at $k_1$ and $k_2$.

\subsection{Inflationary slow-roll parametrization}
\label{sec:SlowrollParametrization}

In the slow-roll two-field inflation approach, we have the following relations
between the slow-roll parameters and spectral indices and $r$, to first order
in slow-roll parameters \cite{Byrnes:2006fr} (see also, e.g.,
\cite{Bartolo:2001rt,Wands:2002bn,vanTent:2003mn,Peterson:2010np},
 and note that in Byrnes \& Wands \cite{Byrnes:2006fr} $n=0$ stands for a
 scale-invariant spectrum whereas we have added the conventional 1 and we use
 a different sign convention for $\cos\Delta$ and hence for $\tan\Delta$ than \cite{Byrnes:2006fr}, since they have a different sign convention for ${\cal R}$)
 \bea
 	\nadI & = & 1 -6\veps + 2\etzz \nn\\
	\nadII & = & 1 -2\veps + 2\etss - 4\etzs\tan\Delta \nn\\
	\niso & = & 1 -2\veps +2\etss \nn\\
	r & = & 16\veps\sin^2\Delta \nn\\
	n_T & = & -2\veps \,,
 \label{eq:BWresults}
 \eea
and to second order in slow-roll parameters, among others,
\beq
q_T = -8\veps^2+4\veps\etzz\,.
\label{eq:runninginfl}
\eeq
Here the correlation angle $\Delta$ is defined by
 \beq
 	\cos\Delta \equiv 
\frac{{\mtc{C}_{\mtc{R}S}}}{\Pow_{\mtc{R}}^{1/2} \Pow_{S}^{1/2}}
= \sgn(A_sB)\sqrt\frac{A_s^2}{A_r^2+A_s^2}
 \eeq
with $0 \leq \Delta \leq \pi$ (so that $\sin\Delta$ is non-negative), and the relation to $\gamma$ is
 \bea
 	\gamma & \equiv & \sgn(\cos\Delta)\cos^2\Delta \nn\\
	\sin^2\Delta & = & 1-|\gamma| \nn\\
	\cos\Delta & = & \sgn(\gamma)\sqrt{|\gamma|} \nn\\
	\tan\Delta & = & \sgn(\gamma)\frac{\sqrt{1-|\gamma|}}{\sqrt{|\gamma|}} \,.
\label{eq:Deltarelations}
 \eea
In principle, the slow-roll parameters are functions of scale $k$, and the parameters we are fitting are the values of the slow-roll parameters when our reference scale $k_0$ exited the horizon.  
We are using these values to construct the approximate forms of the primordial power spectra.

In multi-field inflation, non-zero correlation ($\Delta\neq \pi/2$, $\gamma\neq 0$) results in whenever the background trajectory in the field space is curved between the horizon exit (of the perturbations of cosmologically interesting scales) and the end of inflation.

From Eqs.~(\ref{eq:BWresults}) and (\ref{eq:Deltarelations}) we see that $r_0 = 16\veps(1-|\gamma_0|)$, and  $n_{T0}  = -\frac{r_0}{8(1-|\gamma_0|)}$, 
and furthermore Eq.~(\ref{eq:runninginfl}) can be rewritten as $q_{T0}=-2\veps[-2\veps - (-6\veps+2\etzz)] = n_{T0}[n_{T0} - (\nadI - 1)]$. So, the tensor perturbations satisfy both the first and second consistency relations. 

After doing all our analysis using the phenomenological amplitude parametrization we repeat everything using the theoretically motivated inflationary slow-roll parametrization. 
We summarize the parameters and notation in Table~\ref{tab:parameters}.

\subsection{Specific examples}
\label{sec:TheoryExamples}

Although we will compare to the data the two models described above, and uncorrelated or $\pm$100\% correlated variations of them, we review in this subsection three examples on how a (correlated) mixture of adiabatic and isocurvature perturbations could arise without multiple (dynamical) inflaton fields. In these cases the spectral indices do not necessarily follow exactly the formulas of Eq.~(\ref{eq:BWresults}). Neither are they fully free as in our phenomenological parametrization. So, constraining these specific models accurately would require additional parameter-scan runs. As a fourth example we review one of the simplest two-field inflation models, double quadratic inflation, where the spectral indices obey  Eq.~(\ref{eq:BWresults}).

\subsubsection{Mixed inflaton-curvaton scenario}

In many models the inflationary evolution is dominated by the inflaton field $\phi$, but another light field is present during inflation and entropy perturbations in this field may be converted into curvature perturbations after inflation is over. In the curvaton scenario the curvaton field $\chi$ is constant during inflation but
once the Hubble parameter becomes smaller than the curvaton mass, the curvaton begins oscillating and behaves like dust. Thus its energy density decreases slower than that of
radiation and it can make a significant contribution to the energy density of the universe and to the curvature perturbation. The curvature perturbation is given by~\cite{Langlois:2008vk}
 \beq
  \mathcal{R} = -\frac{1}{\sqrt{2\M^2\veps_*}}\delta\phi_* + \frac{2R}{3}\frac{\delta\chi_*}{\chi_*}
 \eeq
where  $ R \equiv \left(\frac{3\rho_{\chi}}{3\rho_{\chi}+4\rho_{\gamma}}\right)_{\text{dec}}$
is evaluated at the time of curvaton decay and ``$*$'' means that a quantity is evaluated at the time of horizon exit. 

If CDM is created \emph{before curvaton decay}, the entropy perturbation is~\cite{Langlois:2008vk}
 \beq
   S = -2R\frac{\delta\chi_*}{\chi_*}
 \eeq
and we have
 \begin{equation}
    \alpha (k) = \frac{9\lambda}{1 + 10\lambda}\,, \qquad
    \gamma (k) = -\frac{\lambda}{1 + \lambda},
 \end{equation}
where $\lambda \equiv \frac{8}{9}R^2\veps_*\left(\frac{\M}{\chi_*}\right)^2$. Since $\alpha$ is constrained to be small, the only observationally allowed possibility in this scenario is that $\lambda$ is small, which leads to (almost) uncorrelated perturbations.
 
If CDM is created \emph{from curvaton decay} then the entropy perturbation is~\cite{Langlois:2008vk}
 \beq
    S = (1-R)2\frac{\delta\chi_*}{\chi_*}
 \eeq
and
 \begin{equation}
    \alpha (k) = \frac{9(1-R)^2}{R^2(1+\lambda^{-1}) + 9(1-R)^2}\,,\
    \gamma (k) = \frac{\lambda}{1 + \lambda}.
 \end{equation}
Now $\alpha$ can be small either if $\lambda$ is small, which again leads to uncorrelated perturbations, or if $R$ is very close to one, that is, if the curvaton dominates the
energy density of the universe at the time of decay. In that case $\lambda$ can be arbitrarily large,
leading to fully correlated perturbations,
provided that the value of the curvaton field is sufficiently below Planck
scale at the time when observationally relevant scales exit the horizon. The spectral indices are given by
 \begin{equation}
    \nadI \simeq 1 - 6\veps + 2\eta_{\phi\phi}\,; \quad \nadII \simeq \niso \simeq 1 - 2\veps \,,
 \end{equation}
where the parameters are evaluated at horizon exit.

\subsubsection{Modulated reheating with gravitino dark matter}

In the modulated reheating scenario the decay rate $\Gamma = \Gamma(\sigma)$ of the inflaton depends on an additional scalar field $\sigma$ which was light during inflation. The
inflaton decays into radiation when $H\sim \Gamma$, which means that due to entropy perturbations in the field $\sigma$ the inflaton will decay at different times in
different parts of the universe. Assuming that
after the end of inflation the inflaton oscillates and behaves like dust, its energy density decreases slower than that of radiation, which means that the parts of the universe
where it decays later will have an overdensity compared to the parts where it decays earlier. Thus the entropy perturbations in the modulating field are converted into
curvature perturbations, given by 
 \beq
 	\mathcal{R} = -\frac{1}{\sqrt{2\M^2\veps_*}}\delta\phi_*+\frac{1}{6}\left(\frac{d\Gamma/d\sigma}{\Gamma}\right)\delta\sigma_*\,.
 \eeq

If CDM is in the form of gravitinos then the entropy perturbation is given by 
 \beq
 	S = \pm\frac{1}{2}\left(\frac{d\Gamma/d\sigma}{\Gamma}\right)\delta\sigma_* \,,
 \eeq 
where $+$ and $-$ correspond to
thermal and non-thermal production of gravitinos, respectively \cite{Takahashi:2009cx}. Then
 \begin{equation}
    \alpha (k) = \frac{9\xi}{9 +10\xi}\,, \quad
    \gamma (k) = \pm \frac{\xi}{9 + \xi}\,,
 \end{equation}
where 
 \beq
 	\xi \equiv \frac{1}{2}\veps_*\left(\frac{\Gamma'\sigma}{\Gamma}\right)_* ^2\left(\frac{\M}{\sigma}\right)_* ^2\,.
 \eeq 
Again perturbations are constrained to be (almost) uncorrelated.
The spectral indices are the same as in the curvaton case.

\subsubsection{Axion}

In the axion model the axion field is massless during inflation, but obtains a mass after inflation ends, once the temperature of the universe drops below the QCD scale, $\Lambda_\text{QCD}$. When the Hubble
parameter is smaller than the axion mass, the axion behaves like dust and can account for dark matter. The curvature perturbation in the axion model comes from the inflaton, 
 \beq
 	\mathcal{R} = -\frac{1}{\sqrt{2\M^2\veps_*}}\delta\phi_* \,,
 \eeq 
and the isocurvature perturbation from the axion, 
 \beq
 	S = 2\frac{\omega_{\chi}}{\omega_{c}}\frac{\delta \chi_*}{\chi_*} \,,
 \eeq
where $\omega_{\chi}/\omega_c$ is the fraction of axions in CDM. The value of the axion field during inflation can be expressed 
as $\chi_* = f_a \theta_i$, where $f_a$ is the
axion decay constant and $\theta_i$ is the initial misalignment angle. We now have
 \begin{equation}
 	\alpha (k) = \frac{\lambda}{1 +\lambda} \,, \quad
 	\gamma (k) = 0 \,,
 \end{equation}
where 
 \beq
 	\lambda \equiv \frac{8\veps_*}{\theta_i^2}\left(\frac{\omega_{\chi}}{\omega_c}\right)^2\left(\frac{\M}{f_a}\right)^2 \,.
 \eeq 
The final abundance of axions is \cite{Kawasaki:2007mb}
 \begin{equation}
 	\omega_{\chi} = 10^{-3} \times f_d \theta_i^2 \left(\frac{\Lambda_\text{QCD}}{200 \text{MeV}}\right)^{-2/3}\left(\frac{f_a}{10^{10}\text{GeV}}\right)^{7/6} \,,\nonumber
 \end{equation}
where $f_d\le 1$ is a dilution factor. The parameter $f_a$ is cosmologically constrained to be in the region $10^{10}$ GeV $\le f_a \le$ 4.1 $\times 10^{12}$ GeV \cite{Kawasaki:2007mb}. Thus if axions account for
a significant portion of CDM, then $\veps$ is constrained to be extremely small. The spectral indices are given by
\begin{equation}
 \nadI \simeq 1 - 6\veps + 2\eta_{\phi\phi}, \quad \niso \simeq 1 - 2\veps.
\end{equation}

\subsubsection{Double quadratic inflation}

If there are several dynamically important fields during inflation then the direction of the background evolution plays the role of the inflaton. Perturbations in this direction
are adiabatic while perturbations perpendicular to it are entropy perturbations. The simplest such model is double quadratic inflation with two non-interacting fields with
quadratic potentials, 
 \beq
 	V = \frac{1}{2}m_{\phi}^2\phi^2+\frac{1}{2}m_{\chi}^2\chi^2 \,,
 \eeq 
where $m_{\chi}^2 \ge m_{\phi}^2$.
Denoting the perturbations in the direction of the background trajectory by $\delta\sigma$
and ones perpendicular to it by $\delta s$ and assuming that the heavy field decays into CDM and the light field into radiation, the perturbations are \cite{Langlois:2008vk}
 \bea
 	\mathcal{R} & = & -\frac{1}{\sqrt{2\M^2\veps_*}}\delta\sigma_* + \frac{1}{4}\M^{-2}(1-R^{-2})\xi_*\sin2\theta_*\delta s_*
	\nonumber\\
 	S & = & -\frac{4R^2}{\sin2\theta_*\xi_*}\delta s_* \,,
 \eea
where $\tan\theta \equiv \dot{\chi}/\dot{\phi}$ and $R\equiv m_{\chi}/m_{\phi}\ge 1$ and $\xi \equiv \sqrt{\phi^2+R^4\chi^2}$. In this case
 \begin{equation}
 	\alpha (k) = \frac{\beta}{1 + \left[1+\frac{4\veps^2(R^2-1)^2}{\beta^2}\right]\beta} \,,\
 	\gamma (k) = \frac{4\veps^2(R^2-1)^2}{\beta + 4\veps^2(R^2-1)^2} \,, \nonumber
 \end{equation}
where 
 \beq
 	\beta \equiv \frac{32\veps_*}{\sin^22\theta_*}R^4\left(\frac{\M}{\xi}\right)_* ^2 \,. \nonumber
 \eeq 
The isocurvature fraction $\alpha$ can be small either when $\beta \ll 1$, in which case we can have
arbitrary positive correlations, or when $2\veps(R^2-1) \gg \beta \gtrsim 1$, which means fully correlated perturbations. The spectral indices for this model are given by Eq.~(\ref{eq:BWresults}) with $\veps=(\partial_\sigma V)^2/(16\pi G V^2)$,
\begin{eqnarray}
2\etzz & = & \eta_{\phi\phi}(1+\cos2\theta) + \eta_{\chi\chi}(1-\cos2\theta),\nonumber\\
2\etss & = & \eta_{\phi\phi}(1-\cos2\theta) + \eta_{\chi\chi}(1+\cos2\theta),
\end{eqnarray}
and $\etzs\tan\Delta  =  \etss$, where $\eta_{ij}= (\partial_i\partial_j V) / (8\pi G V)$.


\section{About the analysis}

\subsection{Data}
\label{sec:data}

First we use only the CMB temperature and polarization anisotropy data: Seven-Year Wilkinson Microwave Anisotropy Probe (WMAP7) data \cite{Larson:2010gs,Komatsu:2010fb}, the complete Arcminute Cosmology Bolometer Array Receiver (ACBAR) data \cite{Reichardt:2008ay}, and QUEST at DASI (QUaD) data \cite{Brown:2009uy} (QUEST = Q and U Extragalactic Survey Telescope, DASI = Degree Angular Scale Interferometer). Then we add into the analysis either the supernova (SN) data from Supernova Cosmology Project (SCP) Union 2 compilation \cite{Amanullah:2010vv} or the large-scale structure data from Sloan Digital Sky Survey Data Release 7 Luminous Red Galaxies (SDSS DR7 LRG) \cite{Reid:2009xm} that contains baryon acoustic oscillations (BAO) \cite{Percival:2009xn}.

We perform the full Bayesian evidence calculation for each combination of data sets. Although this is computationally demanding, \emph{it is important not to blindly combine different types of data sets} without testing what is the individual information gained from each data set and whether they are consistent with each other. In particular, the black-box method of combining all available data may lead to artificially tight constraints, if there is `tension' between the data sets, and there is between CMB and large-scale structure data. One historical example is using big-bang nucleosynthesis and observed abundances of light elements to constrain $\omega_b$; tension between different elements led to too tight constraints, and a wrong value for $\omega_b$ (see discussion in, e.g., \cite{Kainulainen:1998vh}). Therefore we strongly advocate the approach of adding data sets to the analysis one by one.

{\bf CMB:}
Our choice of the CMB data sets that extend the coverage from WMAP7 to smaller scales and also provide extra polarization data are ACBAR and QUaD. We impose in the CMB likelihood calculations a cut-off multipole $\ell_{\rm max} = 2100$. We do not use at the South Pole Telescope (SPT) data \cite{Shirokoff:2010cs} nor The Atacama Cosmology Telescope (ACT) data \cite{Das:2010ga} that would extend with high precision to even much higher $\ell$, for the following reasons: 1) These data were not publicly available when we started our massive MultiNest runs. 2) In order to employ these data correctly the foreground modelling is of vital importance, requiring marginalization over many ``extra'' nuisance parameters that would slow down the convergence of our parameter scan. 3) Since the tensor spectrum in our analysis is red [$n_T < 0$, see Eq.~(\ref{eq:firstconsistency})], the tensor contribution modifies only the low-$\ell$ part of spectrum. So, the SPT and ACT data would not give extra constraints on tensor-to-scalar ratio. 4) As we will see, in the inflationary parametrization also the isocurvature modifies only the low-$\ell$ spectrum. The only effect of the SPT and ACT data would be possibly a slightly tighter constraint on $\niso$ in the phenomenological parametrization, and somewhat tighter constraints on the background parameters that could indirectly constrain nonadiabaticity.

{\bf SN:}
There are many SN data sets on the market. Based on \cite{Kessler:2009ys}, our aim was to choose a recent set that would be the most ``conservative'', giving the loosest upper bound for the nonadiabatic contribution in the phenomenological mixed model that prefers larger $\Omega_\Lambda$ than the adiabatic model \cite{KurkiSuonio:2004mn,Keskitalo:2006qv,Valiviita:2009bp}. Since the SN data constrain $\Omega_\Lambda$, the largest isocurvature fraction  is allowed by the data that allow for the largest $\Omega_\Lambda$. Even the same ``raw'' SN data give very different results depending on how the light curves are fitted. In general, the Spectral Adaptive Lightcurve Template (SALT-II) fitter \cite{Guy:2007dv} leads to a Hubble diagram that prefers a larger $\Omega_\Lambda$ than  the Multicolor Light Curve Shapes (MLCS2k2) fitter \cite{Jha:2006fm} in the $\Lambda$CDM model \cite{Kessler:2009ys}. Therefore, our choice is SCP Union 2 compilation that has been obtained using SALT-II. In order to be as conservative as possible, we run the likelihood code that comes with CosmoMC with systematic errors turned on.

{\bf MPK:}
It should be noticed that in Sec.~\ref{sec:ExtraData} we use the full matter power spectrum data (MPK) as a function of $k$ that include the baryon wiggles spectrum $\mr{BAO}(k)$, not the simplified BAO distance measure(s) or BAO scale(s). In the \emph{mixed} models, these simplified measures cannot be used as given in literature, since the extraction of these numbers  from the data assumes adiabatic initial conditions, see \cite{Kasanda:2011np} (and \cite{Carbone:2011bx,Mangilli:2010ut,Zunckel:2010mm}) for detailed accounts on the danger of misinterpretation of these measures in the presence of isocurvature modes. Indeed, employing the MPK data with BAO in a right and consistent way in the mixed models is not a trivial task. The detailed procedure and modifications we did to CosmoMC and the MPK likelihood code \cite{Reid:2009xm} will be described elsewhere \cite{Valiviita:inprep}. In short: for each parameter combination, we extracted the oscillations separately for the adiabatic and isocurvature modes with unit primordial amplitude and scale-invariant spectrum by going from Fourier space to real space, i.e., by finding the correlation function. This has a bump that was removed with a technique described in the Appendix A.1 of \cite{Hamann:2010pw}. The bumpless correlation functions were transformed back to $k$-space. These smooth adiabatic and isocurvature matter power spectra were then weighted by the initial amplitudes and  the spectral tilts were added. Finally, the two smooth adiabatic spectra $P_{\mr{ar}}(k)$ and $P_{\mr{as}}(k)$ as well as the smooth $P_{\mr{iso}}(k)$ and $\mr{sign}(\gamma)\sqrt{ P_{\mr{as}}P_{\mr{iso}}}$ (the correlation) were added together. The same was done for the original component spectra giving the unsmooth spectrum with the baryonic wiggles, $P_{\mr{full}}(k)$. After this the MPK likelihood routine was used in the normal way \cite{Reid:2009xm}, i.e., a nonlinearity correction was applied to the smooth spectrum, and, for presentation purposes, BAO was calculated as $\mr{BAO}(k) = \log_{10} [P_{\mr{full}}(k) / P_{\mr{smooth}}(k)]$. When using the MPK data, in order to reliably find $P_{\mr{smooth}}$ we restrict the prior ranges of $\omega_b$ to (0.016, 0.029), $\omega_c$ to (0.05, 0.17), $100\theta$ to (0.90, 1.18), and $\tau$ to (0.02, 0.20), while keeping the priors of all the other parameters the same as indicated in Table~\ref{tab:parameters}.

\subsection{Sampling method}

For the full parameter scan and for
calculating the total model probability ${\cal Z}$ (or Bayesian evidence)
of each variant of our model as well as the reference adiabatic models
we employ MultiNest package \cite{Feroz:2008xx} that is based on an efficient variant of nested sampling algorithm.

After the conceptual introduction by
\cite{Skilling:2004}, this method has been first applied to cosmology
in a simple case
by \cite {Mukherjee:2005wg}. Its most sensitive part, the sampling
technique, has been refined by \cite {Shaw:2007jj}
and \cite{Feroz:2007kg} to minimize the required number of likelihood
evaluations and to deal efficiently with possible
pathologies, such as multi-modal posterior distributions and strongly
curved parameter degeneracies. Finally, an even more robust and
efficient code has been released by \cite{Feroz:2008xx} for
applications in cosmology, astronomy and particle physics. The
package, available for public use from 
\href{http://ccpforge.cse.rl.ac.uk/gf/project/multinest/}{http://ccpforge.cse.rl.ac.uk/gf/project/multinest/},
contains an easily usable interface for the CAMB/CosmoMC cosmology code \cite{CAMB,COSMOMC} that we have modified to handle an arbitrarily correlated mixture of adiabatic and isocurvature perturbations.

The MultiNest user has simply to tune three
parameters: the tolerance (accuracy), the number of live points $N$,
and the maximum efficiency
$e$, which sets how aggressively we want the reduce
the volume of parameter space at each iteration. A  very attractive feature
of MultiNest is that the need of a proposal matrix for the
parameters' covariance, a well known hassle for conventional Markov Chain Monte Carlo (MCMC) users, is now superseded.

In this paper we set
the efficiency parameter in MultiNest to 0.3, the tolerance parameter to 0.5, and the number of live points to $N = 400$. With these
settings the error estimate for the logarithm of the total model probability
turns out to be $|\delta\ln{\cal Z}| = 0.20$ -- $0.26$ for all of our runs. Therefore, when comparing model $q$ to model $q'$, the logarithm of the Bayes factor, i.e., the logarithm of the ratio of the model probabilities, 
 \beq 
  {\cal B}_f \equiv \ln B_{qq'} = \Delta\ln{\cal Z} = \ln ({\cal Z}_q  / {\cal Z}_{q'})\,
\label{eq:Bf}
 \eeq
 has an uncertainty $\delta{\cal B}_f \approx  \pm 0.4$.

\subsection{Bayesian model comparison and Jeffreys' scale}

A brief summary of Bayesian model comparison, calculation of evidence, i.e., the model probability $P={\cal Z}$, the sampling method, and MultiNest can be found in the Appendix of \cite{Valiviita:2009bp}. For a comprehensive review of many aspects of Bayesian cosmology, see \cite{Trotta:2008qt}. 

The Bayes factors are classified with Jeffreys' scale \cite{JeffreysBook} for the strength of evidence. This empirically calibrated scale has thresholds at certain values of odds. If we take the model $q$ in Eq.~(\ref{eq:Bf}) as our null hypothesis and the model $q'$ as an alternative model, then the original Jeffreys' book \cite{JeffreysBook} grades verbally the odds according to Table~\ref{tab:Jeffreys}. Jeffreys calls the simpler model (which is nested, i.e., a subset of the alternative model --- with the extra parameters of the alternative model fixed, often to zero values) the ``null hypothesis'', and the more complicated model that would lead to a new discovery (and determination of extra parameters) an ``alternative'' model. From this set-up it is clear that one can only find evidence against the null hypothesis, not evidence against the alternative model, but merely just an indication that the null hypothesis is supported (while the extra parameters of the alternative model could be so small as not to show up) in light of the current data.

In the recent literature, $q$ and $q'$ have often been considered from symmetrical footing, and indeed this is the only approach one can take in situations where $q$ is not a subset of $q'$, i.e., whenever the two models are totally different and don't share (any) parameters. However, in all our cases we have a clear null hypothesis and hence we take a conservative ``Jeffreys' approach'': When comparing adiabatic and mixed models the null hypothesis is the adiabatic model, where the ``nonadiabaticity'' parameters have null values. When comparing models with and without primordial tensor contribution, the null hypothesis is the model without tensors where $r_0$ or $\veps$ has a null value. In models with and without CMB lensing the number of parameters is the same, but the model with lensing has some extra ingredients, e.g., lensing potential that modifies the unlensed $C_\ell$ spectrum, and in this sense it is natural to choose the model without lensing as our null hypothesis. For classifying the evidence against null hypothesis, we will use simplified thresholds and modernized wording \cite{Trotta:2008qt}, modified in Table~\ref{tab:ModJeffreys} in order to take into account our conservative null hypothesis approach.
\begin{table}
\begin{tabular}{l|l|l}
 $B_{qq'}\equiv P_q/P_{q'}$ & ${\cal B}_f$ & Interpretation \\
\hline
$>1$     & $> 0$ & ``Null hypothesis $q$ supported''\\
1:3.16 ... 1:1 & -1.15 &  ``Evidence against $q$, but not\\ 
 & & \ \ worth more than a bare mention''\\
1:10 ... 1:3.16 & -2.72 & ``Evidence against $q$ substantial''\\
1:31.6 ... 1:10 & -3.45 & ``Evidence against $q$ strong''\\
1:100 ... 1:31 & -4.61 & ``Evidence against $q$ very strong''\\
$< 1:100$ & $< -4.61$ & ``Evidence against $q$ decisive''
\end{tabular}
\caption{A verbal interpretation of various odds ($B_{qq'}$) and the corresponding logarithm of the Bayes factor (${\cal B}_f \equiv \ln B_{qq'}$) at the lower bound according to Jeffreys \cite{JeffreysBook}.}
\label{tab:Jeffreys}
\end{table}
\begin{table}
\begin{tabular}{l|l|l}
$B_{qq'}\equiv P_q/P_{q'}$ & ${\cal B}_f\equiv \ln B_{qq'}$ & Interpretation \\
\hline
$> 3:1$  & $>+1.0$ & ```Null hypothesis $q$ supported''\\
1:3 ... 3:1     & -1.0 ... +1.0 & ``Inconclusive''\\
1:12 ... 1:3  & -2.5 ... -1.0 &  ``Weak evidence against $q$''\\
1:150 ... 1:12 & -5.0 ... -2.5 & ``Moderate evidence against $q$''\\
$<1:150$ & $<-5.0$ & ``Strong evidence against $q$''
\end{tabular}
\caption{Simplified thresholds and modernized verbal interpretation of various odds ($B_{qq'}$) and the corresponding logarithm of the Bayes factor (${\cal B}_f$). The tabulated values of $B_{qq'}$ are rounded, while the values of ${\cal B}_f$ are the actual thresholds used in most of the modern cosmology literature. The actual wordings vary from source to source.}
\label{tab:ModJeffreys}
\end{table}


\subsection{Lensing:  Bayesian evidence and a degeneracy with isocurvature}
\label{sec:lensing}

All runs were made with CMB lensing on, unless stated otherwise. 
Initially, we did most of our runs without lensing, which is by a factor of 4--10 faster. Comparison of lensed and unlensed runs showed that lensing had only a minor effect on most of the 1-d posterior probability densities. 
However, the best-fit $\chi^2$ improve a lot when allowing for lensing ($\Delta\chi^2 \approx -8$
in the adiabatic case, $-6$ in the mixed case), 
and furthermore we find \emph{moderate Bayesian evidence against the unlensed models.} 
For example, in the amplitude parametrization the adiabatic unlensed model (without tensor perturbations) has $\ln{\cal Z} = -3861.3$  while the lensed model has $\ln{\cal Z} = -3856.6$. This gives the for the logarithm of Bayes factor ${\cal B}_f = \ln{\cal Z}_{\rm unlensed} - \ln{\cal Z}_{\rm lensed} = -4.7 \pm 0.4$, or odds $P_{\rm unlensed} / P_{\rm lensed} = \exp(-4.7) = 1:110$.
In the mixed model the odds against the unlensed model are less overwhelming: 
${\cal B}_f=-3.3 \pm 0.4$, i.e., $P_{\rm unlensed} / P_{\rm lensed} = 1:27$.

The reason for this is that in the amplitude parametrization large isocurvature spectral indices $\niso \gtrsim 2$ are favored and this causes the isocurvature and correlation contribution to affect rather uniformly the whole $C_\ell$ spectrum \cite{KurkiSuonio:2004mn}. 
In particular, at high multipoles the almost opposite phase isocurvature (and correlation) components are able to ``fill'' dips and ``eat'' peaks of the adiabatic spectrum, and hence they smear the total $C_\ell$ spectrum compared to the pure adiabatic case. 
The CMB lensing has a similar smearing effect. Since the peaks and dips at high $\ell$ in the data are smoother than the unlensed pure adiabatic $\Lambda$CDM model predicts, the fit to the CMB data improves considerably either by allowing for the isocurvature contribution (in the unlensed case $\Delta\chi^2 \approx -5.5$ for the best-fit mixed model in amplitude parametrization compared to the best-fit adiabatic model) or by turning the lensing on. 
Therefore, there is a degeneracy between lensing and isocurvature, 
and the Bayesian evidence against ``no lensing'' is weaker in the mixed model.

\section{The general case with CMB data}
\label{sec:general_cmb}

\subsection{Phenomenological approach --- Amplitude parametrization}
\label{sec:general_cmb_ampl_par}

\begin{figure}[t]
  \centering
  \includegraphics[width=\columnwidth]{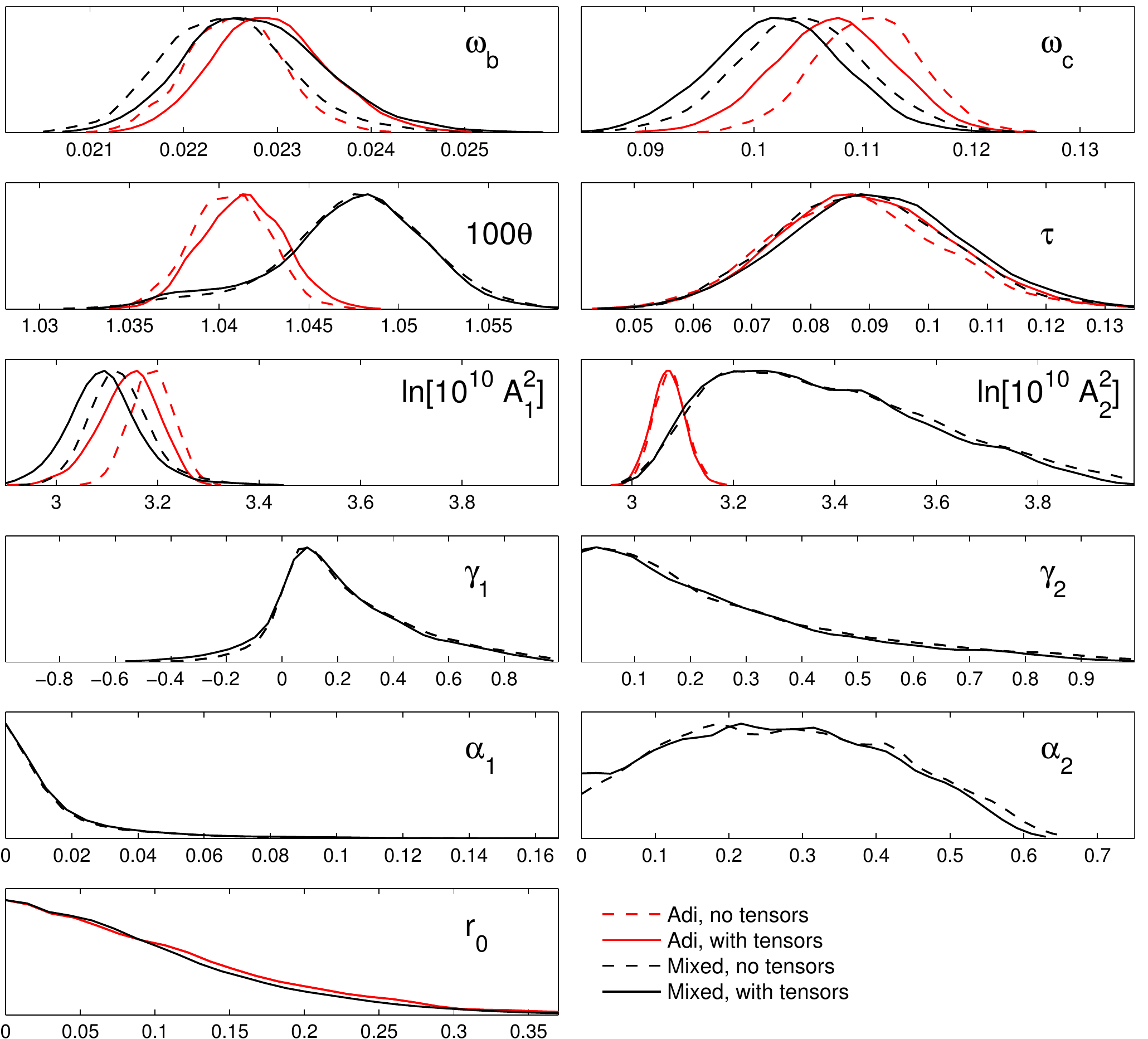}
  \caption{{\bf General case, amplitude parametrization, CMB data.} Marginalized 1-d posterior probability densities of the primary parameters (except $A_{SZ}$ which would be flat) in amplitude parametrization. 
Red lines are for the purely adiabatic model, and black lines for the mixed adiabatic and isocurvature model. The solid lines are for the models with primordial tensor perturbations, while the dashed lines are for models with only scalar perturbations.}
  \label{fig:AmplPrimary}
\end{figure}

\begin{figure}[t]
  \centering
  \includegraphics[width=\columnwidth]{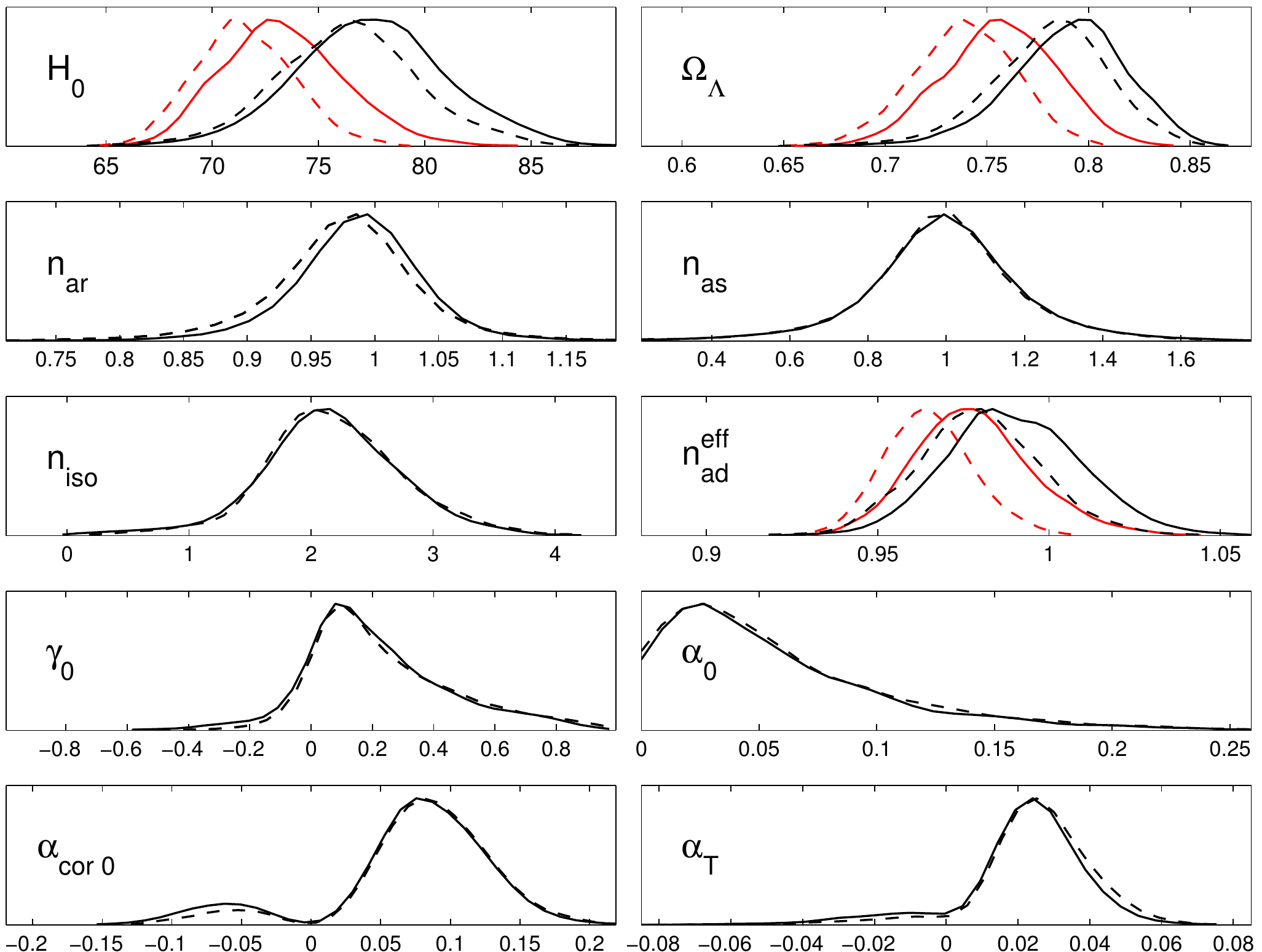}
  \caption{{\bf General case, amplitude parametrization, CMB data.} Marginalized 1-d posterior probability densities of selected derived parameters from runs made in amplitude parametrization. The line styles are the same as in Fig.~\ref{fig:AmplPrimary}.}
  \label{fig:AmplDerived}
\end{figure}

An important question in constraining cosmologies is how much the assumptions made in the analysis affect the values of cosmological parameters obtained from the given data \cite{Trotta:2001yw}. Therefore, we start by comparing the marginalized 1-d posterior probabilities in the mixed adiabatic and isocurvature model to the results obtained for the pure adiabatic model, using the amplitude parametrization.  Fig.~\ref{fig:AmplPrimary} shows these for the primary parameters and Fig.~\ref{fig:AmplDerived} for selected derived parameters for models with and without primordial tensor perturbations. In Appendix \ref{app:BigTables} we tabulate 68\% or 95\% confidence level (C.L.) intervals for selected parameters as well as Bayesian evidences ${\cal Z}$ for each model. 

We notice that the tensor contribution does not significantly affect any of these results, and the maximum allowed primordial tensor-to-scalar ratio, $r_0$, is insensitive to the inclusion of the isocurvature mode:  $r_0<0.249$ in the mixed model and $r_0<0.264$  in the pure adiabatic model at 95\% C.L. The reason for the lack of interplay between tensor and isocurvature constraints is that the consistency relation requires the tensor perturbations to have a red spectrum ($n_T<0$), whereas the isocurvature perturbations favored in our phenomenological approach have a steeply blue spectrum. 
The slight formal tightening of the constraint in generally correlated models is due to the use of the first inflationary consistency relation that with a fixed $r_0$ leads to more negative $n_T$ (whenever $\gamma \neq 0$) than in the adiabatic model, see Eq.~(\ref{eq:firstconsistency}), and hence a larger effect on large scales. This has been noticed also in \cite{Kawasaki:2007mb}. 

There is no evidence for the presence of tensor perturbations. Indeed, Bayesian model selection supports the model without a tensor contribution: in the mixed model the odds are $P_{\rm no\ tensors} / P_{\rm tensors} = 7:1$
(${\cal B}_f= +1.96$), and in the adiabatic model
$11:3 \approx 3.7:1$
(${\cal B}_f= +1.30$)
in support of the model without tensors.
Nevertheless we choose the model \emph{with primordial tensor perturbations} as our baseline case for three reasons: 1) In inflationary models the tensor 
perturbations arise naturally, and hence it is more straightforward to compare our phenomenological model to the inflationary slow-roll model if we include the tensor perturbations. 
2) Given point 1, we have a theoretical prejudice that the tensor perturbations exist, but probably with much smaller amplitude than we guessed when setting our linear flat prior $r_0 < 0.75$, which is an integral part of the ``model'' in the Bayesian evidence calculation and which we had to decide ``before seeing the data''. So, what we found above is that the model without tensors ($r_0=0$) is supported over the model with $r_0 \sim {\cal O}(0.3)$.
3) In the literature and in our previous work
the model without tensor perturbations has already been studied extensively, though with older data sets than here.

\begin{figure}[t]
  \centering
  \includegraphics[width=0.98\columnwidth]{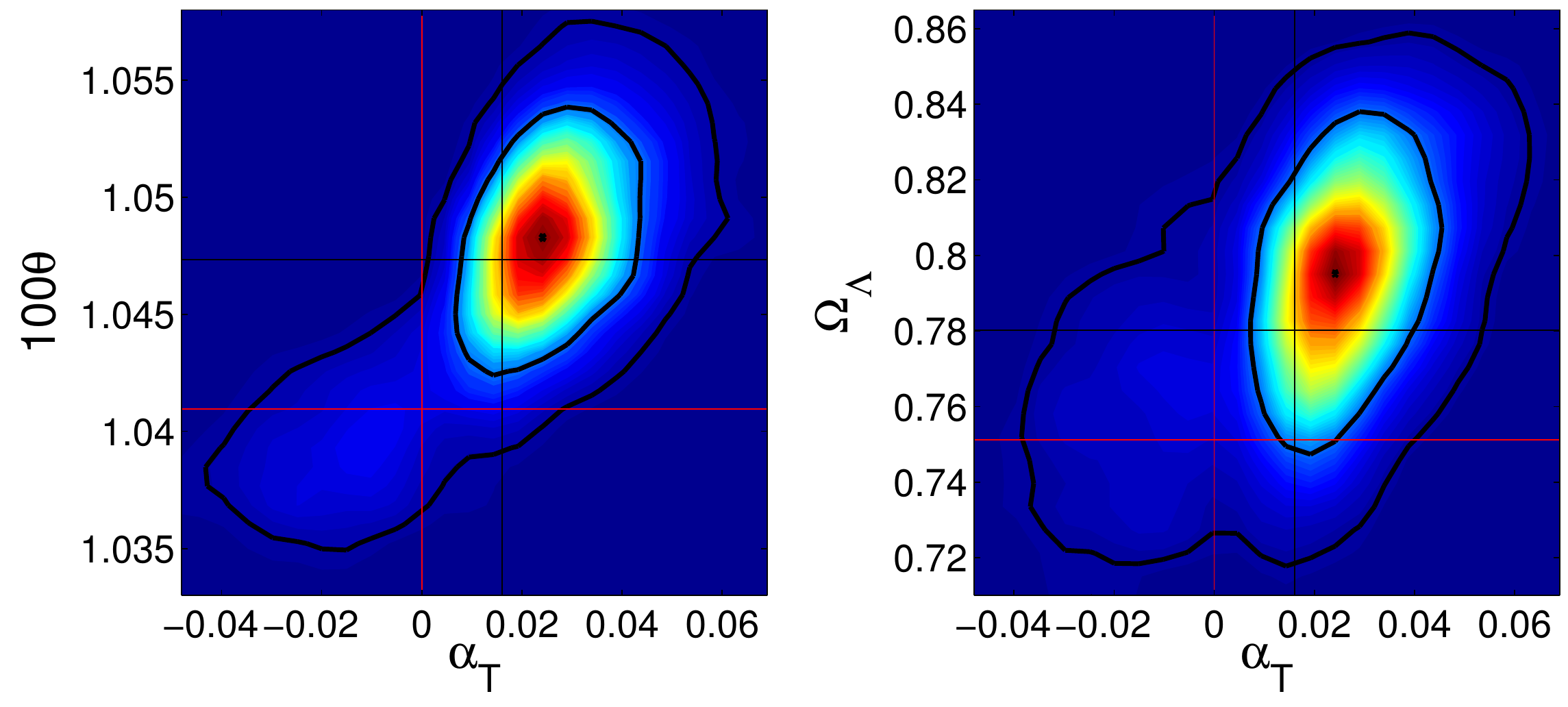}
  \caption{{\bf General case, amplitude parametrization, CMB data.} Marginalized 2-d posteriors of selected parameters for the mixed model with tensors. The thick black curves indicate the 68\% and 95\% C.L. regions. The intersection of thin black (red) lines is the best-fit mixed (adiabatic) model. The difference between these two best-fit models is $\Delta\chi^2\approx -3$ in favour of the mixed model. The colour scale indicates the mean likelihood.}
  \label{fig:Ampl2d}
\end{figure}

The CMB data allow a CDM (baryon) isocurvature fraction $\alpha < 0.064$ (0.42) at scale $k_1$, $\alpha < 0.154$ (0.67) at scale $k_0$, and $\alpha < 0.512$ (0.95) at scale $k_2$, but does not show any clear preference for a nonzero isocurvature contribution.
The transfer function that maps the primordial entropy perturbation $S$ into today's angular power spectrum $C_\ell^{\mr{iso}}$ has an extra factor of $(k_{eq}/k)^2 \propto (\ell_{eq}/\ell)^2$ compared to the transfer function that maps the primordial curvature perturbation into $C_\ell^{\mr{ar}}$ and $C_\ell^{\mr{as}}$ \cite{Kodama:1986fg,Hu:1995en,KurkiSuonio:2004mn,jvPhDthesis}. (Here $k_{eq}$ is the wave number of perturbation that re-enters the Hubble horizon at matter-radiation equality.) Therefore primordial isocurvature perturbations  contribute much more strongly to large-scale (small $k$, small $\ell$) CMB than to small-scale CMB, when compared to adiabatic perturbations of the same primordial magnitude, and hence the limits to $\alpha$ are tighter at large scales than at 
small scales.  Indeed, if no (extra) prior is placed on the isocurvature spectral index, our phenomenological approach formally favors a large $\niso \sim 2~$--$~3$, i.e., a large primordial isocurvature fraction at small scales.  For such isocurvature perturbations, a positive correlation with curvature 
perturbations is favored.  The marginalized posterior probability density for the nonadiabatic contribution to the CMB temperature variance, $\alpha_T$, appears to show a weak detection of a positive value ($\alpha_T > 0$ at 90\% C.L.),
but Bayesian model comparison doesn't support the inclusion of primordial CDM (or baryon) isocurvature mode:
$P_{\rm adiabatic} / P_{\rm mixed} = 240:1$
(${\cal B}_f = +5.48 $) with tensors,
or  $137:1$ (${\cal B}_f = +4.92 $) without tensors.
The last column of the first two rows of Table \ref{tab:IsocParams} indicate that the CMB sets an upper limit of 5\% for $\alpha_T$ which is our pivot-scale free measure of nonadiabaticity.

According to Fig.~\ref{fig:AmplPrimary} the inclusion of isocurvature affects significantly the sound horizon angle $\theta$: mixed models prefer larger $\theta$ than adiabatic models. They also prefer slightly smaller $\omega_c$. 
These changes are reflected in the derived parameters $H_0$ and $\Omega_\Lambda$ in Fig.~\ref{fig:AmplDerived}: mixed models prefer larger values. Fig.~\ref{fig:Ampl2d} demonstrates this degeneracy between the nonadiabatic contribution and $\theta$ or $\Omega_\Lambda$.

The reason for the above degeneracy is extensively explained in \cite{Keskitalo:2006qv} and \cite{Valiviita:2009bp}, see in particular figure 2 and its explanation in \cite{Valiviita:2009bp}. In short: The amplitude and phase of the correlation component is somewhere in between the adiabatic and isocurvature components. Therefore the correlation component gives the main ``nonadiabatic'' contribution to $C_\ell$. When the isocurvature and hence the correlation spectral indices are much larger than 1, the positive correlation component adds some power to the right end of each acoustic peak, moving them slightly to right and making them wider; see figure 2 in \cite{Keskitalo:2006qv}. A larger $\theta$ counteracts this effect by squeezing and moving the whole acoustic peak structure to left, hence retaining the almost ``adiabatic'' shape of the temperature angular power spectrum that fits well the data. The sound horizon angle is defined by $\theta \equiv r_s/D_A$, where $r_s$ is the sound horizon at last scattering and $D_A$ is the angular diameter distance to last scattering. As $r_s$ depends only on $\omega_b$ and $\omega_c$ that are determined  by the relative heights of the acoustic peaks, and as even this dependence is very mild \cite{Hu:2000ti}, the CMB constraint on $\theta$ is directly reflected in the favored $D_A$, which depends on $\Omega_m$, $\Omega_\Lambda$ (or $\Omega_K$ in spatially curved models), and $H_0$. In the curved mixed models a well-fitting $D_A$ is achieved in a slightly closed case \cite{Valiviita:2009bp}, i.e., when $\Omega_m+\Omega_\Lambda \sim 1.05$, but in the flat mixed models studied here, the constraint $\Omega_m+\Omega_\Lambda=1$ dictates that the only way to achieve a well-fitting value for $D_A$ is to increase $\Omega_\Lambda$ (and $H_0$) and decrease correspondingly $\Omega_m$ compared to the well-fitting adiabatic case.

Forcing the background parameters, for example $\Omega_\Lambda$, closer to the ``adiabatic'' values should set \emph{indirect} extra constraints on isocurvature or at least disfavour the positive correlation and large $\niso$. Since $\Omega_\Lambda$ is constrained by the supernova data and $\Omega_m$ ($=1-\Omega_\Lambda$) by the large-scale structure (matter power spectrum) data, we check in Sec.~\ref{sec:ExtraData} how our results are affected by the inclusion of these data. 
However, it should be emphasized that the extra constraints on $\Omega_\Lambda$ may lead to tighter constraints on isocurvature only in the spatially flat case which we consider here, leaving the constraints untouched if spatially curved geometry is allowed \cite{Valiviita:2009bp}.

In addition, the amplitude and phase of baryonic wiggles (BAO) in the matter power spectrum might set \emph{direct} extra constraints on the isocurvature and correlation components if their relative amplitudes at scales probed by MPK were large enough. As the CMB tightly constrains the isocurvature contribution on large and medium scales (wave numbers $k_1$ and $k_0$, respectively), the potential extra constraints from BAO come in cases with a large nonadiabatic component on small scales ($k \gtrsim k_2$), i.e., whenever $\niso$, and hence $\ncor$, are very large, $\niso \gtrsim 2.5$. However, we will show in Sec.~\ref{sec:ExtraData} that we would need more precise data than the current SDSS DR7 LRG data.

\subsection{Two-field inflation approach --- Slow-roll parametrization}
\label{sec:general_cmb_infl_par}

\begin{figure}[t]
  \centering
  \includegraphics[width=\columnwidth]{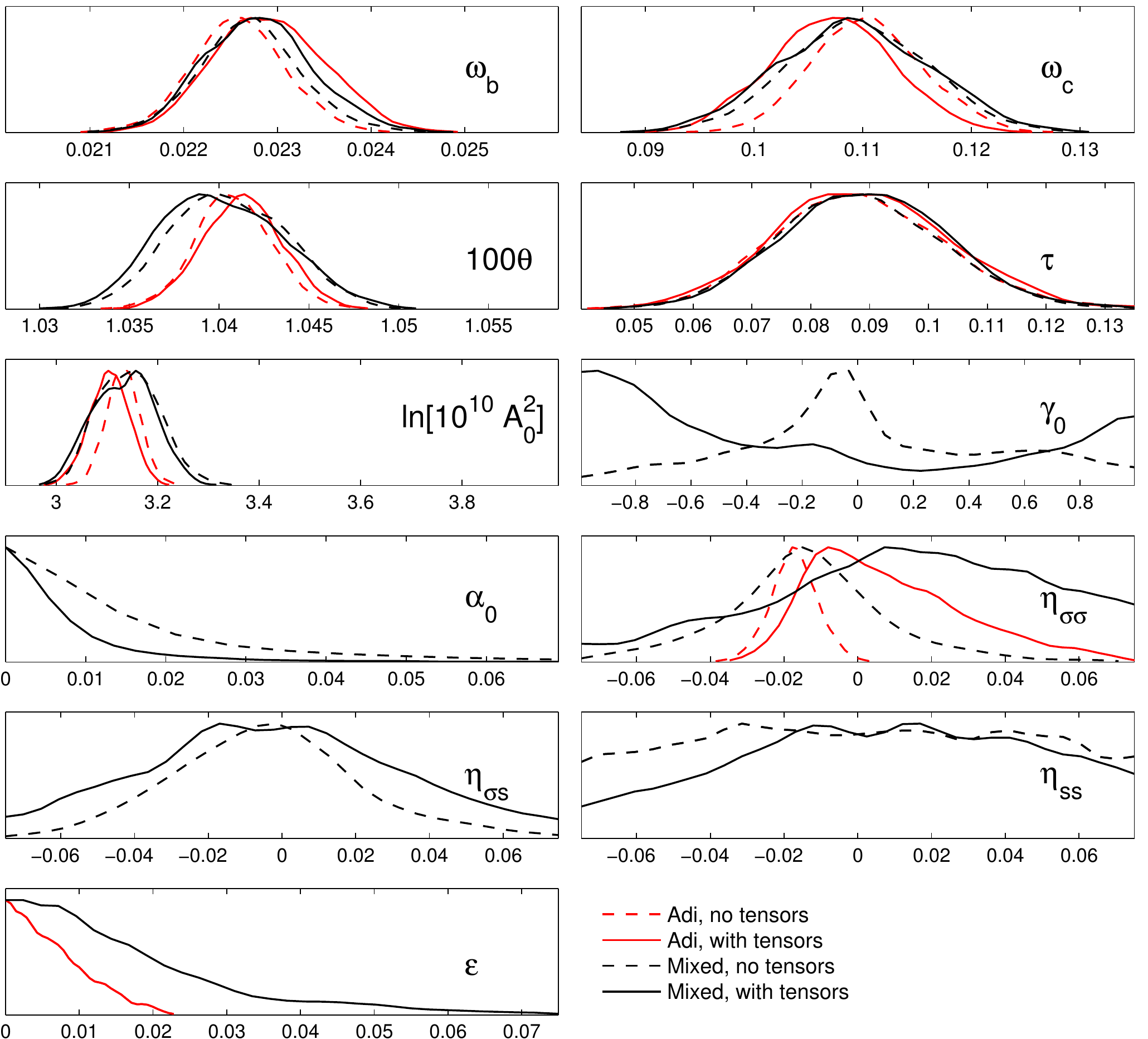}
  \caption{{\bf General case, slow-roll parametrization, CMB data.} Marginalized 1-d posterior probability densities of the primary parameters (except $A_{SZ}$) in slow-roll parametrization. The line styles are the 
same as in Fig.~\ref{fig:AmplPrimary}.}
  \label{fig:InflPrimary}
\end{figure}

\begin{figure}[t]
  \centering
  \includegraphics[width=\columnwidth]{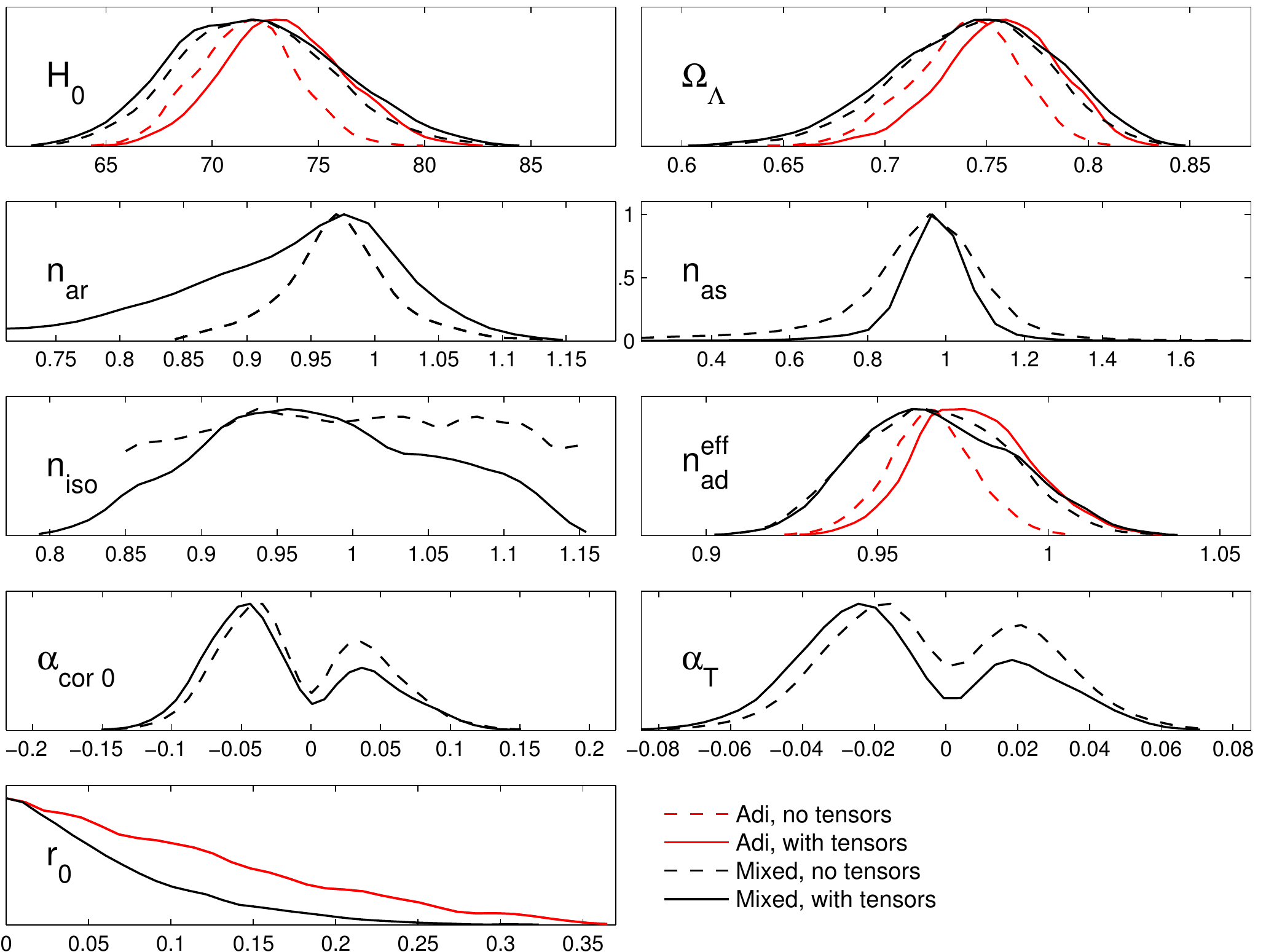}
  \caption{{\bf General case, slow-roll parametrization, CMB data.} Marginalized 1-d posterior probability densities of selected derived parameters from runs made in slow-roll parametrization. The line styles are the same as in Fig.~\ref{fig:AmplPrimary}.}
  \label{fig:InflDerived}
\end{figure}

\begin{figure}[t]
  \centering
 \includegraphics[width=0.98\columnwidth]{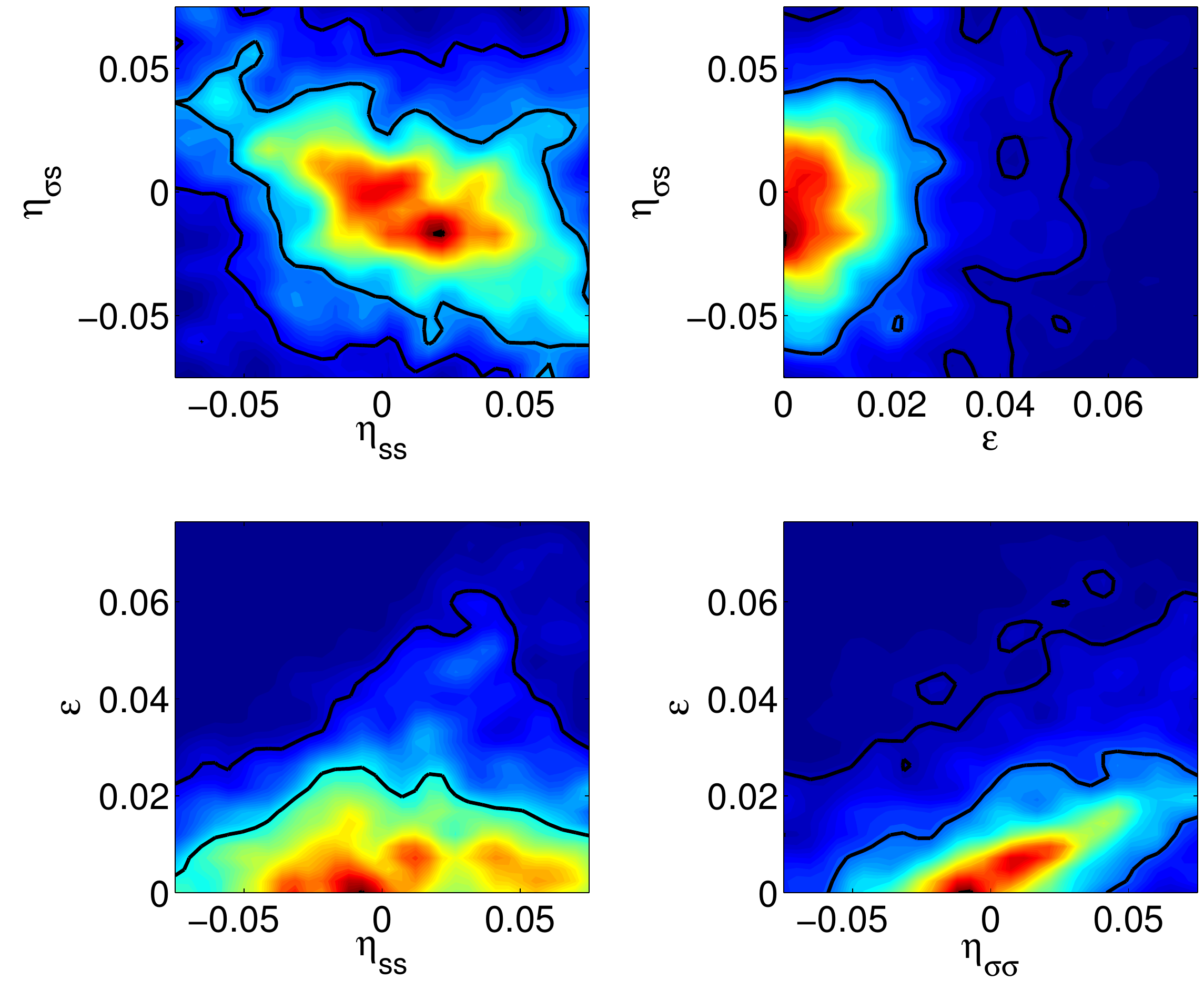}
  \caption{{\bf General case, slow-roll parametrization, CMB data.} Marginalized 2-d posteriors for the two-field inflation slow-roll parameters from MultiNest runs with primordial tensor contribution allowed ($\veps \ge 0$).}
    \label{fig:twod_slow_roll}
\end{figure}

Fig.~\ref{fig:InflPrimary} shows marginalized 1-d posterior probability densities for the primary parameters in the slow-roll parametrization, and Fig.~\ref{fig:InflDerived} selected derived parameters.
We have collected in Fig.~\ref{fig:twod_slow_roll} some 2-d combinations of the four slow-roll parameters in the mixed model with tensors. As indicated in Table~\ref{tab:parameters}, our prior constraint is that the magnitudes of the slow-roll parameters are smaller than 0.075, i.e., $\ll 1$. We see that the data are not able to constrain the slow-roll  parameters this well: the 95\% C.L. contours intersect this prior constraint, except for $\veps$, which we constrain to $\veps < 0.048$ at 95\% C.L. This is three times weaker than the limit $\veps < 0.016$ in the adiabatic model, see Table~\ref{tab:IsocParams} in Appendix~\ref{app:BigTables}. Using wider priors would have the problem, that our approximation of calculating the spectra only to first order in slow-roll parameters (see Sec.~\ref{sec:SlowrollParametrization}) would become questionable; including higher-order terms, on the other hand, would require inclusion of additional slow-roll parameters, whose determination would be infeasible.  Thus the current CMB data are not yet sufficiently accurate for usefully constraining the slow-roll parameters of two-field inflation models, except for $\veps$.


The assumption that the magnitudes of slow-roll parameters are small, requires that all spectra are close to scale invariant [see Eq.~(\ref{eq:BWresults})], except the correlated part of the primordial curvature perturbation in the cases where $\gamma_0$ is close to zero.\footnote{%
\label{foot:nadII} We see from Eq.~(\ref{eq:frac2a}) that the relative amplitudes of the uncorrelated and correlated adiabatic components are $1-|\gamma_0|$ and $|\gamma_0|$ at scale $k_0$. Assuming the slow-roll parameters to be close to zero gives, according to Eq.~(\ref{eq:BWresults}), $\nadI = 1 -6\veps + 2\etzz \sim 1$ and $\nadII = 1 -2\veps + 2\etss - 4\etzs\tan\Delta \sim 1-4\etzs\tan\Delta$. Now the relative amplitudes at another scale, $k$, are $1-|\gamma_0|$ and $|\gamma_0|(k/k_0)^{-4\etzs\tan\Delta}$. Assuming $\gamma_0=-0.0001$ and $\etzs=0.01$ we find $\tan\Delta \approx -100$ and $-4\etzs\tan\Delta \approx 4$. In this case, at $k > 8.4k_0$ (or at $k<0.12k_0$ if $\etzs\times\gamma_0$ had the opposite sign), the correlated adiabatic component would be the dominant component even though, at first sight, we would naively think that $|\gamma_0|=0.0001$ means a negligible contribution from ${\cal P}_{\mr{as}}$. This example shows that there are cases where an extremely small $|\gamma_0|$ leads to a large $\nadII$ and to a large ${\cal P}_{\mr{as}}$ either at the large-scale end or the small-scale end of the CMB angular power spectrum. However, these models are ruled out by the data.
}
These cases are excluded by the data, so in practice also $\nadII$ will be close to 1.
In particular, the isocurvature perturbations are not allowed to have a very different spectral index from the adiabatic perturbations.  
This means, that the isocurvature perturbations contribute much more to large-scale CMB than to small-scale CMB, and the constraints to the isocurvature fraction come mainly from the comparison between the level of large- and small-scale CMB anisotropy, and not from the location of the acoustic peaks in the angular power spectrum.  
The near scale invariance of all spectra also means that the upper limit to $\alpha$ is roughly the same at all scales.
The upper limit $\alpha_0 < 0.026$ turns out to be even tighter than our large-scale ($k = 0.002\,\Mpci$) upper limit $\alpha_1 < 0.064$ in the phenomenological approach.

\begin{figure}[t]
  \centering
 \includegraphics[width=0.98\columnwidth]{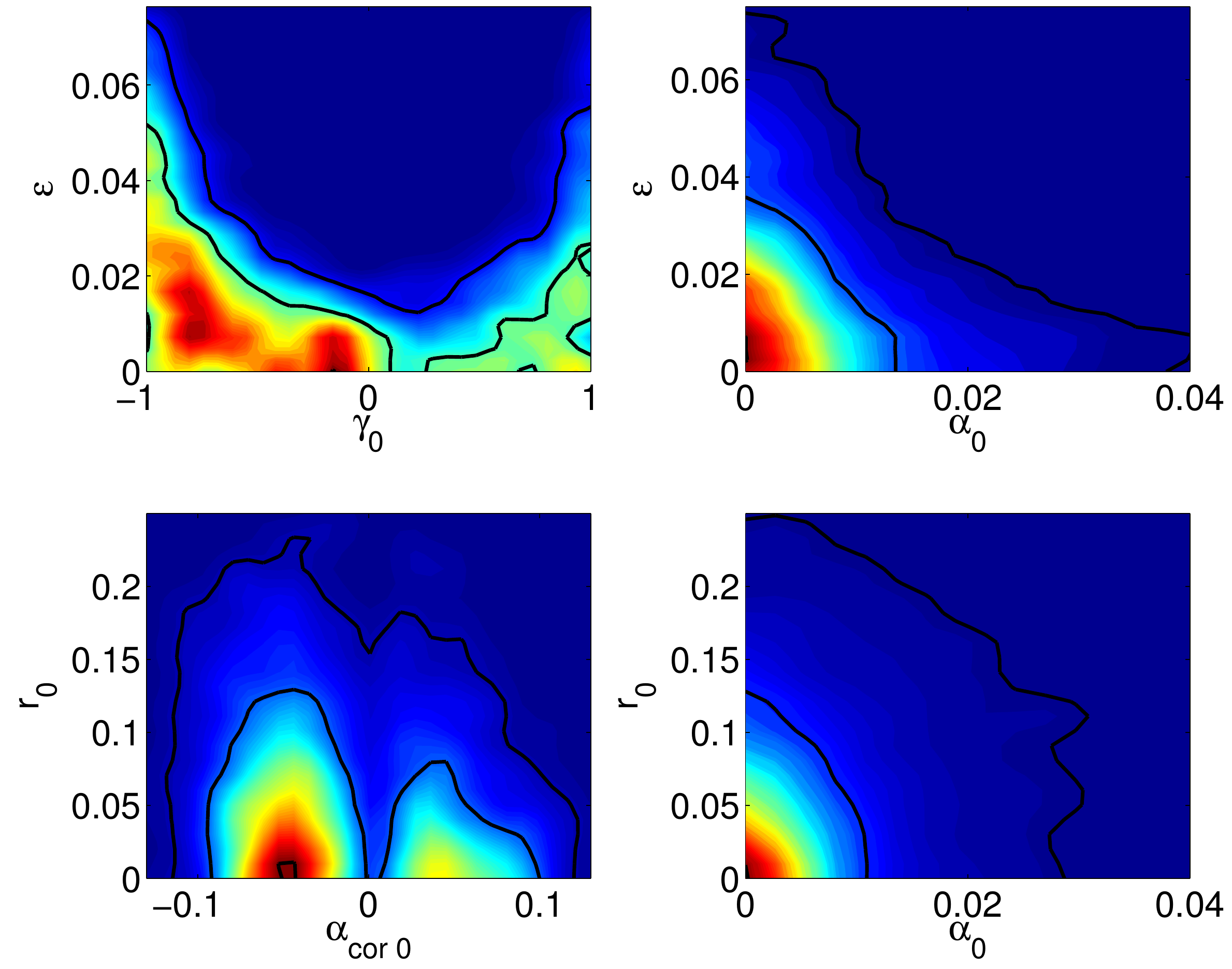}
  \caption{{\bf General case, slow-roll parametrization, CMB data.} Marginalized 2-d posteriors for various combinations of nonadiabaticity ($\gamma_0$, $\alpha_0$, $\alpha_{\mr{cor0}}$) and tensor perturbation parameters ($\veps$, $r_0$).}
    \label{fig:twod_slow_roll_tensor}
\end{figure}

Since tensor perturbations share with isocurvature perturbations the property that, when close to scale invariant, they contribute more to large-scale CMB than to small-scale CMB, we now get a degeneracy between the tensor and nonadiabatic contribution. In Fig.~\ref{fig:twod_slow_roll_tensor} we show how the allowed tensor contribution, parametrized by the primary parameter $\veps$ or the derived parameter $r_0$, depends on the nonadiabaticity parameters $\gamma_0$, $\alpha_0$, and $\alpha_{\mr{cor0}}$. 
Because negative correlations give a negative contribution to large-scale $C_\ell$, whereas tensor perturbations give a positive contribution, a larger tensor contribution is allowed when the isocurvature mode is negatively correlated than when it is positively correlated,
see the left hand side panels of Fig.~\ref{fig:twod_slow_roll_tensor}. Or the other way around, when allowing for the primordial tensor contribution, the 95\% interval of $\alpha_{\mr{cor0}}$ changes from the symmetrical $(-0.092,\,0.092)$ to $(-0.101,\,0.092)$ while the median moves from -0.021 to -0.034, see Table~\ref{tab:IsocParams}. However, since strongly correlated isocurvature perturbations cause a much larger signal for a given $\alpha_0$ than uncorrelated ones, the limit to the CDM isocurvature fraction is formally tighter ($\alpha_0 < 0.026$), when tensor perturbations are allowed, than when they are not ($\alpha_0 < 0.068$). The corresponding numbers for the baryon isocurvature are 0.29 and 0.44.

Allowing for isocurvature perturbations also leads formally to a tighter limit on the amplitude of tensor perturbations ($r_0 < 0.18$) than in the adiabatic case ($r_0 < 0.26$). The main reason for this is the $\tan\Delta$ term in the expression for $\nadII$ in Eq.~(\ref{eq:BWresults}). As explained in footnote \ref{foot:nadII}, if $\gamma_0$ was too close to zero, the ${\cal P}_{\rm{as}}$ spectrum could be steeply red or blue, and hence excluded by the data. This disfavors those values of $\gamma_0$  that would lead to a large $r_0$, since according to Eqs.~(\ref{eq:BWresults}) and (\ref{eq:Deltarelations}) $r_0 = 16\veps (1-|\gamma_0|)$. Moreover, due to the $1-|\gamma_0|$ factor, and due to large (negative) correlations being favored, the constraint $\veps < 0.048$ is much weaker than in the adiabatic case ($\veps < 0.016$).


The $1-|\gamma_0|$ factor in $r_0$ together with the upper left corner plot of Fig.~\ref{fig:twod_slow_roll_tensor} also explains the difference of 1-d posteriors of $\gamma_0$ in Fig.~\ref{fig:InflPrimary}:
The data constrain the primordial tensor amplitude, $r_0$, but whenever $|\gamma_0|$ is close to 1, we obtain an acceptably small value for $r_0$ no matter what value the primary parameter $\veps$ has. Therefore, when marginalizing over $\veps$, there is a large volume of well-fitting models when $|\gamma_0|$ is close to 1. In contrast, when $\gamma_0$ is close to zero, $\veps$ must be small in order to obtain a small $r_0$. Hence the well-fitting parameter space volume is small, if $|\gamma_0|$ is small. This means that the difference between 1-d posteriors of $\gamma_0$ of ``no tensors'' and ``with tensors'' cases is due to a marginalization artifact, and the apparent preference of $\gamma_0 = -1$ with tensors is not strictly speaking physical. From the mean likelihoods (indicated by the colour scale) in Fig.~\ref{fig:twod_slow_roll_tensor} we see that indeed the best-fitting models have only a mildly negative correlation, $\gamma_0 \sim -0.1$, and the good-fit region covers the range $(-0.95,\,0)$, but does not particularly favour $\gamma_0 = -1$.
Indeed, often 1-d marginalized posteriors give limited information on high-dimensional problems, in particular, if some parameters are loosely constrained: although the 1-d Figs.~\ref{fig:InflPrimary} and \ref{fig:InflDerived} give only weak hints, the 2-d Fig.~\ref{fig:twod_slow_roll_tensor} supports the conclusion that in the slow-roll parametrization, negative correlation (negative $\gamma_0$ and $\alpha_{\mr{cor0}}$) is clearly favoured --- slightly more favoured with tensors than without them.

The amplitude parametrization favours a positive correlation. So why does the slow-roll parametrization favour a negative correlation? The explanation lies again in the spectral indices, i.e., what part of the angular power spectrum the nonadiabatic contribution is able to modify. In the amplitude parametrization the width (and location) of the acoustic peaks got slightly modified due to large $\niso$ and $\ncor$. In the slow-roll parametrization the isocurvature mode does not have the ability to modify this region, since $\niso \sim \nadI$. So, the nonadiabatic contribution can significantly modify only the large-scale ($\ell\lesssim 50$) part of $C_\ell$. There the angular power at some multipoles is lower in the data than predicted by the adiabatic $\Lambda$CDM model, most notably the quadrupole, $\ell=2$. An uncorrelated or positively correlated isocurvature contribution is thus disfavoured as they would further increase the discrepancy between the data and theory, but a negative correlation can reduce the power in this region. However, due to the cosmic variance the contribution to $\chi^2$ from this region does not have as much weight as the one from the acoustic peak region, and thus the difference of the best-fit mixed model and adiabatic model is only $\Delta\chi^2 \approx -2$ in favour of the mixed model (while it is $-3$ in the amplitude parametrization).

The shapes of the preferred regions in the slow-roll parameter space can be understood by the preference for certain values of the spectral indices. From Eq.~(\ref{eq:BWresults}) we have that $\nadI \sim 1$ corresponds to $\veps \sim 
\etzz/3$. Since $\nadI$ somewhat less than 1 is preferred by the data, this line is shifted a bit to the left in the $(\etzz,\veps)$ plane, see Fig.~\ref{fig:twod_slow_roll}.  The slope of the good-fit region in the ($\etss,\etzs$) plane is about $-2/3$, i.e., 
$2\etss-3\etzs \approx \mbox{const.}$, which corresponds to $\tan\Delta \approx -3/4$ or $\gamma_0 \approx -0.64$, which is in line with negative correlations being preferred.

The effect of the optical depth $\tau$ on the CMB temperature angular power spectrum is also somewhat degenerate with isocurvature and tensor perturbations, affecting the relative level of low and high $\ell$, but since $\tau$ is mainly determined by low-$\ell$ WMAP7 polarization data, where it causes the distinctive reionization bump, isocurvature perturbations have negligible effect on the determination of $\tau$.

\begin{figure}[t]
  \centering
  \includegraphics[width=\columnwidth]{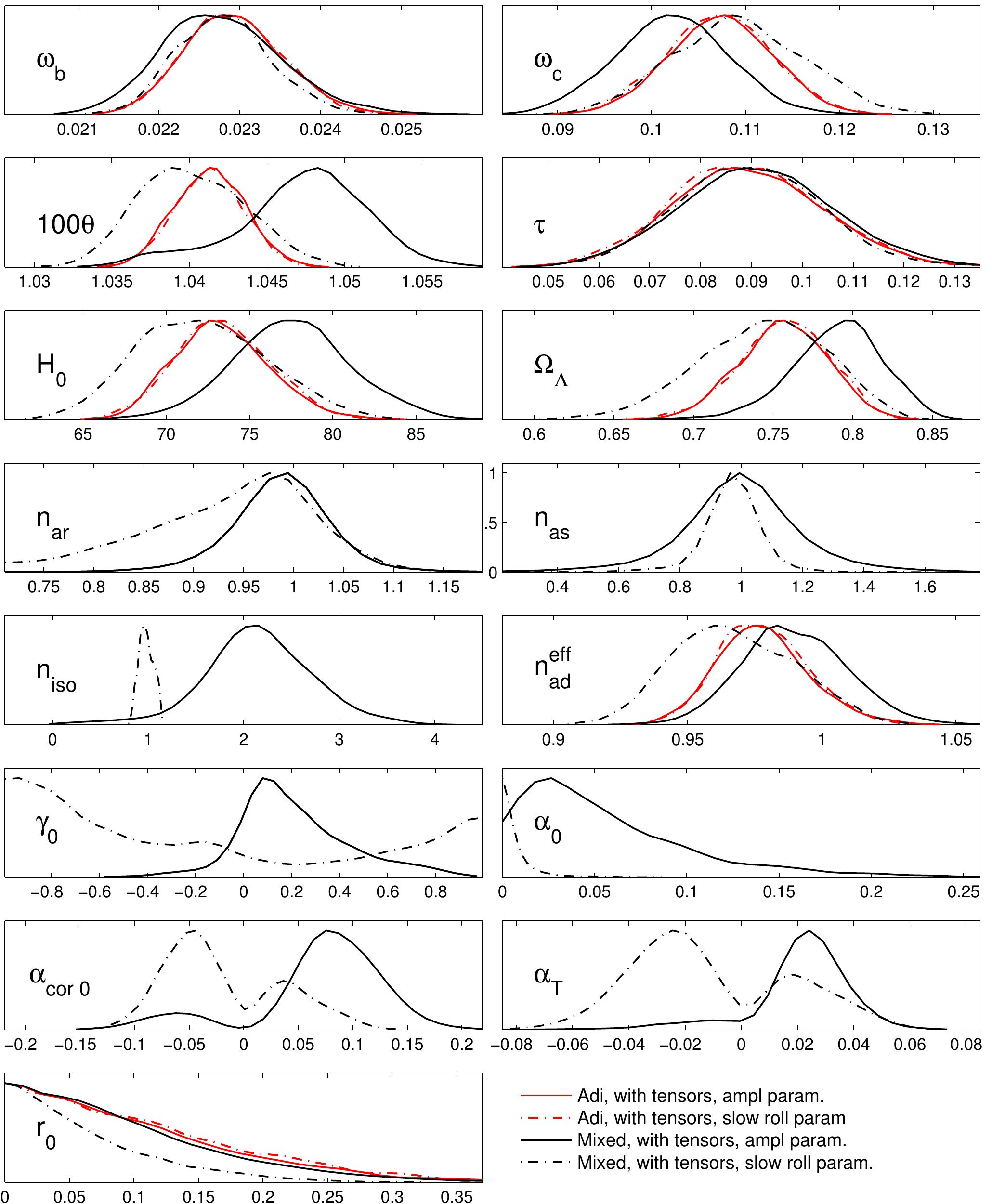}
  \caption{{\bf General case, comparison of amplitude and slow-roll parametrizations, CMB data.} Marginalized 1-d posterior probability densities of selected parameters from runs made in amplitude and slow-roll parametrizations, allowing for a tensor contribution. The solid lines are for the amplitude parametrization and the dot-dashed for the slow-roll parametrization, red color for the adiabatic and black for the mixed model. 
}
  \label{fig:AmplVsInfl}
\end{figure}
\begin{figure}[t]
  \centering
  \includegraphics[width=\columnwidth]{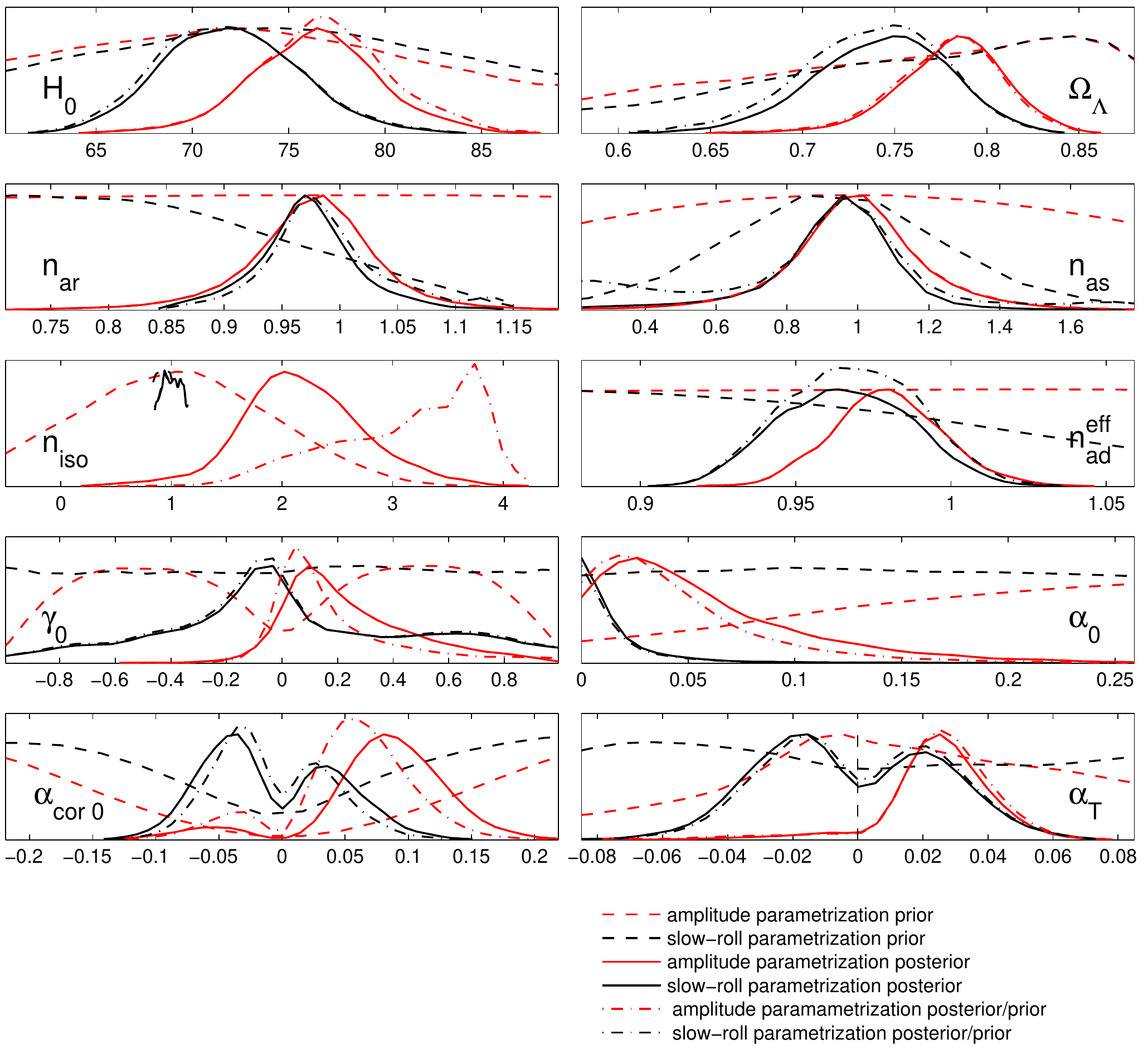}
  \caption{Priors (dashed) and posteriors (solid) of selected derived parameters, and posteriors divided by priors (dot-dashed). The red color is for the amplitude parametrization and the black for the slow-roll parametrization.
}
  \label{fig:Priors}
\end{figure}

Finally, the odds for the adiabatic model compared to the mixed model in the slow-roll parametrization are
$24:1$
(${\cal B}_f = +3.19$) with tensors,
and
$80:1$
(${\cal B}_f = +4.38$) without tensors,
in support of the adiabatic model.
In the adiabatic model, with our priors for the slow-roll parameters, we find again support for the model \emph{ without tensor contribution}, $P_{\rm no\ tensors} / P_{\rm tensors} = 9:1$ 
(${\cal B}_f=+2.23$). In the mixed models this changes to
$P_{\rm no\ tensors} / P_{\rm tensors} = 2.8:1$ 
(${\cal B}_f=+1.04$), but
recalling our error estimate $\delta{\cal B}_f \approx \pm 0.4$,
the last result could fall into the category ``inconclusive''.

\subsection{Comparison of parametrizations, and priors of  derived parameters}

To allow for an easy comparison between our phenomenological and two-field inflation approaches, we repeat the 1-d posteriors of some parameters of the adiabatic and mixed models with tensors in Fig.~\ref{fig:AmplVsInfl}. First we notice that the parameters of the adiabatic model are so well constrained that the two parametrizations lead to almost identical results.

Since the allowed isocurvature contribution is smaller in the two-field inflation approach than in the phenomenological approach, the preferred values for the 
background parameters in the mixed model are closer to the adiabatic case in the slow-roll parametrization (see Fig.~\ref{fig:AmplVsInfl}). Even more important is the preferred negative sign of the correlation. Due to this, the preferred value of the sound horizon angle moves to the other side (smaller) of the adiabatic result, which is then reflected in $\omega_c$, $H_0$, and $\Omega_\Lambda$. Since the negative correlation reduces the power at large scales compared to the small scales, the (slow-roll) models with negative correlation prefer a smaller spectral index $\nad^{\mr{eff}}$.

\begin{figure*}[t]
  \centering
  \includegraphics[width=0.87\columnwidth]{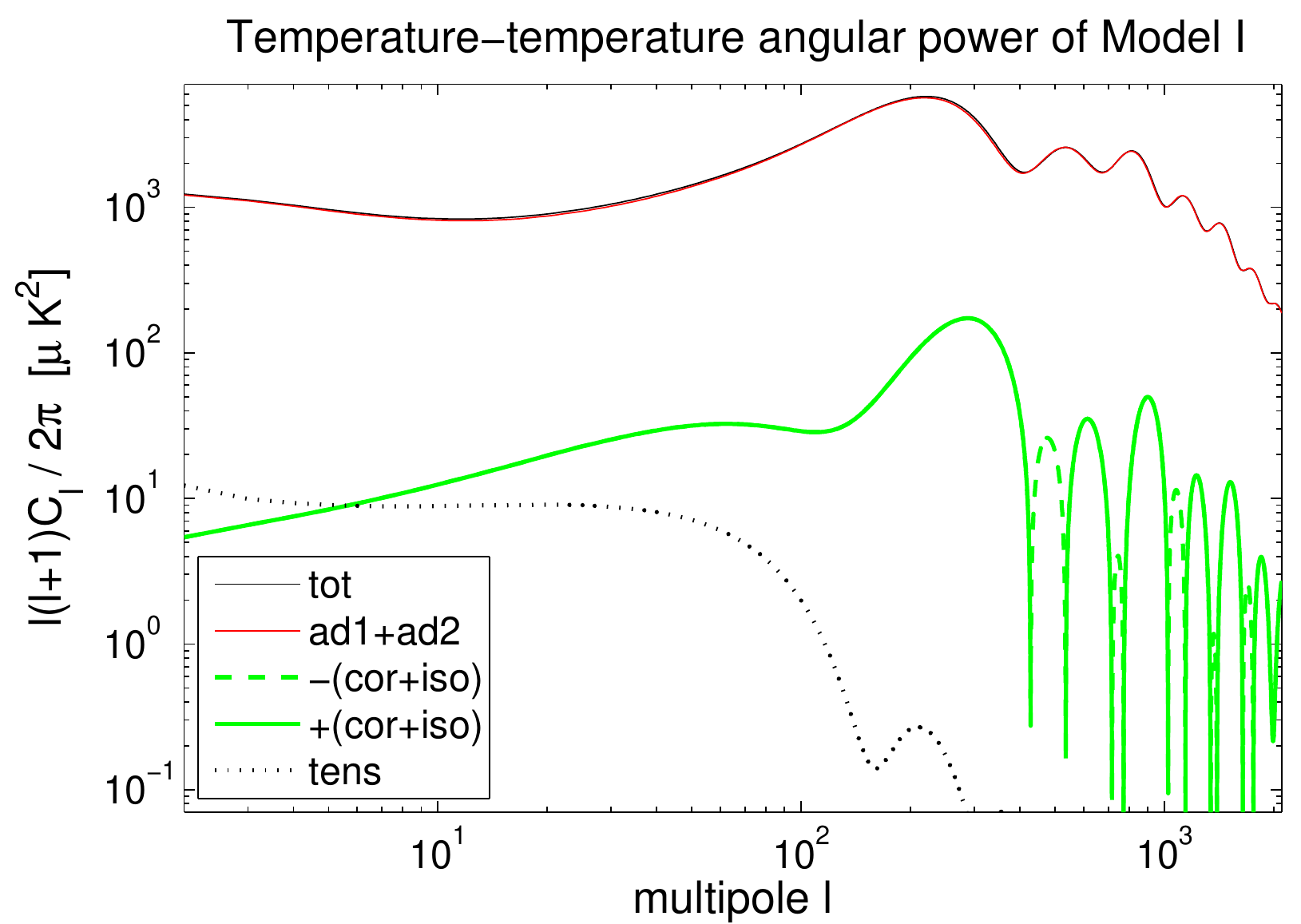}
  \includegraphics[width=0.87\columnwidth]{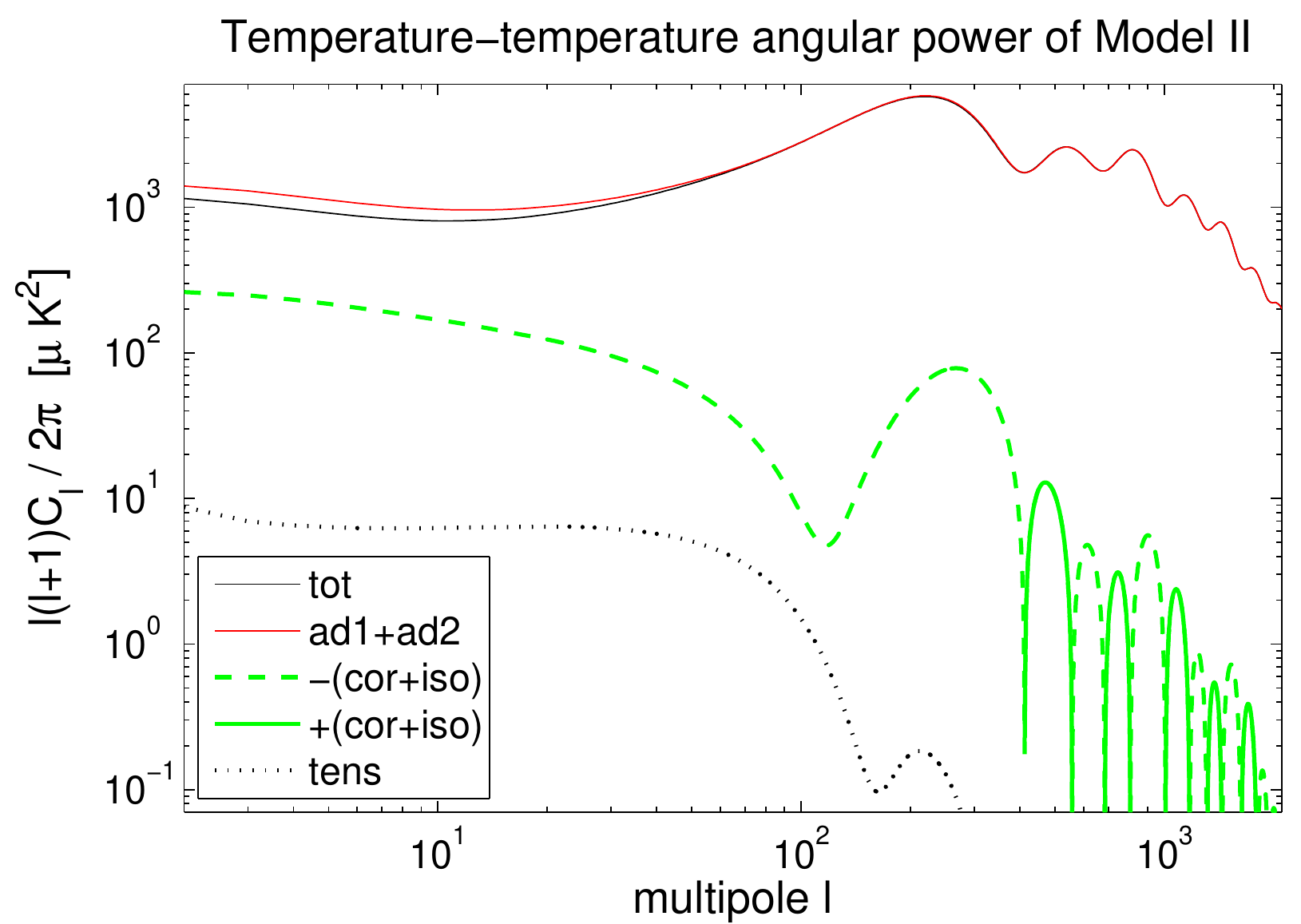}
  \caption{Typical temperature angular power spectra of well-fitting mixed models with a non-negligible nonadiabatic contribution.  Model I ($\alpha_T=+0.025$) is from our amplitude parametrization runs and Model II ($\alpha_T=-0.049$) from our slow-roll parametrization runs. Both models are within $\Delta\chi^2 \approx 4$ from the best-fit adiabatic model. Solid black curves are for the total lensed $\ell(\ell+1)C_\ell/(2\pi)$, the solid red for the adiabatic contribution, the solid green for the positive nonadiabatic contribution, the dashed green for the negative nonadiabatic contribution, and the dotted black for the tensor contribution. Parameters of these models are reported in  Tables \ref{tab:CommonParams} and  \ref{tab:IsocParams} in Appendix~\ref{app:BigTables}.}
  \label{fig:ExampleModels}
\end{figure*}


It should be kept in mind that for the derived parameters the prior is not necessarily flat, and hence the posteriors could in some cases directly reflect the prior or be biased by it. We have checked that this is not the case for our derived parameters. In Fig.~\ref{fig:Priors} we show the prior in both parametrizations for selected derived parameters.  We plot also the posterior divided by the prior, which indicates how the results would look like if we had a flat prior for the derived parameters. Importantly, the prior is almost flat for all of our derived parameters (including the isocurvature parameters) in the region of the peak of their posterior. There is one exception: since the CMB data do not constrain $\niso$, a flat wide range prior for it (like in the spectral index parametrization) would lead to very large values being ``preferred'' \cite{KurkiSuonio:2004mn,Sollom:2009vd}, $\niso \sim 3$--$7$ depending on the chosen pivot scale. (See figure 21 in \cite{KurkiSuonio:2004mn} and figure 5 in \cite{Valiviita:2009bp}. Note from the latter figure also that although $\alpha_T$ is a pivot-scale free measure of non-adiabaticity, its posterior probability density is affected by the choice of pivot scale.). The amplitude parametrization leads naturally to a prior that excludes models with $\niso \gtrsim 3.5$, see Fig.~\ref{fig:Priors}, and the slow-roll parametrization limits $\niso$ close to one: $1 -2\max(\veps) +2\min(\etss) \le \niso \le 1 -2\min(\veps) +2\max(\etss)$.

It is also interesting to compare how the parametrization and prior ranges (that are considered as part of the model in Bayesian thinking) affect the Bayesian evidence. In the adiabatic case with tensors we find ${\cal B}_f = 1.30$ and in the mixed case ${\cal B}_f=3.59$ in favour of the slow-roll parametrization. This is what one would have expected, since the prior ranges are so much narrower in the slow-roll parametrization.

To complete the comparison of parametrizations, we show the temperature $C_\ell$ spectra of two example models in Fig.~\ref{fig:ExampleModels}: Model I is from our amplitude parametrization run and has roughly the maximum 95\% C.L. allowed primordial nonadiabatic contribution at scale $k_2$ ($\alpha_{\mr{cor2}} = 0.42$). Model II is from our slow-roll parametrization run and has roughly the maximum allowed negative nonadiabatic contribution at $k_1$ ($\alpha_{\mr{cor1}} = -0.11$). These models represent the above-described typical features of well-fitting mixed models in amplitude and slow-roll parametrizations, respectively. The parameters of these examples can be found in Tables \ref{tab:CommonParams} and  \ref{tab:IsocParams} in Appendix~\ref{app:BigTables} under names ``Model I'' and ``Model II''.


\section{The general case with additional data}
\label{sec:ExtraData}

\begin{figure}[t]
  \centering
  \includegraphics[width=\columnwidth]{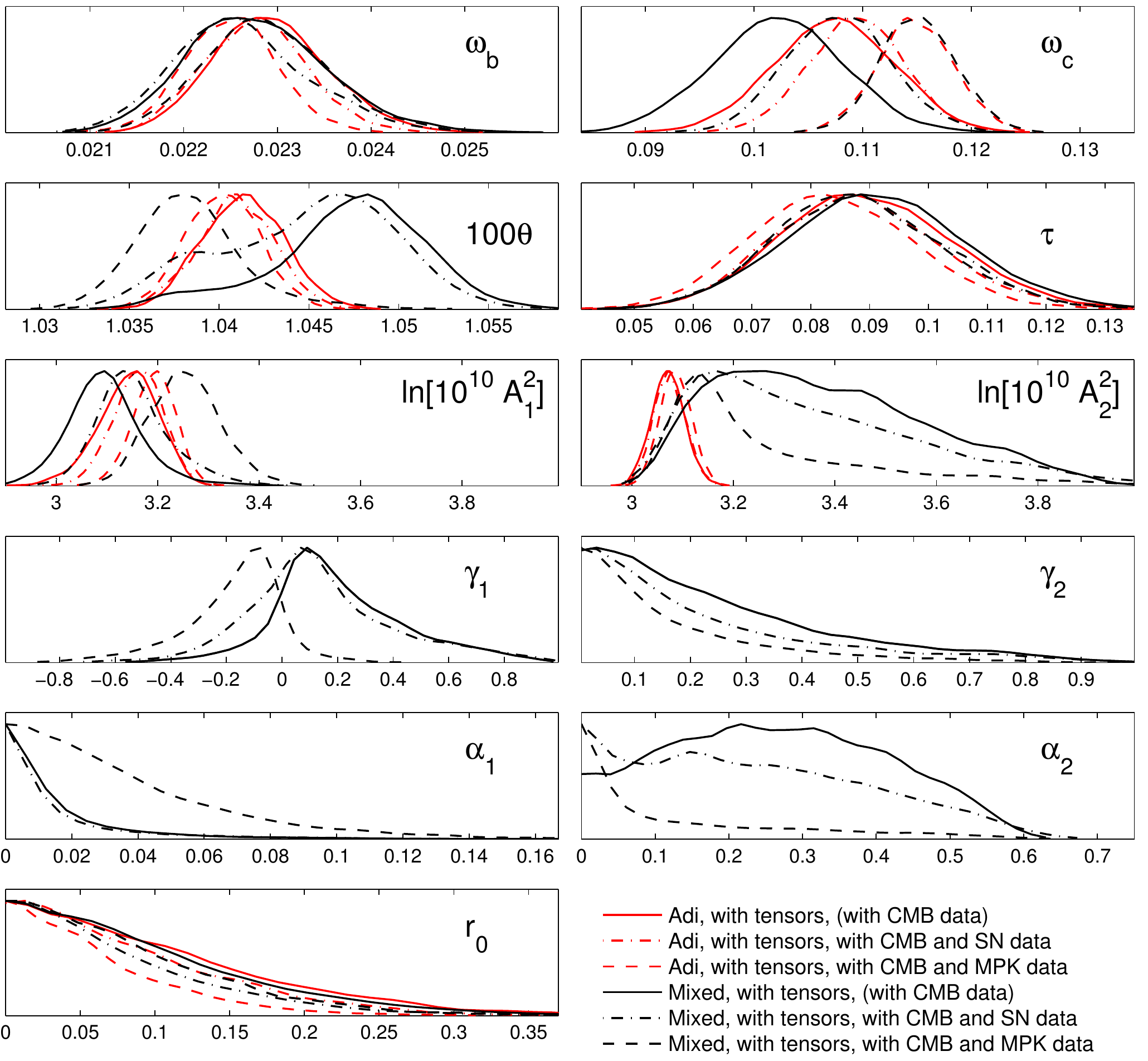}
  \caption{{\bf General case, amplitude parametrization, comparison of CMB, CMB and SN, and CMB and MPK data.}  Marginalized 1-d posterior probability densities of the primary parameters in amplitude parametrization with different data sets: CMB (solid line style), CMB and SN (dot-dashed), or CMB and MPK (dashed). The red color is for the adiabatic and black for the mixed model.
}
  \label{fig:AmplDataSetsPrimary}
\end{figure}

\begin{figure}[h!]
\vspace{-2mm}
  \centering
  \includegraphics[width=\columnwidth]{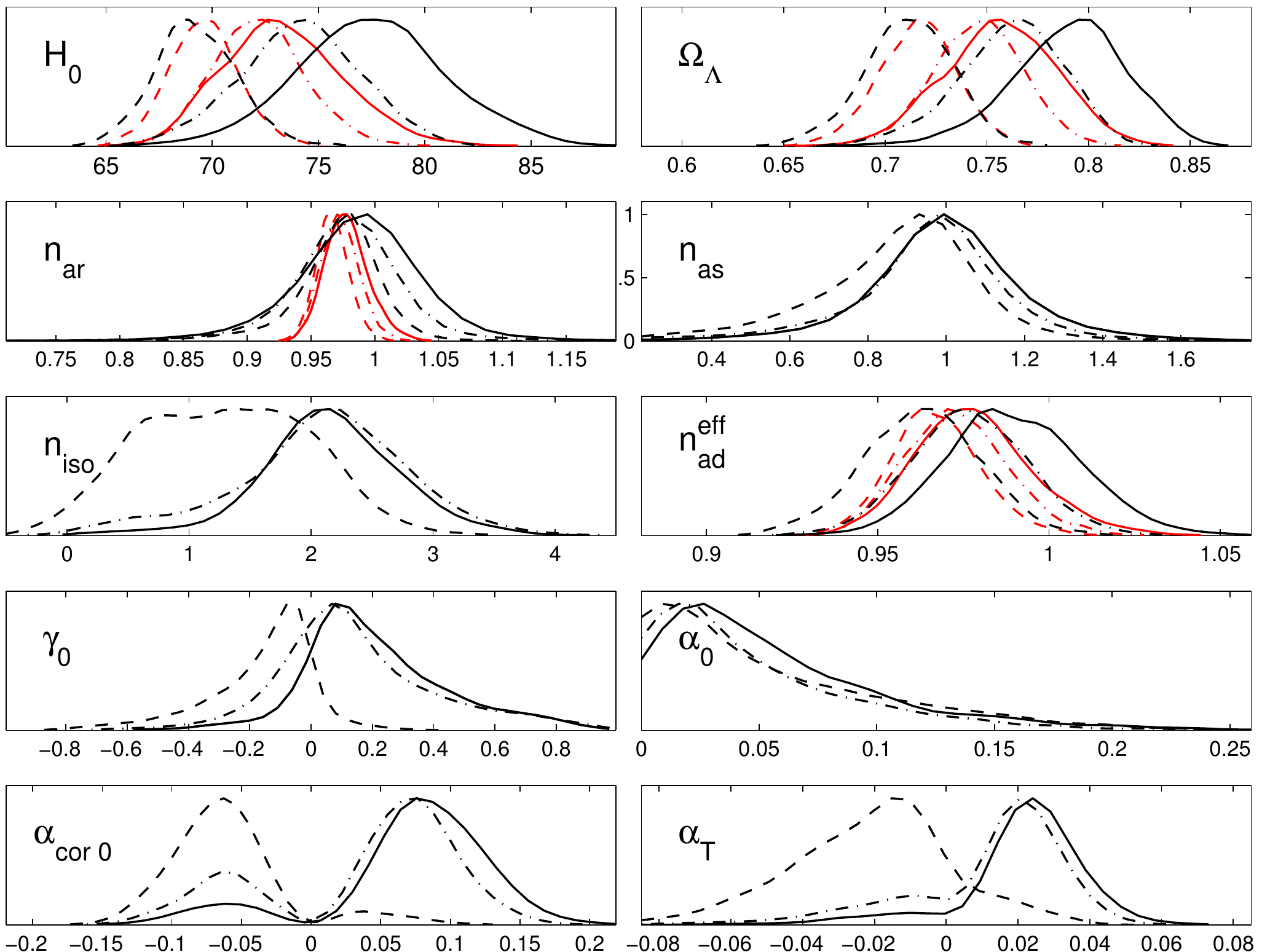}
  \caption{{\bf General case, amplitude parametrization, comparison of CMB, CMB and SN, and CMB and MPK data.} Marginalized 1-d posterior probability densities of selected derived parameters from runs made in amplitude parametrization with different data sets. The line styles are the same as in Fig.~\ref{fig:AmplDataSetsPrimary}.}
  \label{fig:AmplDataSetsDerived}
\end{figure}

\begin{figure}[t]
  \centering
  \includegraphics[width=\columnwidth]{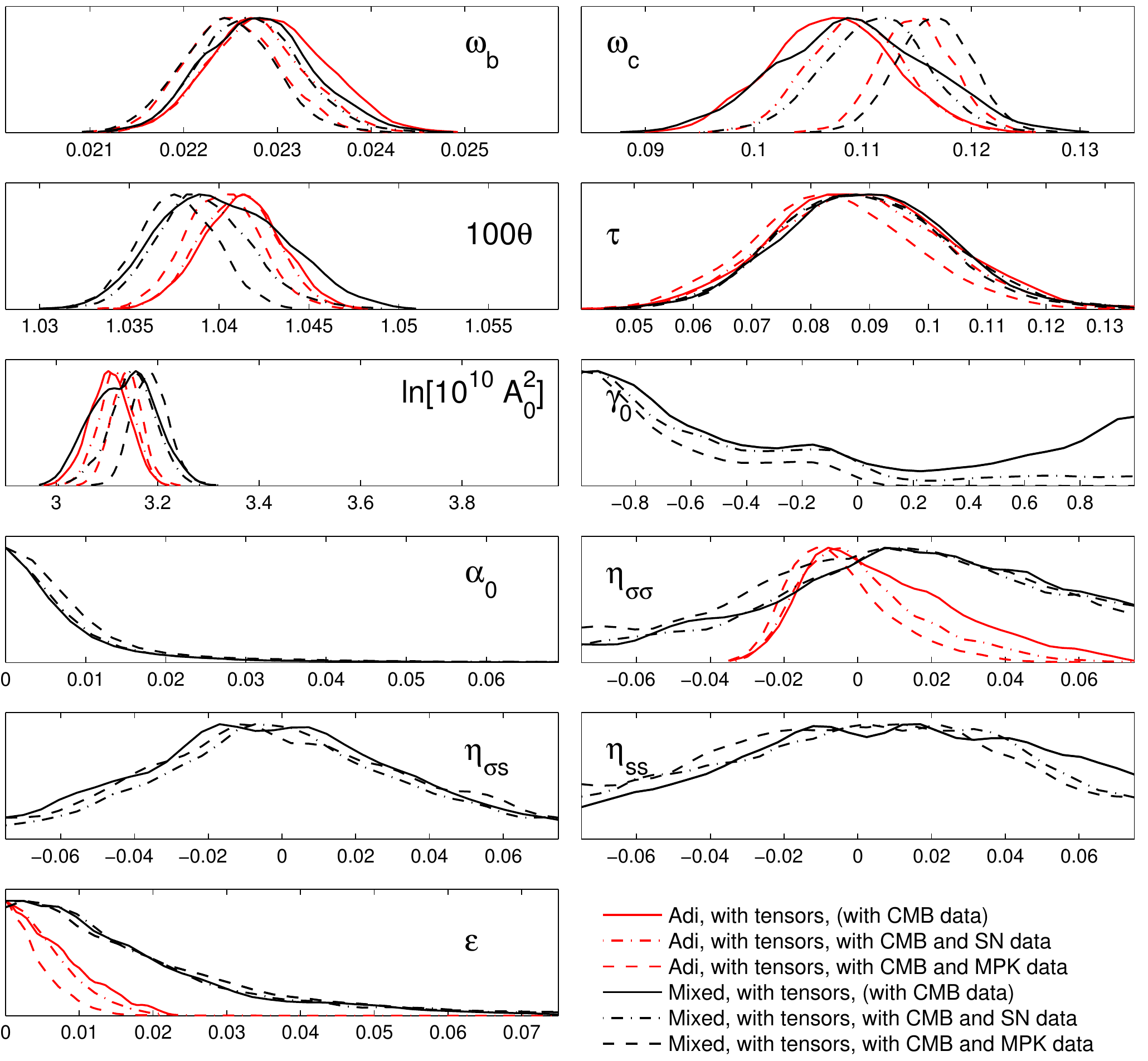}
  \caption{{\bf General case, slow-roll parametrization, comparison of CMB, CMB and SN, and CMB and MPK data.} Marginalized 1-d posterior probability densities of the primary parameters in slow-roll parametrization with different data sets. The line styles are the same as in Fig.~\ref{fig:AmplDataSetsPrimary}.}
  \label{fig:InflDataSetsPrimary}
\end{figure}

\begin{figure}[t]
\vspace{3.5mm}
  \centering
  \includegraphics[width=\columnwidth]{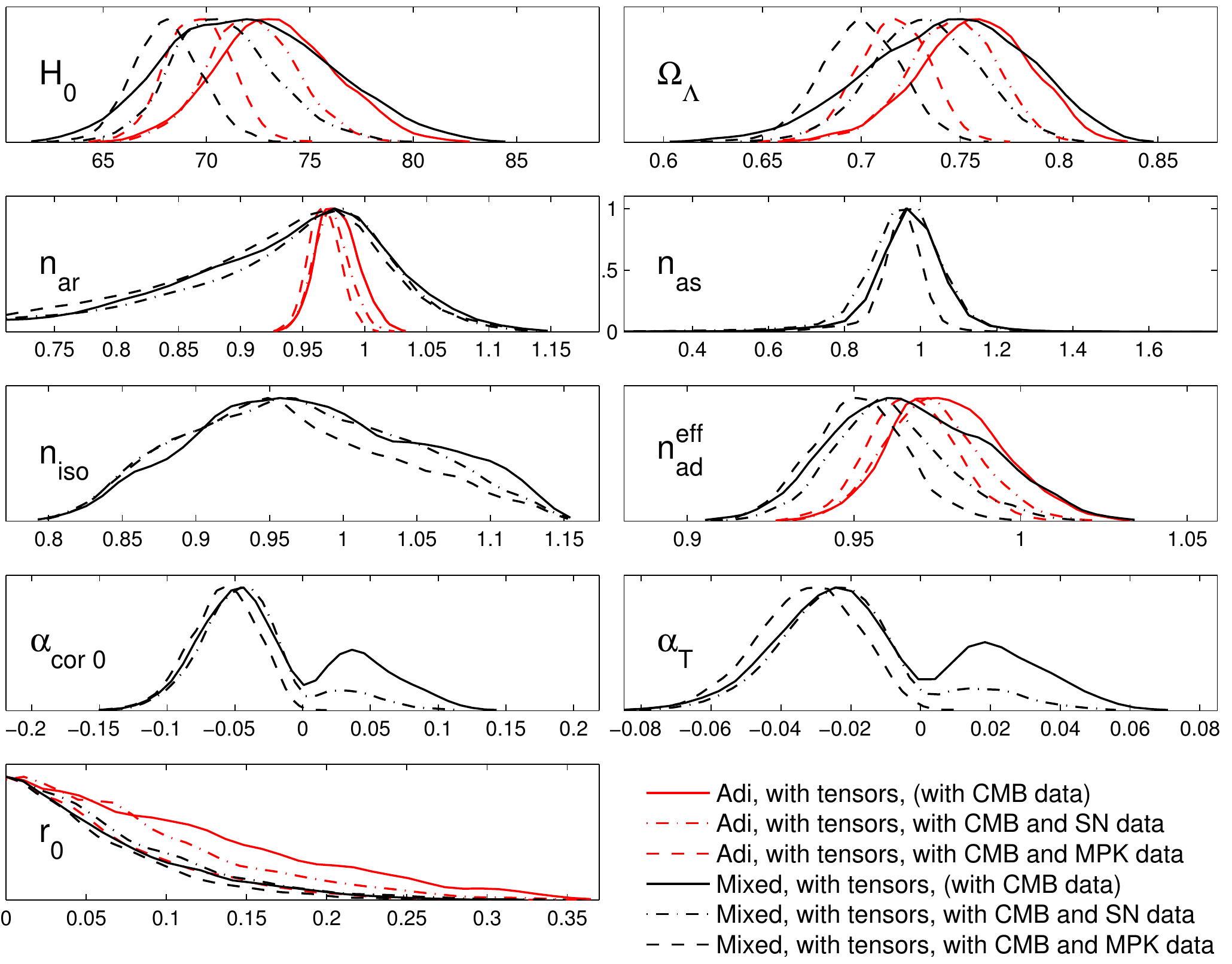}
  \caption{{\bf General case, slow-roll parametrization, comparison of CMB, CMB and SN, and CMB and MPK data.} Marginalized 1-d posterior probability densities of selected derived parameters from runs made in slow-roll parametrization with different data sets.  The line styles are the same as in Fig.~\ref{fig:AmplDataSetsPrimary}.}
  \label{fig:InflDataSetsDerived}
\end{figure}

We study how the results are affected by using supernova (SN) \cite{Amanullah:2010vv} or matter power spectrum (MPK) \cite{Reid:2009xm} data in addition to the same CMB data as in the previous sections. For brevity and clarity, we show figures only for models with tensors. Fig.~\ref{fig:AmplDataSetsPrimary} shows the 1-d posteriors of the primary parameters of amplitude parametrization for different data sets: 1) with CMB only, 2) with CMB and SN and 3) with CMB and MPK. Fig.~\ref{fig:AmplDataSetsDerived} gives the  same sets for derived parameters. The analysis is repeated in the slow-roll parametrization in Figs.~\ref{fig:InflDataSetsPrimary} and \ref{fig:InflDataSetsDerived}.



In the phenomenological approach, the additional data exclude those mixed models that have a large $\Omega_\Lambda$ (see Fig.~\ref{fig:AmplDataSetsDerived}). These models have a large positive nonadiabatic contribution, which thus gets less favoured (see $\gamma_0$, $\alpha_{\mr{cor0}}$, and $\alpha_T$). Therefore, by constraining the background parameters (see also $H_0$ in  Fig.~\ref{fig:AmplDataSetsDerived} and $\omega_c$, $\theta$ in  Fig.~\ref{fig:AmplDataSetsPrimary}) closer to the ``adiabatic values'' or to less favorable values to isocurvature, the positive nonadiabatic contribution gets \emph{indirectly} constrained tighter than with the CMB data alone: the upper limit for $\alpha_T$ changes from 4.9\% (CMB) to 4.2\% (CMB and SN) or to 2.4\% (CMB and MPK).

The additional data have less effect in the slow-roll approach, since the background parameters were already closer to the adiabatic case. Actually the additional data prefer a smaller $\Omega_\Lambda$ than the CMB data in the adiabatic model, which is the same direction where the isocurvature contribution pulls in the slow-roll approach. Therefore, including the additional data make the mixed model more favourable than with the CMB data alone. The upper limit to the CDM isocurvature fraction relaxes from $\alpha_0 < 0.026$ (CMB) to $0.032$ (CMB and SN) or to $0.034$ (CMB and MPK). From the $\alpha_T$ plot in Fig.~\ref{fig:InflDataSetsDerived} it is clear that the preference for small $\Omega_\Lambda$ leads to a very strong preference for a negative nonadiabatic contribution. The 95\% C.L. range for $\alpha_T$ in the slow-roll parametrization changes from $\alpha_T \in (-5.8\%,4.5\%)$ with CMB to  $(-5.6\%,3.0\%)$ with CMB and SN or to $(-6.3\%,-0.8\%)$ with CMB and MPK. So, in the slow-roll parametrization the MPK data exclude positive nonadiabatic contribution with more than 95\% C.L.

The results with SN data are between the results of CMB alone and the ones with CMB and MPK data for most of the parameters. This happens, since the SN constrain $\Omega_\Lambda$ only mildly and are in a relatively good agreement with the CMB. In contrast, the MPK data prefer a significantly smaller $\Omega_\Lambda$ (larger $\Omega_m$) than the CMB \cite{Reid:2009xm,Percival:2009xn}. 
If this discrepancy in the determination of $\Omega_m$ persists in the future, then MPK data will continue to have a dramatic \emph{indirect} effect on the determination of isocurvature.

In the amplitude parametrization the MPK data might lead to also a \emph{direct} extra constraint on nonadiabatic \emph{perturbations}. At first sight, the $\niso$ plot of Fig.~\ref{fig:AmplDataSetsDerived} seems to support this: the MPK data exclude the largest isocurvature spectral indices which the CMB data are not able to constrain. More clearly this effect can be seen in the $\alpha_2$ plot of Fig.~\ref{fig:AmplDataSetsPrimary}. However, as we will see below, in practice the extra constraint on $\alpha_2$ comes almost fully from the background constraint on $\Omega_m$. This can be also realized by noticing that the SN data (that are purely background data) have a similar, but milder, effect.

\begin{figure*}[t]
  \centering
  \includegraphics[width=\columnwidth]{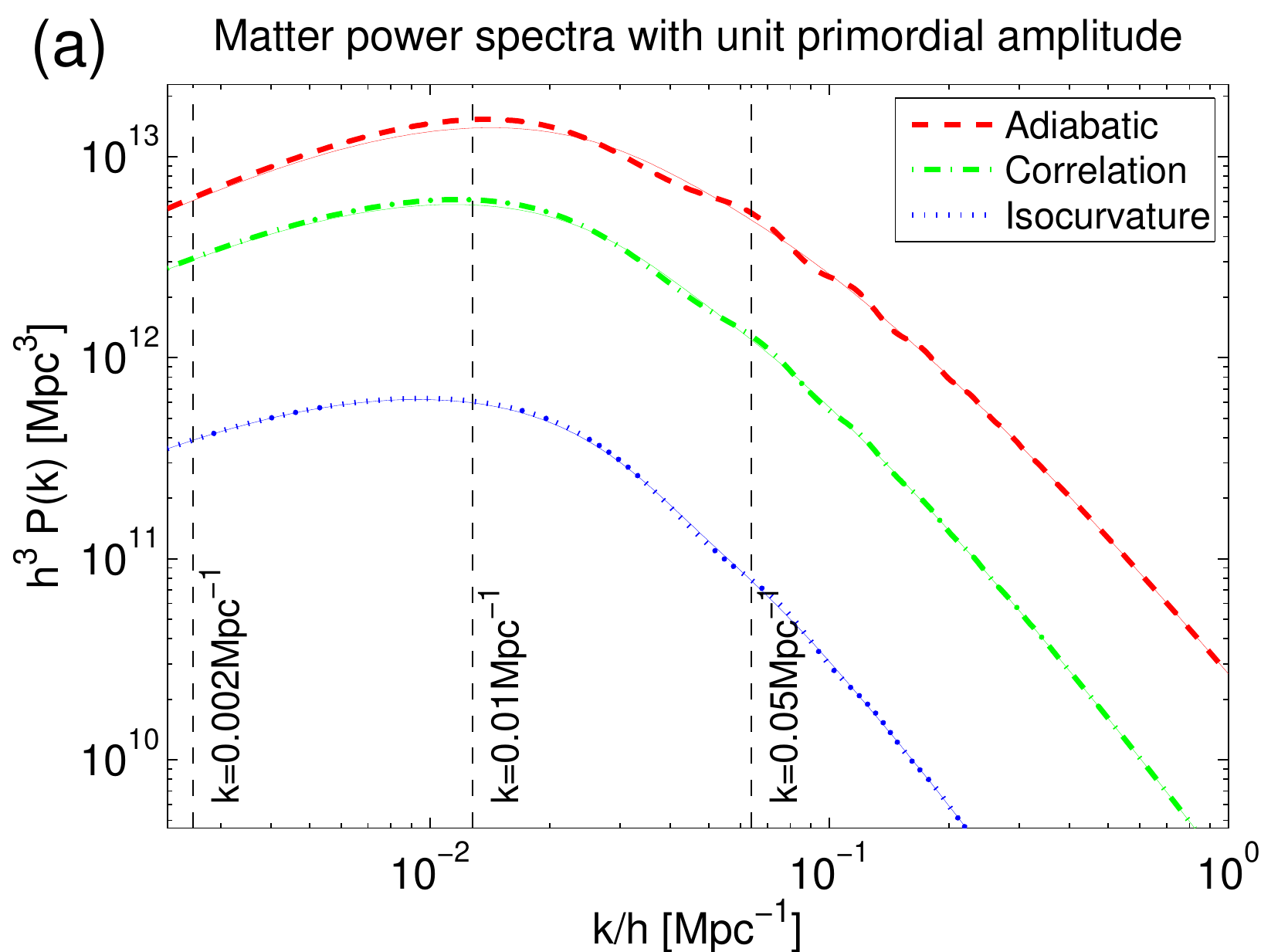}
  \includegraphics[width=\columnwidth]{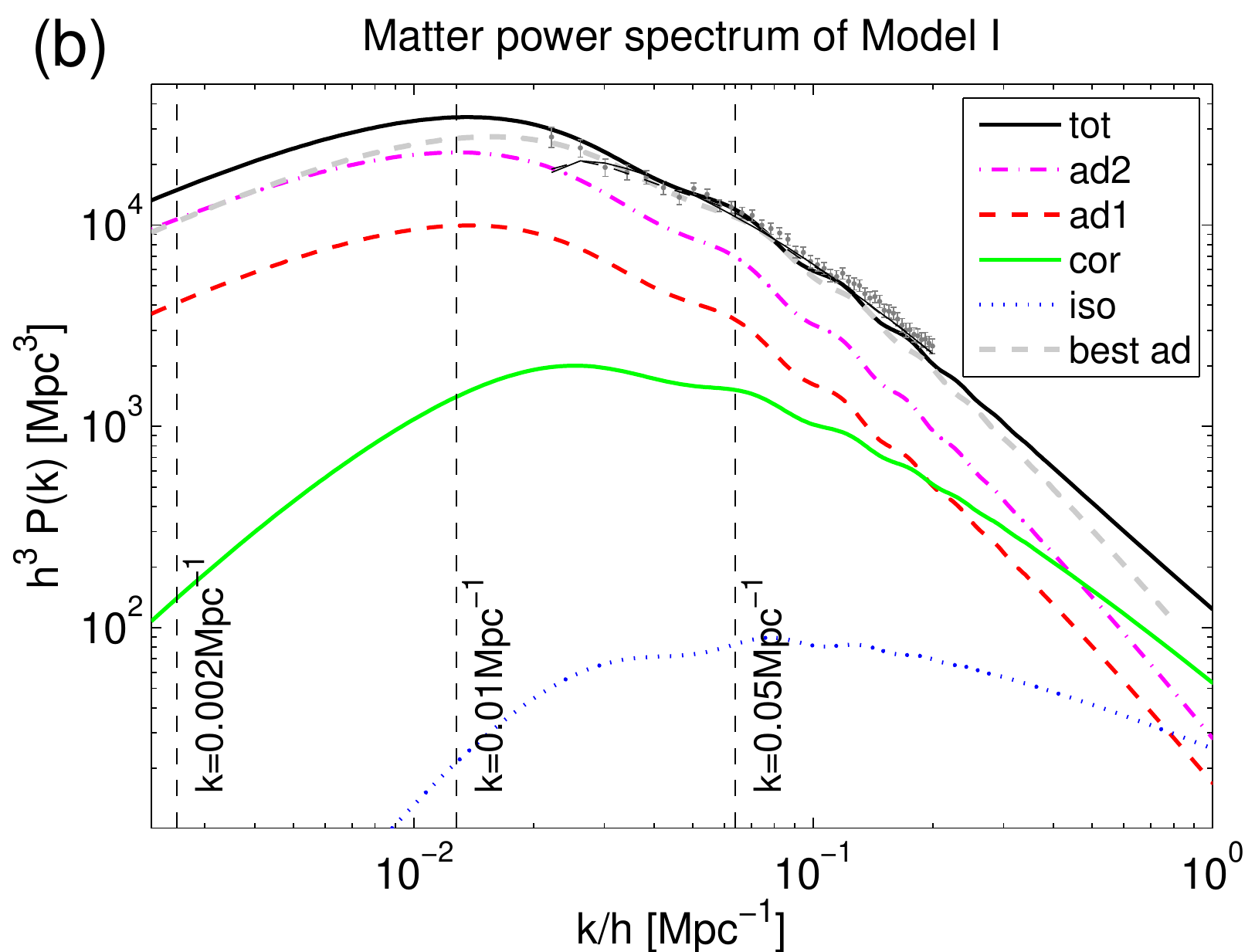}
  \includegraphics[width=\columnwidth]{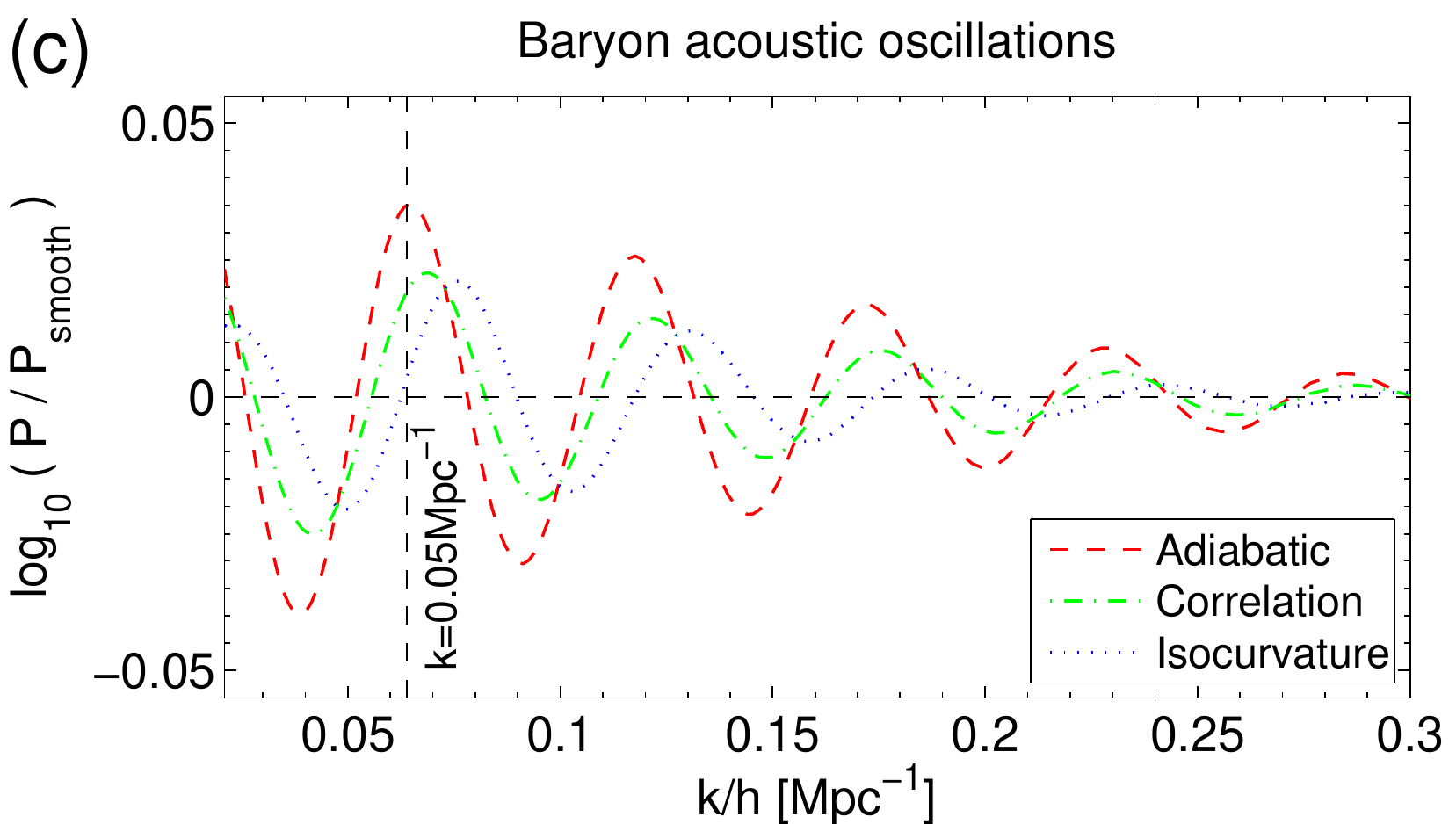}
  \includegraphics[width=\columnwidth]{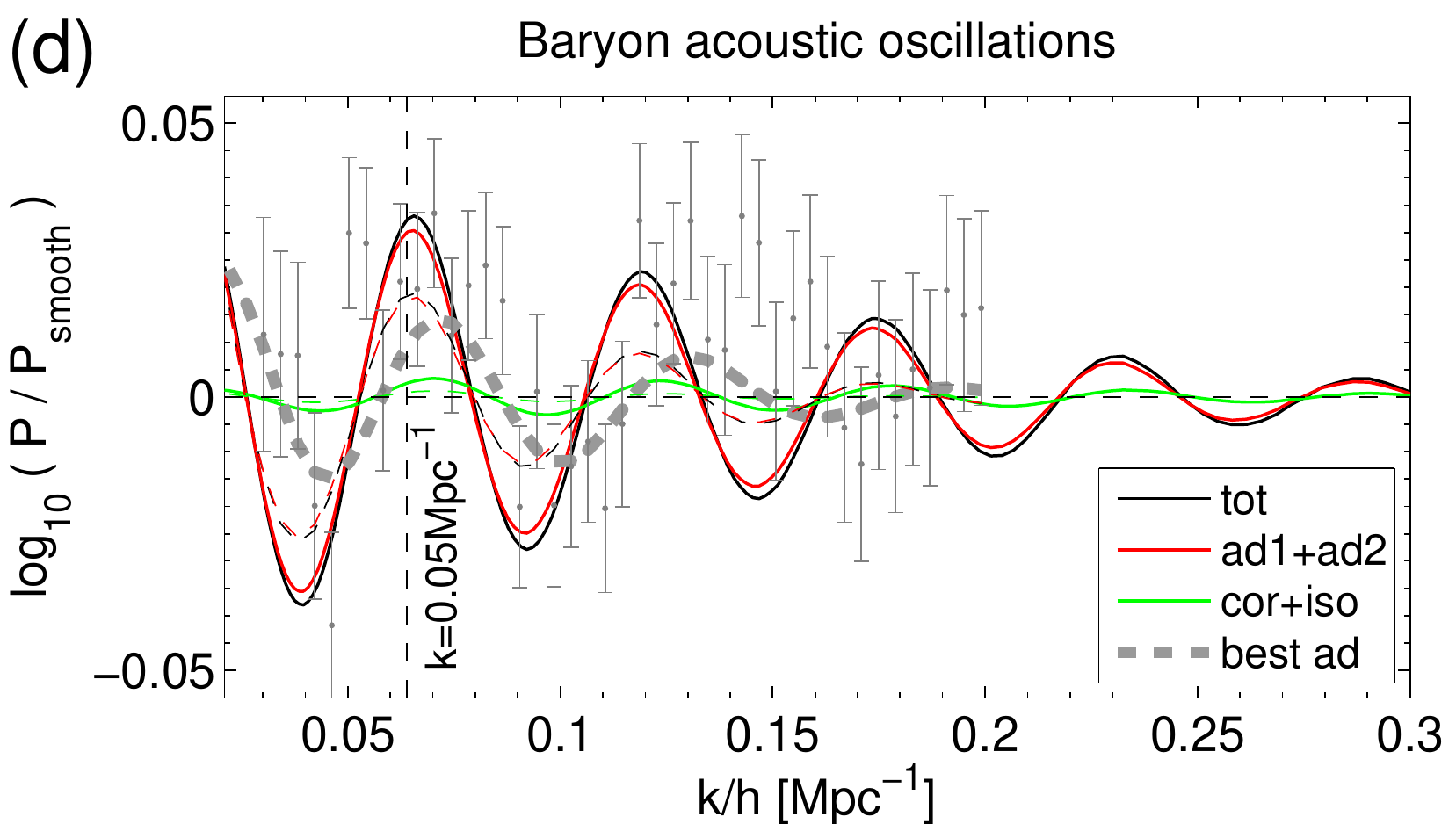}
  \caption{{\bf (a)} Linear matter power spectra at redshift $z=0$ (today) resulting from scale-invariant primordial spectra with unit amplitude. The dashed red curve is for the adiabatic primordial perturbations ($A_i^2=1$, $\gamma_i=0$, $\alpha_i=0$, $\nadI =1$), the dotted blue for the isocurvature ($A_i^2=1$, $\gamma_i=0$, $\alpha_i=1$, $\niso =1$), and the green for their correlation.
The corresponding thin solid curves are the smooth spectra from where the baryonic wiggles have been removed as described in Sec.~\ref{sec:data}.
{\bf (b)}  Linear matter power spectra at redshift $z=0$ for Model I of Tables~\ref{tab:CommonParams} and \ref{tab:IsocParams} in Appendix~\ref{app:BigTables}.
The gray dots with error bars are the SDSS DR7 LRG data \cite{Reid:2009xm} as supplied with CosmoMC, and the dashed thin black curve is the linear power spectrum ('tot') modified to take into account the BAO damping, the uncertainty in the non-linear modelling, and finally convolved with the observational window function (see equation (5) in \cite{Reid:2009xm}) in order to be compared directly to the data. The solid thin black curve is the smooth part of this. 
Note that plot (d) provides a ``zoom-up'' of the difference between these dashed (wiggly) and solid (smooth) thin black curves that are not easily visible in this plot.
{\bf (c)} The linear baryon acoustic oscillations of plot (a): the full oscillating curves divided by the smooth thin solid curves. 
{\bf (d)} The baryon acoustic oscillations of plot (b): the solid black, red and green curves are the linear BAO($k$) at redshift $z=0$ for the total, adiabatic and non-adiabatic components of Model I, respectively. The corresponding dashed curves that have smaller amplitude are obtained from the linear curves at three different redshifts modified to take into account the BAO damping, the uncertainty in the non-linear modelling, and finally convolved with the observational window function. For comparison, the dashed thick gray curve indicates BAO($k$) for the best-fit (to CMB and MPK) pure adiabatic model. The gray dots with error bars are the SDSS DR7 LRG data divided by the smooth part of the best-fit adiabatic model. The first two data points [see plot (b)] fall outside of the scale of this plot, the first being $0.22$ and the second one $0.10$.
It should be noticed that plot (d) is only for illustration purposes, and the actual MPK likelihood is calculated comparing the theoretical convolved matter power to the data, as shown in plot (b), and taking into account the full covariance matrix of the MPK data that accounts for the strong correlations between the data points.} 
  \label{fig:BAO}
\end{figure*}

The MPK constraints on nonadiabaticity in the slow-roll parametrization are (and will always be) purely indirect background constraints, since the nearly scale-invariant isocurvature perturbations with the tiny amplitude allowed by the large-scale CMB are not able to modify the matter power spectrum at all.  
In contrast, in our phenomenological approach, which (after the CMB constraints) allows \emph{large primordial} isocurvature and correlation components on small scales, the slope of the matter power spectrum at large $k$ can be affected \cite{KurkiSuonio:2004mn,Sollom:2009vd} by the correlation component (and by the isocurvature component on very small scales) \footnote{%
Our phenomenological model with a free $\niso$ differs from, e.g., Ref.~\cite{Trotta:2002iz} where the adiabatic and isocurvature components share the same spectral index --- the situation that is close to our two-field slow-roll inflation approach. As the CMB data prefer predominantly adiabatic nearly scale-invariant perturbations, the common spectral index is forced near to 1, which in \cite{Trotta:2002iz} leads to a conclusion that the isocurvature would not affect the matter power spectrum. This is in agreement with what we concluded in the case of slow-roll parametrization. 
}.
However, here the MPK likelihood code tries to soften the effect of the overall shape (by marginalizing over two nuisance parameters) as the exact shape is not known due to the nonlinear effects.  Therefore we would not expect tight extra constraints from the slightly modified slope, but instead from the phase and amplitude of the baryon acoustic oscillations (BAO). Unfortunately, primordial isocurvature perturbations of the same amplitude as the adiabatic ones lead to 20 -- 150 times smaller matter power spectrum today over the observable scales. This is clearly seen in Fig.~\ref{fig:BAO}(a) where we plot the resulting matter power spectra from scale-invariant primordial spectra with primordial amplitude 1. The background parameters in Fig.~\ref{fig:BAO}(a) are the same as in Model I.

In Fig.~\ref{fig:BAO}(b) we show a break down of the total matter power spectrum into its components for the mixed Model I of Tables~\ref{tab:CommonParams} and \ref{tab:IsocParams}. This model has a maximum allowed (at 95\% C.L. by CMB alone) primordial non-adiabatic contribution at scale $k=k_2=0.05\,$Mpc$^{-1}$ that falls in the middle of the $\log(k)$-range of the MPK data: $\alpha_{\rm cor2}=0.42$, $\gamma_2=0.73$, $\alpha_2=0.43$, and $\niso=3.3$. We choose this model in order to study why the MPK data rule it out. For this model the isocurvature (correlation) contribution to the matter power at $k_2$ is 120 (6.7) times smaller than the sum of the two adiabatic components. As seen in Fig.~\ref{fig:BAO}(b), the correlation component makes the matter power spectrum to decrease less steeply compared to the adiabatic spectrum at highest wave numbers probed by the data. (Note that the observational window functions related to each data point are rather broad, extending well beyond the last data point.)

From Fig.~\ref{fig:BAO}(c) we see that, although the amplitude of the isocurvature BAO is damped compared to the adiabatic BAO, the isocurvature would distinctively modify the phase of BAO if the isocurvature contribution in the matter power spectrum was of the same order or larger than the adiabatic contribution. As seen above, after the CMB constraints, the maximum non-adiabatic contribution is $\sim 10\%$. Moreover, this comes mostly from the correlation component whose phase is rather close to the adiabatic component. According to Fig.~\ref{fig:BAO}(d), the maximum non-adiabatic contribution to the total BAO is $\lesssim 5\%$ (compare the green and black curves). This is not detectable by the SDSS DR7 data.
To set direct extra constraints on isocurvature, the matter power data should be about an order of magnitude more accurate, which could be within the reach of Euclid \cite{EditorialTeam:2011mu}. Anyway the future LSS surveys will be important in constraining the isocurvature modes by breaking degeneracies that the CMB leaves between the background parameters and nonadiabaticity.\footnote{%
Detailed forecasts for combining the near future Planck \cite{Ade:2011ah,Mennella:2011ay} CMB data with the predicted LSS data from Baryon Oscillation Spectroscopy Survey (BOSS, see e.g. \cite{Schlegel:2009hj,Eisenstein:2011sa}) and Advanced Dark Energy Physics Telescope (ADEPT, see e.g. \cite{Sefusatti:2007ih}) or Euclid-like survey are provided in \cite{Kasanda:2011np} and \cite{Carbone:2011bx}. Using the LSS data together with Planck is expected to bring down the uncertainty of the isocurvature fraction from a few percentage points to better than 1\%.
}

As the non-adiabatic contribution to BAO($k$) does not explain the very  bad MPK likelihood of Model I, the main reason has to be the wrong turn-over scale [compare the position of the peak of the best-fit (to CMB and MPK) adiabatic matter power spectrum to that one of Model I in Fig.~\ref{fig:BAO}(b)]. In the $\Lambda$CDM model the turn-over scale is determined by the scale that enters the Hubble horizon at the time when the matter and radiation energy densities are equal. This scale is $k_{eq}=\omega_m \times 0.0729\,$Mpc$^{-1}$, i.e.,
$k_{eq} / h =  h\Omega_m \times 0.0729\,$Mpc$^{-1}$. For Model I, that has
$\Omega_m = 0.203$ and $h=0.783$, this gives $k_{eq}/h=0.0116\,$Mpc$^{-1}$ whereas for the best-fit adiabatic model we have $\Omega_m = 0.293$ and $h=0.688$ leading to $k_{eq}/h=0.0147\,$Mpc$^{-1}$ in agreement with Fig.~\ref{fig:BAO}(b). The rest of the bad MPK likelihood of Model I comes from the wrong phase of (even the adiabatic component of) BAO caused by the ``non-adiabatic'' values of the background parameters, i.e., the very small $\Omega_m$.

To test the above observations further we report several MPK $\chi^2$: The best-fit adiabatic model to CMB and MPK has an MPK $\chi^2 = 47.0$ ($60.0$) with (without) baryonic wiggles. Model I has $\chi^2=60.5$ ($67.8$) with (without) wiggles. Furthermore, Model I has $\chi^2 = 60.8$ without non-adiabatic wiggles. This is about the same as with all the wiggles. So, as expected, the ``wrong'' phase of (the less than 5\% contribution of) non-adiabatic wiggless does not affect $\chi^2$. Interestingly, dropping the non-adiabatic contributions from Model I gives $\chi^2=58.2$ ($ 62.3$) with (without) wiggless, i.e., improves the fit only by $\Delta\chi^2 = -2.3$ ($-5.5$). This gives an indication of the magnitude of the effect from the modified slope by the non-adiabatic contribution. Finally, changing the 4 backgound parameters to the values of the best-fit adiabatic model, the mixed model gives $\chi^2 = 48.8$. This is only $1.8$ worse than for the best-fit adiabatic model, and shows clearly that the bad MPK $\chi^2$ of high isocurvature fraction mixed models come almost solely from the unfavourable values of the background parameters while only $\Delta\chi^2 \sim 1$--$2$ comes from the too gradual slope and in practise nothing from the phase information of the non-adiabatic BAO component.


\section{Special cases with CMB data}

\subsection{Uncorrelated perturbations}

In the uncorrelated case there is no $\Pow_{\mr{as}}(k)$ spectrum, and therefore we have 
only 9 independent parameters:  the usual 4 background parameters
 \beq
 	\omega_b\,,\ \omega_c\,,\ \theta\,,\ \tau\,,
 \eeq
and 5 perturbation parameters
 \beq
 	\ln A_0^2\,,\ \nadI\,,\ \alpha_0\,, \niso\,,\ r_0\,,
 \eeq
where
 \bea 
 	\nadI & = & 1 - 6\veps+2\etzz \nn\\  
	\niso & = & 1 - 2\veps+2\etss \nn\\
	         r_0 & = & 16\veps\,.
\label{eq:nocor}
 \eea

\begin{figure}[t]
  \centering
  \includegraphics[width=\columnwidth]{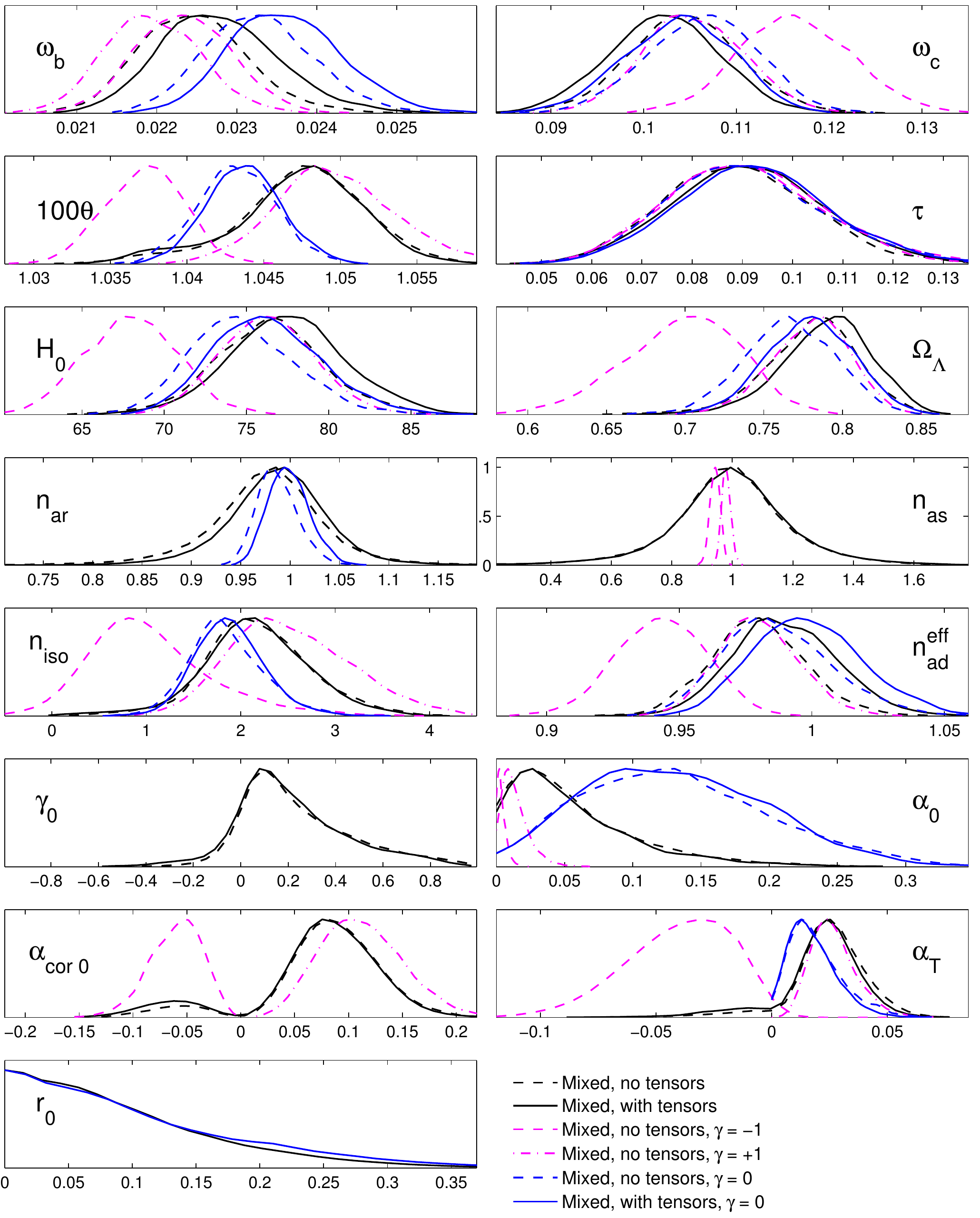}
  \caption{{\bf Special cases, amplitude parametrization, CMB data, comparison of uncorrelated or fully correlated models to the generally correlated model.}
The solid lines are for models with tensors. The black color is for the generally correlated model, the magenta dashed lines for fully anticorrelated ($\gamma=-1$), magenta dot-dashed for fully correlated ($\gamma=+1$), and blue for uncorrelated ($\gamma=0$) model.  
}
  \label{fig:AmplSpecialCases}
\end{figure}

\begin{figure}[t]
  \centering
  \includegraphics[width=\columnwidth]{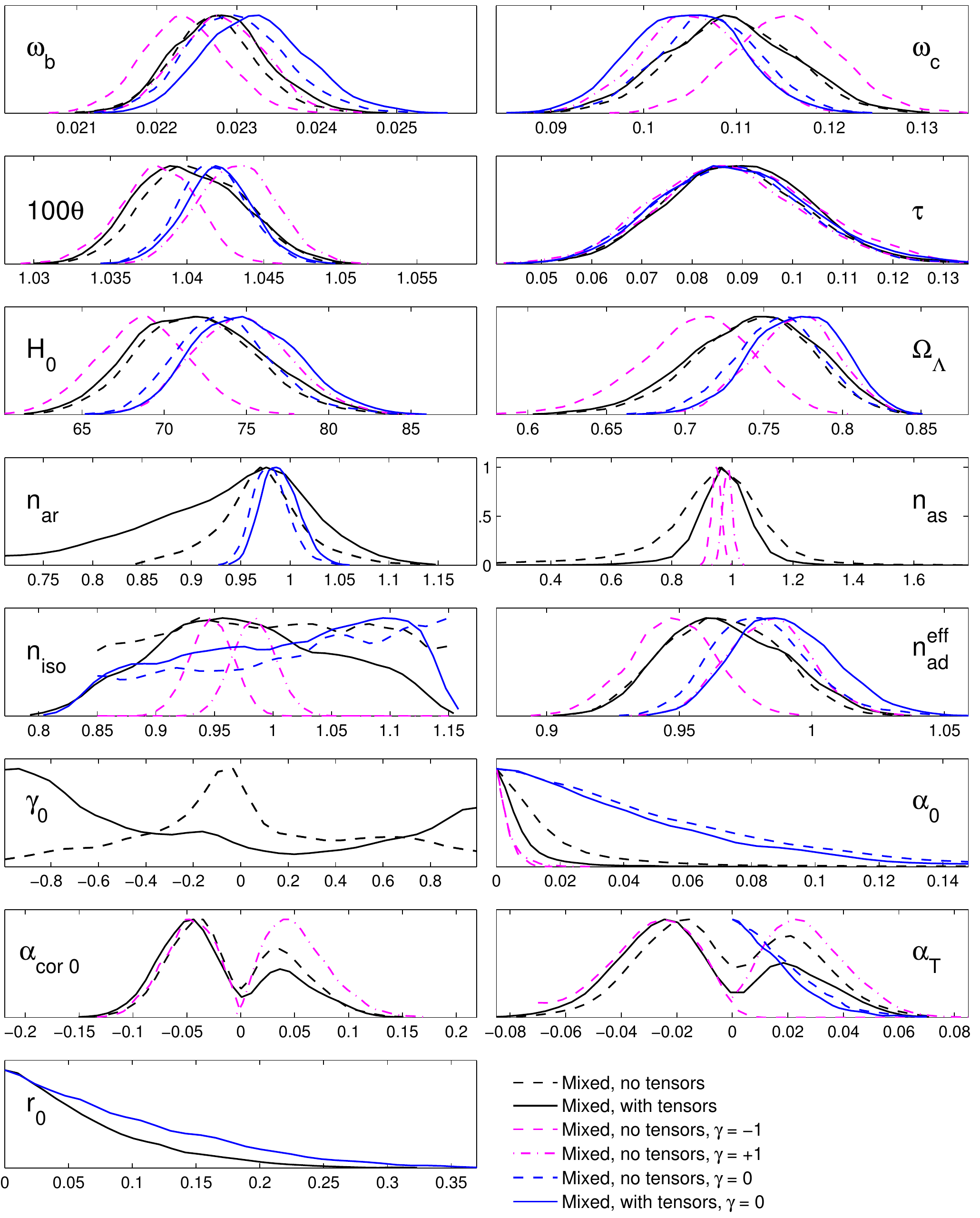}
  \caption{{\bf Special cases, slow-roll parametrization, CMB data, comparison of uncorrelated or fully correlated models to the generally correlated model.}
The line styles are the same as in Fig.~\ref{fig:AmplSpecialCases}.}
  \label{fig:InflSpecialCases}
\end{figure}

\begin{figure}[t]
  \centering
  \includegraphics[width=\columnwidth]{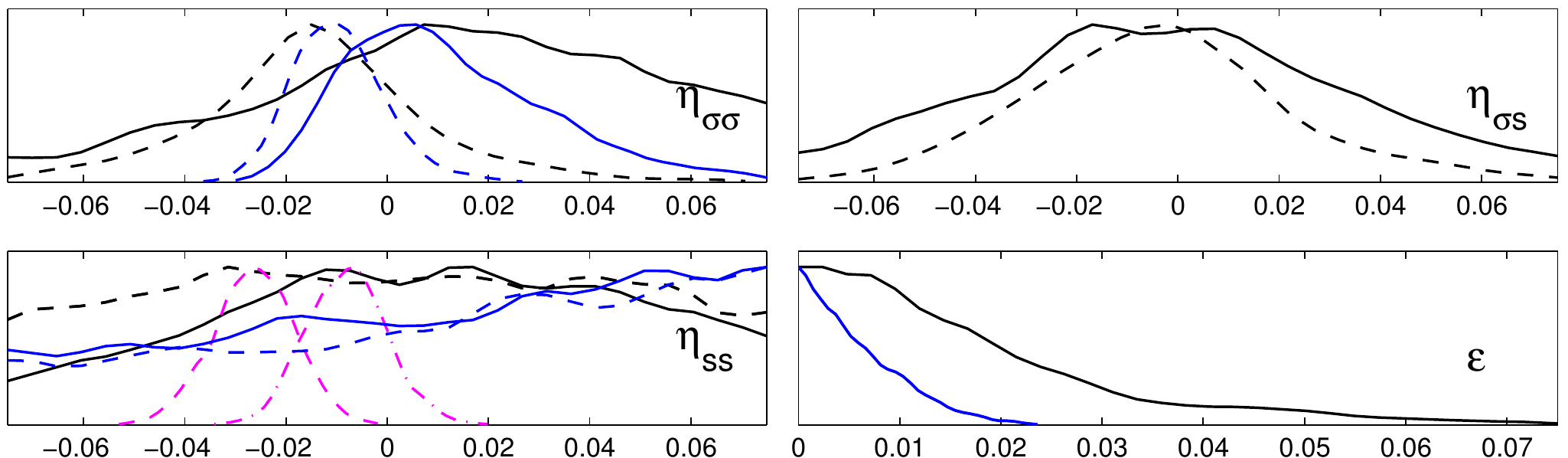}
  \caption{{\bf Special cases, slow-roll parametrization, CMB data, comparison of uncorrelated or fully correlated models to the generally correlated model.}
The line styles are the same as in Fig.~\ref{fig:AmplSpecialCases}.}
  \label{fig:InflSpecialCasesSlowRoll}
\end{figure}

In the amplitude parametrization, the primary perturbation parameters are
 \beq
 	\ln A_1^2\,,\ \ln A_2^2\,,\ \alpha_1\,,\ \alpha_2\,,\ r_0 \,,
 \eeq
and in the slow-roll parametrization
 \beq
    \ln A_0^2\,,\ \alpha_0\,,\ \etzz\,,\ \etss \, \ \veps\,.
 \eeq
The marginalized 1-d posteriors for this model with the CMB data are indicated by the solid blue $\gamma=0$ curves in Fig.~\ref{fig:AmplSpecialCases} for the amplitude parametrization, and in Figs.~\ref{fig:InflSpecialCases} and \ref{fig:InflSpecialCasesSlowRoll} for the slow-roll parametrization. The black curves show for comparison the general case studied in Secs.~\ref{sec:general_cmb_ampl_par} and \ref{sec:general_cmb_infl_par}. The dashed lines are for the same models but without tensors, i.e., $r_0=0$ or $\veps = 0$.


Let us first comment on Fig.~\ref{fig:AmplSpecialCases} (amplitude parametrization).
Since it was the correlation component that modified the acoustic peak structure in the generally correlated case, and this component is missing in the uncorrelated case, we do not have a mechanism that would favour a small $\omega_c$ and large $\theta$, $H_0$ and $\Omega_\Lambda$. Therefore these background parameters are now closer to their ``adiabatic'' values. Also smaller $\niso$ is now preferred. This is compensated by a slightly larger $\nad^{\mr{eff}}$. Much larger $\alpha_0$ is allowed but, as seen from the $\alpha_T$ plot, this leads to a smaller overall nonadiabatic contribution to $C_\ell$, since the correlation component is missing. The determination of $r_0$ is unaffected for the same reason as in the general case: isocurvature with a large $\niso$ modifies the small-scale end and the tensor contribution with nearly scale-invariant spectrum the large-scale end of the $C_\ell$ spectrum. Taking the numbers from Table~\ref{tab:CommonParams} 
we find that, compared to the mixed model, the adiabatic model has odds $4.9:1$ 
(${\cal B}_f=+1.59$)
with tensors, and $3.9:1$ 
(${\cal B}_f=+1.35$)
without tensors. 
Since the CMB data are compatible with the pure adiabatic model and the uncorrelated mixed model has two parameters less than the generally correlated model, it is natural that the uncorrelated model is much less disfavoured than the generally correlated model that gave ${\cal B}_f \approx +5$ in support of the adiabatic model.


Our result $\alpha_1 < 0.112$ with tensors or $0.127$ without tensors in Table ~\ref{tab:IsocParams} agrees well with the WMAP7 result \cite{Larson:2010gs,Komatsu:2010fb} $\alpha_1 < 0.13$ (their "Axion" model). For the nonadiabaticity we find $\alpha_T < 3.7\%$ with tensors and $4.3\%$ without tensors.

In the slow-roll parametrization (Figs.~\ref{fig:InflSpecialCases} and \ref{fig:InflSpecialCasesSlowRoll}) $\veps$ receives a tighter constraint than in the general case, since the $\sin^2\Delta$ factor is now 1 in Eq.~(\ref{eq:BWresults}). The constraint on $r_0$ loosens. 

In the general case we had a tight connection between $\niso$ and $\nadII$ [see Eq.~(\ref{eq:BWresults})] and the data (loosely) constrained $\nadII$. This led to a similar constraint on $\niso$. However, now $\nadII$ is absent, and hence $\niso$ does not receive any constraint, see Fig.~\ref{fig:InflSpecialCases}. As the data seem to prefer the isocurvature contribution not showing up at any scale (which is achieved the easier the closer to 3 $\niso$ is; recall the left panel of Fig.~\ref{fig:ExampleModels} where $\niso = 3.3$), the posterior of $\niso$ has a peak almost at the upper boundary of its prior range, $\niso=1.15$.

The odds in favor of the corresponding adiabatic models compared to the uncorrelated mixed model in the slow-roll parametrization are
 $17.6:1$ 
(${\cal B}_f=+2.87$)
with tensors, and $80:1$ 
(${\cal B}_f=+4.38$)
without tensors.

Interestingly, the slow-roll parametrization gives very similar constraints on $\alpha$ --- that are almost independent of the scale, since $\niso\sim 1$ --- ($\alpha_0<0.110$ with and $0.125$ without tensors) as the amplitude parametrization gives for $\alpha_1$.
The 95\% C.L. upper limits for $\alpha_T$ are also very similar, although the shape of the posterior is different, the slow-roll parametrization preferring more clearly a zero nonadiabatic contribution. 

\subsection{Fully (anti)correlated perturbations}

In the fully correlated cases there is no $\Pow_{\mr{ar}}(k)$ spectrum ($\gamma=-1$ or $\gamma=+1$).
In the standard two-field inflation approach, we get this situation, when the curvature perturbations are completely dominated by the part that evolved from the entropy perturbations after horizon exit, and the tensor perturbations that were generated at horizon exit together with the uncorrelated part of the curvature perturbations are negligible.

Therefore, in this case we assume no tensor perturbations, and we have just 8  independent parameters:  background parameters
 \beq
 	\omega_b\,,\ \omega_c\,,\ \theta\,,\ \tau\,,
 \eeq
and 4 perturbation parameters
 \beq
 	\ln A_0^2\,,\ \nadII\,, \ \niso \,,\ \alpha_0\,. 
 \eeq
According to Eq.~(\ref{eq:BWresults}), the two-field slow-roll inflation gives
 \beq 
 	\nadII = \niso \ = \ 1 + 2\etss \,,
 \eeq
which reduces the number of independent perturbation parameters to 3 in
the slow-roll parametrization. 

In the amplitude parametrization, the primary perturbation parameters are
 \beq
 	\ln A_1^2\,,\ \ln A_2^2\,,\ \alpha_1\,,\ \alpha_2\,,
 \eeq
and in the slow-roll parametrization
 \beq
    \ln A_0^2\,,\ \alpha_0\,,\ \etss \,.
 \eeq 
The marginalized 1-d posteriors for this model with the CMB data are indicated by the dashed magenta ($\gamma=-1$, 100\% anticorrelation) and dot-dashed magenta ($\gamma=+1$, 100\% correlation)  curves in Fig.~\ref{fig:AmplSpecialCases} for the amplidude parametrization, and in Figs.~\ref{fig:InflSpecialCases} and \ref{fig:InflSpecialCasesSlowRoll} for the slow-roll parametrization.

As we would expect, in the amplitude parametrization (see Fig.~\ref{fig:AmplSpecialCases}), the $\gamma=+1$ case leads to very similar results as we got with the general correlation without tensors (dashed black curves). Some of the features of the well-fitting phenomenological mixed model are amplified, such as the preference for very large $\niso$. Due to the 100\% correlation, for a given $\alpha$ the nonadiabatic contribution modifies the $C_\ell$ spectrum in the maximal way. Therefore $\alpha_0$ is formally constrained very tightly, but the peak of $\alpha_{\mr{cor0}}$ moves to slightly larger value than in the general case. The posterior of $\alpha_T$ is practically unaffected. On the other hand, the $\gamma=-1$ case differs significantly from the general case. Indeed, the amplitude parametrization results are now very close to the slow-roll parametrization results. This is natural, since the slow-roll parametrization favoured the negative correlation. The only clearly noticeable difference lies in the lower limit for $\alpha_T$ that is now $-0.08$. Our result $\alpha_1<0.011$ co-incides with the WMAP7 result \cite{Larson:2010gs,Komatsu:2010fb}, see their "Curvaton" model. 

The odds for the adiabatic model compared to the fully correlated mixed model are $206:1$ 
(${\cal B}_f=+5.33$) and the fully anticorrelated model
$11048:1$ 
(${\cal B}_f=+9.31$). Here the overwhelmingly small odds of the mixed models are partially due to our choice to keep $\alpha_1$ and $\alpha_2$ independent, i.e., allowing for $\niso \neq \nadII$.

In the slow-roll parametrization this extra freedom is missing and we find
that the adiabatic model has odds
 $240:1$ 
(${\cal B}_f=+5.48$)  when compared to the fully correlated model, and
$257:1$ 
(${\cal B}_f=+5.55$) when compared to the fully anticorrelated mixed model.

\section{Comparison to older results}

Some of the earliest constraints
\cite{Stompor:1995py,Hu:1995xs,Hu:1994tj,Chiba:1993px}
on isocurvature were obtained soon after COBE \cite{Wright:1992tf} in the mid 1990s.
Setting tight observational constraints on isocurvature \cite{Enqvist:2000hp} became possible when Boomerang \cite{deBernardis:2000gy} and Maxima \cite{Hanany:2000qf} firmly observed the first acoustic peak in the CMB in the year 2000. Pre-WMAP constraints were derived, e.g., in \cite{Gordon:2002gv,Amendola:2001ni,Langlois:2000ar,Pierpaoli:1999zj}, some papers focusing on the determination of the cosmological constant in the presence of primordial isocurvature modes \cite{Trotta:2003ng,Trotta:2002iz}, while \cite{Enqvist:2001fu} finally ruled out a \emph{pure} CDM isocurvature perturbation even in a spatially curved Universe. Constraints from the Boomerang 2003 flight were derived in \cite{MacTavish:2005yk} and from  Cosmic Background Imager (CBI) observations in \cite{Sievers:2005gj}. The WMAP first-year data \cite{WMAP1} were used in \cite{Beltran:2005gr,Beltran:2005xd,Andrade:2005gw,Lazarides:2004we,KurkiSuonio:2004mn,Beltran:2004uv,Parkinson:2004yx,Moodley:2004nz,Ferrer:2004nv,Valiviita:2003tu,Valiviita:2003ty,Dunkley:2004sv,Andrade:2003xb,Gordon:2003hw,Crotty:2003rz,Peiris:2003ff,Bennett:2003bz} to constrain various mixtures of adiabatic and isocurvature perturbations in a spatially flat Universe.  The focus in \cite{Dunkley:2005va} was in testing how much the (possible) presence of isocurvature modes affects the determination of the geometry (spatial curvature) of the Universe.

The WMAP 3-year data \cite{WMAP3} were employed in \cite{Kawasaki:2007mb,Keskitalo:2006qv,Trotta:2006ww,Seljak:2006bg,Lewis:2006ma} and \cite{Bean:2006qz} (note the post-publication corrections to \cite{Bean:2006qz} in \url{http://arxiv.org/pdf/astro-ph/0606685v3} for the case of a varying isocurvature spectral index). The most recent observational constraints and model selection with WMAP 5-year data \cite{WMAP5} combined with SN and SDSS matter power spectrum data were presented in \cite{Sollom:2009vd} for a spatially flat Universe. In \cite{Valiviita:2009bp} a comprehensive Bayesian comparison of flat and curved Universes both in pure adiabatic and in mixed cases was performed, finding that the spatial curvature is disfavoured roughly by the same amount as the isocurvature compared to the flat adiabatic model.

In \cite{Kawasaki:2011rc} the possibility of extra radiation (a radiation component other than photons or 3 standard species of neutrinos) carrying isocurvature is studied concluding that the current data allow the existence of an extra radiation component but do not favor its isocurvature mode. Ref. \cite{Liu:2010ba} shows that constraints on dark energy isocurvature are very weak.

Finally, forecasts for future experiments have been worked out, e.g., in \cite{Easson:2010uw,Easson:2010zy,Hamann:2009yf,Gordon:2009wx,Baumann:2008aq,Bucher:2000hy,Bucher:2000kb,Enqvist:1999vt}.

For a closer comparison to our results we pick six publications where the pivot scale(s) and parametrizations are the most similar
to our work. In our earlier work \cite{KurkiSuonio:2004mn}, \cite{Keskitalo:2006qv} and \cite{Valiviita:2009bp}, where we used 1-year, 3-year, and 5-year WMAP data, respectively, and older small-scale CMB data, we did not include tensor perturbations.
 In the slow-roll parametrization this would correspond to assuming $\veps = 0$, but as all those work were in a phenomenological parametrization, we pick from Table~\ref{tab:IsocParams} the ``Ampl.\ par.'' results. Our new result (with CMB data) is $\alpha_0 < 0.158$, while in \cite {KurkiSuonio:2004mn} we obtained $\alpha_0 < 0.18$; in \cite{Keskitalo:2006qv} $\alpha_0 < 0.169$; and in \cite{Valiviita:2009bp} $\alpha_0 < 0.22$. The actual limits on $\alpha$ are very sensitive to the precise amount of correlation preferred by the data, since the same $\alpha$ leads to a larger modification of the $C_\ell$ spectrum if $|\gamma|$ differs significantly from zero. The nonadiabatic contribution to the CMB temperature variance is less sensitive to this ``arbitrariness'', and has evolved consistently toward tighter limits: from $-0.075 < \alpha_T < 0.075$ \cite {KurkiSuonio:2004mn}, $0.017 < \alpha_T < 0.073$ \cite{Keskitalo:2006qv},  and $-0.031 < \alpha_T < 0.067$ \cite{Valiviita:2009bp} to our new constraint $-0.021 < \alpha_T < 0.053$ without and  $-0.030 < \alpha_T < 0.049$ with tensors. The WMAP 1-year data had relatively large error bars and did not extend to the third acoustic peak. As seen above, the range for  $\alpha_T$ was symmetric about 0. The later data map more precisely the acoustic peak structure and, as explained earlier, there is a persistent \emph{formal} preference for a positively correlated isocurvature component with a large $\niso$ in the phenomenological approach.

Of other publications, the most similar to our work are \cite{Kawasaki:2007mb,Li:2010yb}, although they do not perform Bayesian model comparison, but just parameter estimation. Ref.~\cite{Kawasaki:2007mb} uses WMAP 3-year data together with or without the SDSS DR4 LRG matter power spectrum, and includes tensor perturbations fixing the tensor spectral index $n_T$ by the first consistency relation. In addition, although their parametrization is phenomenological, they impose an ``inflationary constraint'' that $\niso = \nadII = 1 + n_T$. Therefore their work is closer to our slow-roll case than to our phenomenological approach: although we don't have  $\niso = \nadII$ in the slow-roll, our slow-roll approach guarantees that $\niso$ is nearly scale invariant like theirs. Since they use spectral index parametrization with pivot scale $k=0.05\,$Mpc$^{-1}$, we need to map our slow-roll results to spectral indices and then from our pivot scale $k_0=0.01\,$Mpc$^{-1}$ to their pivot scale. Furthermore, we need to map to their isocurvature and correlation parameters: $B_a = \sqrt{\alpha/(1-\alpha)}$ and $\cos\theta_a = \sgn(\gamma)\sqrt{|\gamma|}$, respectively. After these mappings, our general slow-roll case with (without) tensors, using CMB data, gives $B_a < 0.25$ (0.26) at 95\% C.L. and $-0.79 < \cos\theta_a < 0.73$ ($-0.67 < \cos\theta_a < 0.72$). Adding the SDSS DR7 LRG data into the analysis, these change to $B_a < 0.23$ and $-0.89 < \cos\theta_a < 0.21$. Their results with old WMAP and SDSS DR4 LRG are $B_a < 0.28$ (0.33) and  $-0.77 < \cos\theta_a < 0.27$ ($-0.40 < \cos\theta_a < 0.28$).
For tensor-to-scalar ratio at $k=0.05\,$Mpc$^{-1}$ we find $r < 0.24$ with CMB and 0.17 with CMB and MPK data, while their upper bound is 0.32.
The general agreement is good when keeping in mind the differences in the data sets and parametrizations. For the special cases (uncorrelated model with tensors, uncorrelated model without tensors, fully anticorrelated, fully correlated) we find with CMB data $B_a <$ 0.32, 0.35, 0.083, 0.077, while they find with the old WMAP data  --- in remarkable agreement --- 0.31, 0.33, 0.080, 0.087, respectively. For the tensor-to-scalar ratio at $k=0.05\,$Mpc$^{-1}$ in the uncorrelated case we find $r < 0.25$ while their constraint is 0.26.
 
Ref.~\cite{Li:2010yb} uses the same CMB, SN and MPK data sets as we (except QUaD replaced by CBI and Boomerang).
Unfortunately, they don't specify how the BAO part of the MPK theory and likelihood code was modified in order to take into account the isocurvature BAO when using the LRG sample, or whether the MPK code was not modified relying on the fact that BAO from subdominant isocurvature component would not change the results significantly with the current accuracy of data, as we have explicitly shown here.  (BAO in the presence of isocurvature modes is further discussed in \cite{Kasanda:2011np,Carbone:2011bx,Mangilli:2010ut,Zunckel:2010mm,Valiviita:inprep}.) Ref.~\cite{Li:2010yb} employs the phenomenological amplitude parametrization without tensor contribution and therefore the corresponding case in our study is ``Ampl. par.'' (no tensors) with either CMB and SN or CMB and MPK data. As they have included both SN and MPK simultaneously, we would expect their results to lie somewhere between our CMB\&SN and CMB\&MPK results, and be slightly tighter. Indeed, taking an ``average'' of the dashed and dot-dashed curves in our Figs.~\ref{fig:AmplDataSetsPrimary} and \ref{fig:AmplDataSetsDerived}, noticing that $\alpha_{\mr{their}} = 1 - \alpha_{\mr{our}}$, and comparing to their figures 2 and 3 shows a good qualitative agreement, except for the exact shape of $\alpha_2$. Note that their $\cos\Delta$ equals $\sgn{(\gamma})\sqrt{|\gamma|}$. They do not specify their sign convention, but we deduce from their figure 1 that it is the same as ours. This is supported by their figure 4, ($\cos\Delta,\Omega_m$) plot, that indicates the same degeneracy as we have seen, i.e., positive correlation prefers a large $\Omega_\Lambda$. Finally, their numerical 95\% C.L. constraint on $\alpha_1$ ($\alpha_0$) is 6.1\% (14.6\%) while we find 6.7\% (13.7\%) with CMB and SN and 9.8\% (16.7\%) with CMB and MPK. We cannot do exact one-to-one comparison as they have one parameter less in their model, since they assume $\nadII=\nadI$.
 
Finally, uncorrelated models are constrained in \cite{Castro:2009ej} which assumes $\nadI=\niso$. Closest of our cases is the uncorrelated slow-roll model without tensors. However, even if $\niso$ is forced close to one in this model, it is completely free from $\nadI$, see Eq.~(\ref{eq:nocor}) remembering that $\veps = 0$. As the CMB data are very ``adiabatic'' and hence require that the isocurvature contribution is subdominant at all multipoles, they prefer spectral index $\niso \sim 3$. However, our slow-roll prior requires $\niso \leq 1.15$. Therefore the blue dashed curve in our Fig.~\ref{fig:InflSpecialCases} peaks at $\niso = 1.15$. This means that our model is quite different from the model of \cite{Castro:2009ej} where the posterior peaks at $\niso=\nadI\sim0.987$. We find $\alpha_0 < 0.125$. Mapping into the spectral index parametrization and to pivot scale  $k=0.002\,$Mpc$^{-1}$ used in \cite{Castro:2009ej} we obtain $\alpha < 0.117$, whereas their constraint 0.19 with QUaD and WMAP 5-years data is much looser, showing --- once again --- that the formal constraints on undetected isocurvature parameters can depend strongly on the theoretical assumptions, here whether one imposes $\nadI=\niso$ or uses slow-roll parametrization that makes them independent.


\section{Conclusions}

Apart from using more recent CMB data, the main new results of this paper compared to \cite{Valiviita:2009bp}, which also utilizes MultiNest and Bayesian model comparison, are: 1) Showing that allowing for a primordial tensor contribution has a negligible effect on the determination of the nonadiabatic contribution and that the possible presence of a CDM or baryon isocurvature mode has a minor effect on the determination of the tensor-to-scalar ratio, as long as the tensor spectral index obeys the first inflationary consistency relation. 2) Showing that the results change considerably compared to \cite{Valiviita:2009bp} if, instead of the phenomenological parametrization, one uses inflationary slow-roll parametrization that forces all the spectra near to scale invariance. 3) Including one of the most important complementary (to CMB) data sets into the analysis, namely the matter power spectrum, and showing that its current accuracy is not enough to give direct extra constraints on isocurvature. However, even the current MPK data change the results considerably due to the tension between the low matter density preferred by the CMB data and the high matter density preferred by the MPK data.
4) In \cite{Valiviita:2009bp} (and \cite{Sollom:2009vd}) all the runs were done without CMB lensing, because already computationally intensive runs turn by an order of magnitude heavier if lensing is applied. Ignoring the lensing was justified by the fact that the 1d posteriors stayed rather unaffected. However, in this paper, in Sec.~\ref{sec:lensing} we have noticed that comparing the unlensed mixed model to the unlensed adiabatic model may make the mixed model seem slightly more favourable than it actually is, in particular what comes to the Bayesian evidences. Therefore, the CMB lensing is now included in the analysis. In the end the effect on model comparison is mild: for example, ${\cal B}_f = +5.48$ with and ${\cal B}_f = +4.18$ without lensing when comparing the adiabatic model to the mixed one (phenomenological model with tensors).

We conclude that current cosmological data give no evidence of an isocurvature perturbation contribution.  Thus we can only set upper limits to it.  Since isocurvature perturbations contribute to the large-scale CMB more strongly than to 
small-scale CMB, in comparison to the adiabatic perturbations, the tightest limits to isocurvature perturbations are at the large scales.  In the absence of a detection of an isocurvature contribution, the phenomenological approach to constraining the parameters then leads to a preference of a steeply blue primordial isocurvature perturbation spectrum (large $\niso$) unless a prior constraint on spectral indices is imposed. The limits to an isocurvature contribution then come mainly from the relative locations of the acoustic peaks in the CMB angular power spectrum.

In addition to a phenomenological approach, we considered an approach motivated by two-field inflation, where the four first-order slow-roll parameters were fitted to the data, with the prior assumption that the slow-roll parameters are small.  This means that all primordial spectra are constrained to be close to scale invariant.  Since in this case isocurvature perturbations contribute much more to large CMB scales than to small scales, when compared to the adiabatic perturbations, the isocurvature contribution is mainly constrained by the relative levels of the low-$\ell$ and high-$\ell$ parts of the observed CMB angular power spectrum.  This leads to a tighter limit to isocurvature perturbations than when their spectral index is not constrained a priori.  When spectra are close to scale invariant, the contribution of isocurvature perturbations to $C_\ell$ are somewhat similar to that of tensor perturbations---both contribute mainly to low $\ell$.  If isocurvature perturbations are anticorrelated with adiabatic perturbations, they have a negative contribution at low $\ell$, and thus their contribution can be partly canceled by tensor perturbations.  Thus the combination of negatively correlated isocurvature perturbations and tensor perturbations fit the data better than positively correlated isocurvature perturbations and tensor perturbations.  However, since for a given isocurvature fraction $\alpha$, correlated perturbations give a much stronger signal in the CMB than uncorrelated perturbations, allowing for tensor perturbations does not relax the upper limit to $\alpha$.

In the two-field slow-roll inflation approach it is not feasible to go beyond the level of first order in slow-roll parameters, since the number of parameters would then become too large.  In any case, the slow-roll approximation is only valid, when the parameters are small.  Therefore we assumed a prior constraint that their magnitudes are smaller than 0.075.
With this prior assumption, we obtained a somewhat tighter posterior limit to the slow-roll parameter related to the first derivatives of the inflation potential $V(\phi_1,\phi_2)$, namely $\veps < 0.048$, but 
no useful constraint to the second-derivative slow-roll parameters $\eta_{ij}$.
Thus to make progress on constraining two-field inflation, we need more accurate data, e.g., from the Planck satellite.

\vspace{-0.5cm}
\begin{acknowledgments}
 This work was supported by the Academy of Finland grant 121703. 
 We thank the DEISA Consortium (www.deisa.eu), co-funded through 
 the EU FP6 project RI-031513 and the FP7 project RI-222919, 
 for support within the DEISA Virtual Community Support Initiative. 
 This work was granted access to the HPC resources of CSC made
 available within the  Distributed European Computing Initiative by
 the PRACE-2IP, receiving funding from the European Community's Seventh
 Framework Programme (FP7/2007-2013) under grant agreement RI-283493.
 We thank the CSC - Scientific Computing Ltd (Finland) for computational
 resources. 
 JV was supported by the Research Council of Norway, 
 MS by the V\"{a}is\"{a}l\"{a} Foundation, and 
 SR by the Jenny and Antti Wihuri Foundation.
 We acknowledge travel support from the Magnus Ehrnrooth Foundation.  
 JV acknowledges Jan Hamann for making available the code described
 in Appendix A.1 of \cite{Hamann:2010pw} that allows for a robust
 identification of the baryonic wiggles in the theoretical matter
 power spectrum. We thank Tone Melv{\ae}r Ruud for testing the
 robustness of this and various other wiggles identification techniques,
 in particular when applied to the CDM isocurvature mode.
\end{acknowledgments}

\vspace{-0.5cm}
\bibliography{isoc_wmap7_arxiv_v1}

\appendix
\section{Tables}
\label{app:BigTables}
\begin{turnpage}
\begin{table*}
\scriptsize
\centering
\begin{tabular}{|l|l|l|l|l|l|l|l|l|l|l|}
\hline
 {\bf Ampl.\ par.} & \normalsize $\omega_b$ & \normalsize $\omega_c$ & \normalsize $100\theta$ & \normalsize $\tau$ & \normalsize $H_0$ & \normalsize $\Omega_\Lambda$ &  \normalsize $\ln (10^{10} A_0^2)$ & \normalsize $n_{\mathrm{ad}}^{\mathrm{eff}}$ & \normalsize $r_0$ & \normalsize ${}-\ln{\cal Z}$ \\
\hline
   \ Mixed model &  \hfill (68\% C.L.)&  \hfill (68\% C.L.)  &   \hfill (68\% C.L.)&  \hfill (68\% C.L.) &  \hfill (68\% C.L.)&   \hfill (68\% C.L.)&  \hfill (68\% C.L.)&  \hfill (68\% C.L.)& 95\% C.L.& \\
\hline
\ \ \ \  CMB& 0.0227 (0.0221 0.0235) & 0.102  (0.096 0.108) & 1.048 (1.043 1.051) & 0.091 (0.076 0.106) &   77 (  74   81) & 0.79 (0.76 0.82) & 3.11 (3.07 3.17) & 0.989 (0.971 1.009) &   \hfill $<$0.25 &  3863.45\\ 
\hline
\ \ \ \  & 0.0224 (0.0218 0.0231) & 0.104  (0.098 0.110) & 1.048 (1.044 1.051) & 0.089 (0.075 0.104) &   76 (  73   80) & 0.78 (0.75 0.81) & 3.13 (3.08 3.19) & 0.979 (0.963 0.997) &      &  3861.49\\ 
\hline
\ \ \ \  CMB\&SN& 0.0226 (0.0219 0.0233) & 0.108  (0.103 0.112) & 1.046 (1.039 1.050) & 0.088 (0.074 0.103) &   74 (  71   77) & 0.76 (0.74 0.79) & 3.14 (3.09 3.20) & 0.976 (0.960 0.994) &   \hfill $<$0.21 &  4129.81\\ 
\hline
\ \ \ \  & 0.0223 (0.0217 0.0230) & 0.109  (0.105 0.114) & 1.046 (1.040 1.050) & 0.086 (0.073 0.101) &   74 (  71   76) & 0.76 (0.73 0.78) & 3.14 (3.10 3.20) & 0.970 (0.954 0.986) &      &  4127.80\\ 
\hline
\ \ \ \   CMB\&MPK & 0.0229 (0.0222 0.0235) & 0.115  (0.112 0.119) & 1.038 (1.036 1.041) & 0.087 (0.074 0.102) &   69 (  67   71) & 0.71 (0.69 0.73) & 3.21 (3.16 3.27) & 0.963 (0.946 0.980) &   \hfill $<$0.22 &  3882.84\\ 
\hline
\ \ \ \  & 0.0226 (0.0219 0.0233) & 0.116  (0.112 0.119) & 1.039 (1.036 1.044) & 0.086 (0.072 0.101) &   69 (  67   71) & 0.71 (0.69 0.73) & 3.21 (3.15 3.28) & 0.959 (0.943 0.975) &      &  3880.82\\ 
\hline
\ \ \ \   CMB  \hfill $\gamma=0$ & 0.0236 (0.0229 0.0243) & 0.104  (0.098 0.110) & 1.044 (1.041 1.046) & 0.092 (0.077 0.107) &   76 (  73   80) & 0.78 (0.75 0.81) & 3.21 (3.15 3.28) & 0.998 (0.978 1.019) &   \hfill $<$0.28 &  3859.56\\ 
\hline
\ \ \ \   \hfill $\gamma=0$& 0.0233 (0.0226 0.0240) & 0.106  (0.100 0.112) & 1.043 (1.041 1.046) & 0.090 (0.075 0.106) &   75 (  72   78) & 0.77 (0.74 0.80) & 3.23 (3.17 3.30) & 0.985 (0.967 1.008) &      &  3857.92\\ 
\hline
\ \ \ \   \hfill $\gamma=-1$& 0.0223 (0.0217 0.0229) & 0.116  (0.110 0.122) & 1.037 (1.035 1.040) & 0.091 (0.076 0.107) &   68 (  65   71) & 0.70 (0.66 0.73) & 3.20 (3.15 3.24) & 0.943 (0.926 0.960) &      &  3865.88\\ 
\hline
\ \ \ \   \hfill $\gamma=+1$& 0.0219 (0.0213 0.0225) & 0.105  (0.100 0.111) & 1.049 (1.046 1.054) & 0.089 (0.075 0.105) &   76 (  73   79) & 0.78 (0.75 0.80) & 3.10 (3.06 3.13) & 0.979 (0.965 0.995) &      &  3861.90\\ 
\hline
%
\ \ \ \  Model I & 0.0220 & 0.103 & 1.052 & 0.094 &  78 & 0.80 & 3.08 & 0.978 &  0.12    &  --- \\ 
\hline
 \ Adiabatic  & & & & & & & & & & \\

\hline
\ \ \ \  CMB & 0.0229 (0.0223 0.0235) & 0.107  (0.102 0.113) & 1.041 (1.039 1.044) & 0.089 (0.075 0.105) &   73 (  70   76) & 0.76 (0.73 0.78) & 3.11 (3.07 3.15) & 0.977 (0.961 0.995) &   \hfill $<$0.26 &  3857.97\\ 
\hline
\ \ \ \  & 0.0225 (0.0221 0.0230) & 0.111  (0.106 0.115) & 1.041 (1.038 1.043) & 0.088 (0.073 0.103) &   71 (  69   74) & 0.74 (0.71 0.76) & 3.13 (3.10 3.16) & 0.964 (0.952 0.977) &      &  3856.57\\ 
\hline
\ \ \ \ CMB\&SN& 0.0228 (0.0222 0.0233) & 0.109  (0.105 0.114) & 1.041 (1.039 1.043) & 0.088 (0.074 0.102) &   72 (  70   74) & 0.75 (0.72 0.77) & 3.12 (3.08 3.15) & 0.972 (0.958 0.988) &   \hfill $<$0.22 &  4123.95\\ 
\hline
\ \ \ \  & 0.0225 (0.0220 0.0230) & 0.111  (0.107 0.116) & 1.041 (1.038 1.043) & 0.087 (0.074 0.101) &   71 (  69   73) & 0.74 (0.71 0.76) & 3.13 (3.10 3.17) & 0.963 (0.952 0.975) &      &  4121.96\\ 
\hline
\ \ \ \ CMB\&MPK& 0.0225 (0.0220 0.0230) & 0.115  (0.111 0.118) & 1.040 (1.038 1.042) & 0.083 (0.070 0.098) &   70 (  68   71) & 0.72 (0.70 0.73) & 3.14 (3.11 3.17) & 0.966 (0.954 0.980) &   \hfill $<$0.17 &  3877.99\\ 
\hline
\ \ \ \  & 0.0223 (0.0219 0.0228) & 0.116  (0.112 0.119) & 1.040 (1.038 1.042) & 0.083 (0.070 0.098) &   69 (  67   71) & 0.71 (0.69 0.73) & 3.15 (3.12 3.18) & 0.961 (0.949 0.973) &      &  3875.85\\ 
\hline

\hline 
\hline
{\bf Slow-roll par.}  & \normalsize $\omega_b$ & \normalsize $\omega_c$ & \normalsize $100\theta$ & \normalsize $\tau$ & \normalsize $H_0$ & \normalsize $\Omega_\Lambda$ &  \normalsize $\ln (10^{10} A_0^2)$ & \normalsize $n_{\mathrm{ad}}^{\mathrm{eff}}$ & \normalsize $r_0$ & \normalsize ${}-\ln{\cal Z}$ \\
\hline

   \ Mixed model &  \hfill (68\% C.L.)&  \hfill (68\% C.L.)  &   \hfill (68\% C.L.)&  \hfill (68\% C.L.) &  \hfill (68\% C.L.)&   \hfill (68\% C.L.)&  \hfill (68\% C.L.)&  \hfill (68\% C.L.)& 95\% C.L.& \\
\hline
\ \ \ \ CMB& 0.0228 (0.0222 0.0234) & 0.109  (0.102 0.116) & 1.040 (1.037 1.044) & 0.090 (0.075 0.103) &   72 (  68   76) & 0.75 (0.70 0.78) & 3.14 (3.07 3.19) & 0.966 (0.945 0.991) &   \hfill $<$0.18 &  3859.86\\ 
\hline
\ \ \ \ & 0.0227 (0.0222 0.0232) & 0.109  (0.103 0.116) & 1.040 (1.037 1.044) & 0.088 (0.075 0.103) &   72 (  69   76) & 0.74 (0.71 0.78) & 3.14 (3.08 3.21) & 0.966 (0.944 0.988) &      &  3858.82\\ 
\hline
\ \ \ \  CMB\&SN& 0.0227 (0.0222 0.0232) & 0.111  (0.106 0.116) & 1.039 (1.037 1.042) & 0.089 (0.075 0.104) &   71 (  69   73) & 0.73 (0.71 0.76) & 3.15 (3.11 3.20) & 0.960 (0.944 0.977) &   \hfill $<$0.19 &  4125.68\\ 
\hline
\ \ \ \  CMB\&MPK& 0.0225 (0.0220 0.0230) & 0.117  (0.113 0.120) & 1.038 (1.036 1.040) & 0.088 (0.075 0.102) &   68 (  66   70) & 0.70 (0.68 0.72) & 3.18 (3.15 3.22) & 0.951 (0.937 0.965) &   \hfill $<$0.16 &  3878.24\\ 
\hline
\ \ \ \  & 0.0224 (0.0219 0.0229) & 0.117  (0.113 0.120) & 1.038 (1.036 1.040) & 0.086 (0.073 0.100) &   68 (  66   70) & 0.70 (0.68 0.72) & 3.19 (3.16 3.23) & 0.949 (0.936 0.963) &      &  3877.14\\ 
\hline
\ \ \ \  CMB \hfill $\gamma= 0$ & 0.0232 (0.0226 0.0239) & 0.105  (0.099 0.110) & 1.042 (1.040 1.044) & 0.088 (0.074 0.104) &   75 (  72   78) & 0.77 (0.74 0.80) & 3.12 (3.08 3.16) & 0.989 (0.971 1.009) &   \hfill $<$0.24 &  3859.54\\ 
\hline
\hfill $\gamma= 0$ & 0.0230 (0.0224 0.0236) & 0.107  (0.101 0.113) & 1.042 (1.039 1.044) & 0.087 (0.073 0.102) &   74 (  71   77) & 0.76 (0.73 0.79) & 3.14 (3.11 3.18) & 0.980 (0.963 0.999) &      &  3856.66\\ 
\hline
 \hfill $\gamma=-1$ & 0.0223 (0.0218 0.0229) & 0.115  (0.109 0.121) & 1.038 (1.036 1.041) & 0.089 (0.075 0.106) &   69 (  66   71) & 0.71 (0.67 0.74) & 3.18 (3.14 3.23) & 0.948 (0.931 0.963) &      &  3859.92\\ 
\hline
 \hfill $\gamma=+1$ & 0.0228 (0.0223 0.0234) & 0.105  (0.100 0.111) & 1.043 (1.041 1.046) & 0.086 (0.072 0.101) &   75 (  72   78) & 0.77 (0.74 0.80) & 3.09 (3.05 3.13) & 0.984 (0.968 0.999) &      &  3859.99\\ 
\hline
\ \ \ \ Model II& 0.0221 & 0.115 & 1.037 & 0.087 & 68 & 0.70 & 3.20 & 0.927 & 0.015 & --- \\
\hline
 \ Adiabatic  & & & & & & & & & & \\
\hline
\ \ \ \ CMB& 0.0229 (0.0223 0.0235) & 0.107  (0.102 0.113) & 1.041 (1.039 1.044) & 0.088 (0.075 0.104) &   73 (  71   76) & 0.76 (0.73 0.78) & 3.10 (3.07 3.14) & 0.978 (0.962 0.995) &   \hfill $<$0.26 &  3856.67\\ 
\hline
\ \ \ \  & 0.0226 (0.0221 0.0231) & 0.110  (0.105 0.116) & 1.041 (1.039 1.043) & 0.088 (0.074 0.103) &   72 (  69   74) & 0.74 (0.71 0.76) & 3.13 (3.10 3.17) & 0.965 (0.953 0.978) &      &  3854.44\\ 
\hline
\ \ \ \  CMB\&SN& 0.0228 (0.0222 0.0233) & 0.109  (0.105 0.114) & 1.041 (1.039 1.043) & 0.088 (0.074 0.102) &   72 (  70   74) & 0.75 (0.72 0.77) & 3.12 (3.08 3.15) & 0.973 (0.959 0.987) &   \hfill $<$0.22 &  4122.71\\ 
\hline
\ \ \ \ CMB\&MPK& 0.0225 (0.0220 0.0230) & 0.115  (0.112 0.118) & 1.040 (1.038 1.042) & 0.083 (0.070 0.097) &   70 (  68   71) & 0.72 (0.70 0.73) & 3.14 (3.11 3.17) & 0.967 (0.955 0.980) &   \hfill $<$0.17 &  3876.69\\ 
\hline
\ \ \ \  & 0.0224 (0.0219 0.0229) & 0.116  (0.113 0.120) & 1.040 (1.038 1.042) & 0.083 (0.070 0.098) &   69 (  67   71) & 0.71 (0.69 0.73) & 3.15 (3.12 3.18) & 0.961 (0.949 0.973) &      &  3874.03\\ 
\hline
\end{tabular}
\caption{The median values and 68\% C.L. intervals (in parenthesis) with various combinations of data for a selection of parameters that exist both in the mixed model and in the pure adiabatic model. Since there is no dectection of primordial tensor perturbations, 95\% C.L. upper limits are reported for $r_0$.
The last column gives ${}-\ln{\cal Z}$, where ${\cal Z}$ is the Bayesian evidence that is proportional to the total probability of the model. Whenever there are two lines for the given model and data sets, the first line represents the results with and the second line without primordial tensor perturbations, as is obvious from the $r_0$ column. For $\gamma=\pm 1$ we have automatically $r_0=0$, so these models are without tensor contribution. 
Models I and II are example models with high nonadiabaticity (see Table~\ref{tab:IsocParams}), but within ``$2\sigma$'' from the best-fit adiabatic model. Their $C_\ell$ spectra are shown in Fig~\ref{fig:ExampleModels}, and matter power spectrum of Model I in Fig.~\ref{fig:BAO}(b).
 \label{tab:CommonParams}}
\end{table*}
\end{turnpage}


\begin{turnpage}
\begin{table*}
\scriptsize
\centering
\begin{tabular}{|l|l|l|l|l|l|l|l|l|l|l|l|}
\hline
 {\bf Ampl.\ par.} & \normalsize $100 \gamma_1$ & \normalsize $100 \gamma_2$ & \normalsize $100 \alpha_1$ & \normalsize $100 \alpha_2$ & \normalsize $100 \gamma_0$ & \normalsize $100 \alpha_0$ &  \normalsize $n_{\mathrm{ar}}$ & \normalsize $n_{\mathrm{as}}$ & \normalsize $n_{\mathrm{iso}}$ & \normalsize $100 \alpha_{\mathrm{cor0}}$  & \normalsize $100 \alpha_{\mathrm{T}}$\\
 \hline

 \ Mixed model &  \hfill (95\% C.L.)& 95\% C.L. & 95\% C.L.&  \hfill (95\% C.L.)&  \hfill (95\% C.L.)& 95\% C.L.&  \hfill (95\% C.L.)&  \hfill (95\% C.L.)&  \hfill (95\% C.L.)& \hfill (95\% C.L.)& \hfill (95\% C.L.)\\
\hline
\ \ \  CMB  & 16.5(-22.9 75.2) &   \hfill$<$69.0 &   \hfill$<$6.4 &   \hfill$<$51.2 &  16.8(-21.1 76.6) &   \hfill$<$15.4 & 0.989(0.878 1.082) & 0.997(0.505 1.434) & 2.146(0.871 3.304) &   8.0(-8.8 16.2) &   2.4(-3.0  4.9) \\

\hline
\ \ \ \hfill$r=0$   & 18.1(-13.7 80.7) &   \hfill$<$72.9 &   \hfill$<$6.1 &   \hfill$<$53.6 &  18.2(-9.7 81.4) &   \hfill$<$15.8 & 0.979(0.844 1.096) & 0.992(0.507 1.407) & 2.157(1.032 3.376) &   8.4(-7.2 17.0) &   2.6(-2.1  5.3) \\

\hline
\ \ \  CMB\&SN   & 10.4(-33.3 76.8) &   \hfill$<$68.7 &   \hfill$<$7.0 &   \hfill$<$52.3 &  10.7(-32.4 77.6) &   \hfill$<$13.7 & 0.979(0.877 1.065) & 0.973(0.392 1.330) & 2.138(0.418 3.434) &   5.9(-10.7 13.7) &   1.8(-4.8  4.2) \\

\hline
\ \ \ \hfill$r=0$    & 13.9(-24.3 77.7) &   \hfill$<$68.3 &   \hfill$<$6.7 &   \hfill$<$47.7 &  14.1(-21.8 77.0) &   \hfill$<$13.7 & 0.973(0.871 1.063) & 0.965(0.347 1.331) & 2.159(0.627 3.379) &   6.4(-9.0 14.0) &   2.0(-4.1  4.3) \\

\hline
\ \ \  CMB\&MPK   &-13.8(-61.3 12.0) &   \hfill$<$52.5 &   \hfill$<$10.2 &   \hfill$<$47.5 & -11.2(-61.1 11.4) &   \hfill$<$16.0 & 0.975(0.883 1.028) & 0.901(0.310 1.247) & 1.291(0.046 2.584) &  -6.4(-12.9  6.5) &  -1.9(-7.1  2.4) \\

\hline
\ \ \ \hfill$r=0$    & -8.3(-41.8 43.2) &   \hfill$<$42.0 &   \hfill$<$9.8 &   \hfill$<$48.2 &  -5.9(-41.0 42.7) &   \hfill$<$16.7 & 0.969(0.897 1.014) & 0.900(0.190 1.276) & 1.615(0.199 2.811) &  -5.4(-11.8  9.3) &  -1.1( -6.0  3.0) \\

\hline
 \ \ \   CMB  \hfill$\gamma=0$  & 0  & 0 &     \hfill$<$11.2 &  38.5(7.5 62.3) &     0               &   \hfill$<$24.9 & 0.998(0.962 1.042) &                    & 1.869(1.221 2.694) &                    &   \hfill$<$3.7 \\

\hline
 \ \ \   \hfill$r=0, \gamma=0$  & 0  & 0 &     \hfill$<$12.7 &   37.9(6.4 60.8) &    0                &   \hfill$<$25.8 & 0.985(0.950 1.032) &                    & 1.808(1.141 2.742) &                    &   \hfill$<$4.3 \\

\hline
\ \ \   \hfill$r=0,\gamma=-1$   &  -100 & -100 &     \hfill$<$1.1 &   \hfill$<$3.3 &      -100              &   \hfill$<$1.1 &                    & 0.943(0.909 0.975) & 0.957(-0.02 2.649) &  -5.8(-11.4 -1.9) & -3.8(-8.8 -0.7) \\

\hline
\ \ \   \hfill$r=0,\gamma=+1$   &  +100 & +100&     \hfill$<$0.8 &   \hfill$<$40.6 &     +100               &   \hfill$<$3.4 &                    & 0.979(0.952 1.011) & 2.477(1.411 3.885) &  11.0(4.7 19.5) &  \hfill$<$4.5 \\ 

%
\hline
\ \ \  Model I  & 66.2 & 73.0 & 0.0413 &  43.2 & 69.7  &  1.75 & 0.908 &  1.009 &  3.315 &  10.9  & 2.5 \\ 

\hline
 \ Adiabatic  & & & & & & & & & & &\\

\hline
\ \ \  CMB   &   & &             &           &                    &              & 0.977(0.947 1.016) &                    &                    &                    &                    \\

\hline
\ \ \ \hfill$r=0$    &   & &             &           &                    &              & 0.964(0.941 0.992) &                    &                    &                    &                    \\

\hline
\ \ \  CMB\&SN   &   & &             &           &                    &              & 0.972(0.946 1.004) &                    &                    &                    &                    \\

\hline
\ \ \ \hfill$r=0$     &   &  &             &           &                    &              & 0.963(0.941 0.986) &                    &                    &                    &                    \\

\hline
\ \ \  CMB\&MPK   &   & &             &           &                    &              & 0.966(0.943 0.994) &                    &                    &                    &                    \\

\hline
\ \ \ \hfill$r=0$   &   &  &             &           &                    &              & 0.961(0.937 0.983) &                    &                    &                    &                    \\

\hline
\hline
 {\bf Slow-roll par.} & \normalsize $100 \eta_{\sigma\sigma}$ & \normalsize $100 \eta_{\sigma s}$ & \normalsize $100 \eta_{s s}$ & \normalsize $100 \veps$ & \normalsize $100 \gamma_0$ & \normalsize $100 \alpha_0$ &  \normalsize $n_{\mathrm{ar}}$ & \normalsize $n_{\mathrm{as}}$ & \normalsize $n_{\mathrm{iso}}$ & \normalsize $100 \alpha_{\mathrm{cor0}}$  & \normalsize $100 \alpha_{\mathrm{T}}$\\
\hline

 \ Mixed model &  \hfill (95\% C.L.)&  \hfill (95\% C.L.)  &   \hfill (95\% C.L.)& 95\% C.L.  &  \hfill (95\% C.L.)& 95\% C.L.&  \hfill (95\% C.L.)&  \hfill (95\% C.L.)&  \hfill (95\% C.L.)& \hfill (95\% C.L.)& \hfill (95\% C.L.)\\

\hline
\ \ \  CMB  & \hfill p.r.  &  \hfill p.r. &   \hfill p.r. &   \hfill$<$4.8 &  \hfill p.r. &   \hfill$<$2.6 & 0.946(0.686 1.070) & 0.973(0.726 1.153) & 0.975(0.845 1.121) &  -3.4(-10.1  9.2) &  -1.7(-5.8  4.5) \\

\hline
\ \ \ \hfill$\veps=0$    & -1.4( -6.1  3.5) &  -0.5(-5.3  5.4) & \hfill p.r. &   0   &  \hfill p.r.  &   \hfill$<$6.8 & 0.971(0.878 1.071) & 0.957(0.314 1.243) & \hfill ``p.r.'' &  -2.1( -9.2  9.2) &  -0.6( -5.0  4.7) \\

\hline
\ \ \  CMB \&SN   &  \hfill p.r. & \hfill p.r.  &   \hfill p.r. &   \hfill$<$5.0 & \hfill$<$65.9 &   \hfill$<$3.2 & 0.949(0.672 1.059) & 0.962(0.663 1.117) & 0.967(0.841 1.115) &  -4.3(-10.4  5.8) &  -2.3(-5.6  3.0) \\

\hline
\ \ \ CMB\&MPK   &  \hfill p.r. &  \hfill p.r. &  \hfill p.r. &   \hfill$<$5.0 & \hfill$<$-12.9  &   \hfill$<$3.4 & 0.932(0.638 1.058) & 0.954(0.752 1.052) & 0.954(0.841 1.111) &  -5.7(-11.1 -1.6) &  -3.2(-6.3 -0.8) \\ 

\hline
\ \ \  \hfill$\veps=0$    &  -1.7(-5.8  3.6) &  -1.3(-6.0 2.7)  &  \hfill p.r. &  0  & -25.1(-89.4 0.1)  &   \hfill$<$7.3 & 0.965(0.885 1.073) & 0.919(0.396 1.147) & \hfill ``p.r.'' &  -5.1(-10.3  0.2) &  -2.5(-5.8  0.8) \\

\hline
\ \ \   CMB  \hfill$\gamma=0$   &  0.9(-1.8  5.9) &           &   \hfill p.r. &   \hfill$<$1.5 &      0              &   \hfill$<$11.0 & 0.989(0.956 1.030) & & 1.020(0.849 1.136) &                    &   \hfill$<$3.9 \\

\hline
\ \ \   \hfill$\veps=0, \gamma=0$   & -1.0(-2.6  1.1) &           &  \hfill p.r. &  0    &   0                 &   \hfill$<$12.5 & 0.980(0.948 1.022) &  &  \hfill ``p.r.'' &                    &   \hfill$<$4.1 \\ 

\hline
\ \ \   \hfill$\veps=0, \gamma=-1$   &    &           & -2.6(-4.3 -1.1)  &  0    &  -100                  &   \hfill$<$0.9 &                    & 0.948(0.915 0.978) &  0.948(0.915 0.978) &  -4.7(-9.9 -0.9) & -2.8(-6.4 -0.5) \\ 

\hline
\ \ \   \hfill$\veps=0, \gamma=+1$   &  &           &  -0.8(-2.3 0.8)&  0    &    +100                &   \hfill$<$1.2 &                    & 0.984(0.954 1.016) & 0.984(0.954 1.016) &   4.9(0.9 12.1) &   \hfill$<$5.2 \\ 

\hline
\ \ \ Model II & 1.8 & -0.9 & -7.2 & 0.18 & -46.1 & 1.02 & 1.026  &  0.812 &    0.852 & -6.84 & -4.9 \\
\hline
 \ Adiabatic  & & & & & & & & & & &\\

\hline
\ \ \ CMB   &  0.5(-2.2  5.9) &           &           &   \hfill$<$1.6 &                    &              & 0.978(0.947 1.013) &                    &                    &                    &                    \\

\hline
\ \ \  \hfill$\veps=0$   & -1.7(-3.0 -0.5) &           &           &   0   &                    &              & 0.965(0.939 0.991) &                    &                    &              
      &                    \\

\hline
\ \ \  CMB\&SN   & -0.1(-2.3  4.4) &           &           &   \hfill$<$1.4 &                    &              & 0.973(0.947 1.002) &                    &                    &                    &                    \\

\hline
\ \ \  CMB\&MPK   & -0.7(-2.5  3.0) &           &           &   \hfill$<$1.1 &                    &              & 0.967(0.942 0.994) &                    &                    &                    &                    \\ 

\hline
\ \ \  \hfill$\veps=0$    &   -2.0 ( -3.1 -0.8)  &           &           &  0 &                    &              & 0.961(0.938 0.983) &                    &                    &                    &                    \\

\hline

\end{tabular}
\caption{The median values and 95\% C.L. intervals (in parenthesis) or upper
  limits with various combinations of data for a selection of parameters
  especially needed in the mixed model. The spectral index of the
  (uncorrelated part of) curvature perturbation $\nadI$ and the slow-roll
  parameters $\eta_{\sigma\sigma}$ and $\veps$ are present also in the pure
  adiabatic model. Whenever there are two lines for the given model and data
  sets, the first line represents the results with and the second line without
  primordial tensor perturbations. In some cases the data do not constrain
  some parameters, so the whole prior range is allowed. This is indicated by
  ``p.r.'', and the prior range is given in Table~\ref{tab:parameters}.
Models I and II are example models with high nonadiabaticity, but within
``$2\sigma$'' from the best-fit adiabatic model.
\label{tab:IsocParams}}
\end{table*}
\end{turnpage}

\end{document}